# A Semantic Analysis of Key Management Protocols for Wireless Sensor Networks*

Francesco Ballardin, Damiano Macedonio, Massimo Merro, Mattia Tirapelle

*Dipartimento di Informatica, Università degli Studi di Verona, Italy*

**Abstract**

We propose a simple *timed broadcasting process calculus* for modelling wireless network protocols. The operational semantics of our calculus is given in terms of a labelled transition semantics which is used to derive a standard (weak) *bi-simulation theory*. Based on our simulation theory, we reformulate Gorrieri and Martinelli's *timed Generalized Non-Deducibility on Compositions* (*tGNDC*) scheme, a well-known general framework for the definition of timed properties of security protocols. We use *tGNDC* to perform a semantic analysis of three well-known *key management protocols* for wireless sensor networks: $\mu$TESLA, LEAP+ and LiSP. As a main result, we provide a number of *attacks* to these protocols which, to our knowledge, have not yet appeared in the literature.

## 1 Introduction

Wireless sensors are small and cheap devices powered by low-energy batteries, equipped with radio transceivers, and responding to physical stimuli, such as pressure, magnetism and motion, by emitting radio signals. Such devices are featured with *resource constraints* (involving power, storage and computation) and low transmission rates. *Wireless sensor networks* (WSNs) are large-scale networks of sensor nodes deployed in strategic areas to gather data. Sensor nodes collaborate using wireless communications with an asymmetric many-to-one data transfer model. Typically, they send their sensed events or data to a specific node, called sink node or base station, which collects the requested information. WSNs are primarily designed for monitoring environments that humans cannot easily reach (e.g., motion, target tracking, fire detection, chemicals, temperature); they are used as embedded systems (e.g., biomedical sensor engineering, smart homes) or mobile applications (e.g., when attached to robots, soldiers, or vehicles).

An important issue in WSNs is *network security*: Sensor nodes are vulnerable to several kinds of threats and risks. Unlike wired networks, wireless devices use radio frequency channels to broadcast their messages. An adversary can compromise a sensor node, alter the integrity of the data, eavesdrop on messages, inject fake messages, and waste network resource.

*The second author is supported by research fellowship n. AdR1601/11, funded by Dipartimento di Informatica Verona.

Thus, one of the challenges in developing trustworthy WSNs is to provide high-security features with limited resources.

Generally, in order to have a secure communication between two (or more) parties, a secure association must be established by sharing a secret. This secret must be created, distributed and updated by one (or more) entity and it is often represented by the knowledge of a *cryptographic key*. The management of such cryptographic keys is the core of any security protocol. Due to resource limitations, all key management protocols for WSNs, such as $\mu$TESLA [32], LiSP [31], LEAP [43], PEBL [5] and INF [1], are based on *symmetric cryptography* rather than heavy public-key schemes, such as Diffie-Hellman [7] and RSA [34].

In this paper, we adopt a process calculus approach to formalise and verify real-world key management protocols for WSNs. A process calculus is a formal and concise language that allows us to express system behaviour in the form of a process term. In the last years, a number of distributed process calculi have been proposed for modelling different aspects of wireless systems [21, 28, 35, 24, 14, 10, 25, 13]. Except for [28], none of these calculi performs any security analysis. On the other hand, some process algebras, such as CryptoCCS and tCryptoSPA [15] have already been used in [15, 16] to study network security protocols, also in a wireless scenario. These calculi are extensions of Milner's CCS [26], where node distribution, local broadcast communication, and message loss are codified in terms of point-to-point transmission and a (discrete) notion of time.

We propose a simple *timed broadcasting process calculus*, called aTCWS, for modelling wireless network protocols. Our broadcast communications span over a limited area, called *transmission range*. The time model we use is known as the *fictitious clock* approach (see e.g. [17]): A global clock is supposed to be updated whenever all nodes agree on this, by globally synchronising on a special timing action $\sigma$.[1] Both transmission and internal actions are assumed to take no time. This is a reasonable assumption whenever the duration of those actions is negligible with respect to the chosen time unit. The operational semantics of our calculus is given in terms of a labelled transition semantics in the SOS style of Plotkin. The calculus enjoys standard time properties, such as: *time determinism*, *maximal progress* and *patience* [17]. The labelled transition semantics is then used to derive a standard (weak) *bisimulation theory*.

Based on our simulation theory, we reformulate Gorrieri and Martinelli's *timed Generalized Non-Deducibility on Compositions* (*tGNDC*) scheme [15, 16], a well-known general framework for the definition of timed security properties. We concentrate on two particular timed security properties expressed as instances of *tGNDC*: *timed integrity*, which guarantees on the freshness of authenticated packets; and *timed agreement*, for which agreement between two parties must be reached within a certain deadline. A nice aspect of these two properties is that whenever they do not hold then it is possible to build a specific *attacker* that invalidates the property under examination.

We use our calculus to provide a formal specification of three well-known key management protocols for WSNs: (i) $\mu$TESLA [32], which achieves *authenticated broadcast*; (ii) the *Localized Encryption and Authentication Protocol*, LEAP+ [43], intended for large-scale wireless sensor networks; (iii) the *Lightweight Security Protocol*, LiSP [31], that, through an efficient mechanism of re-keying, provides a good trade-off between resource consumption and network security.

[1] Time synchronisation relies on the presence of some clock synchronisation protocol for sensor networks [37].

**Table 1** Syntax of aTCWS.

| | | | |
|---|---|---|---|
| *Networks:* | | | |
| $M, N$ | $::=$ | $\mathbf{0}$ | empty network |
| | $\mid$ | $M_1 \mid M_2$ | parallel composition |
| | $\mid$ | $n[P]^{\nu}$ | node |
| *Processes:* | | | |
| $P, Q$ | $::=$ | $\mathsf{nil}$ | termination |
| | $\mid$ | $!\langle u \rangle.P$ | broadcast |
| | $\mid$ | $\lfloor ?(x).P \rfloor Q$ | receiver with timeout |
| | $\mid$ | $\lfloor \sum_{i \in I} \tau.P_i \rfloor Q$ | internal choice with timeout |
| | $\mid$ | $\sigma.P$ | sleep |
| | $\mid$ | $[u_1 = u_2]P; Q$ | matching |
| | $\mid$ | $[u_1 \ldots u_n \vdash_r x]P; Q$ | deduction |
| | $\mid$ | $H\langle \tilde{u} \rangle$ | guarded recursion |

As a main result of the paper, we formally prove that the *bootstrapping phase* of $\mu$TESLA enjoys the timed integrity property, while it does not satisfy timed agreement as it is exposed to a *replay attack*. Once bootstrapping is terminated, the core of the protocol, i.e. the *authenticated-broadcast phase*, enjoys both timed integrity and timed agreement. Then, we prove that the *single-hop pairwise shared key* mechanism of LEAP+ enjoys timed integrity, while it does not respect timed agreement due to the presence of another replay attack, despite the security assessment of [43]. Finally, we prove that the LiSP protocol does not satisfy neither timed integrity nor timed agreement. Again, our proof relies on the exhibition of a replay attack to the protocol. To our knowledge all these attacks are new and they have not yet appeared in the literature.

We end this introduction with an outline of the paper. In Section 2, we provide syntax, operational semantics and behavioural semantics of aTCWS. In the same section we prove that our calculus enjoys time determinism, maximal progress and patience. In Section 3, we adapt Gorrieri and Martinelli's *tGNDC* framework to aTCWS. In Sections 4, 5 and 6 we provide a security analysis of the three key management protocols mentioned above. The paper ends with a section on conclusions, future and related work.

## 2 The Calculus

In Table 1, we provide the syntax of our *applied Timed Calculus for Wireless Systems*, in short aTCWS, in a two-level structure: A lower one for *processes* and an upper one for *networks*.

We assume a set *Nds* of logical node names, ranged over by letters $m, n$. *Var* is the set of *variables*, ranged over by $x, y, z$. We define *Val* to be the set of values, and *Msg* to be the set of *messages*, i.e., closed values that do not contain variables. Letters $u, u_1 \ldots$ range over *Val*, and $v, w \ldots$ range over *Msg*. We assume a class of *message constructors* ranged over by $\mathrm{F}^i$.

Both syntax and operational semantics of aTCWS are parametric with respect to a given *decidable* inference system, i.e. a set of rules to model operations on messages by using constructors. For instance, the rules

$$(\text{pair})\ \frac{w_1 \quad w_2}{\text{pair}(w_1, w_2)} \qquad (\text{fst})\ \frac{\text{pair}(w_1, w_2)}{w_1} \qquad (\text{snd})\ \frac{\text{pair}(w_1, w_2)}{w_2}$$

allow us to deal with pairs of values. We write $w_1 \ \ldots \ w_k \vdash_r w_0$ to denote an application of rule $r$ to the closed values $w_1 \ldots w_k$ to infer $w_0$. Given an inference system, the *deduction function* $\mathcal{D} : 2^{Msg} \to 2^{Msg}$ associates a (finite) set $\phi$ of messages to the set $\mathcal{D}(\phi)$ of messages that can be deduced from $\phi$, by applying instances of the rules of the inference system.

In aTCWS, networks are collections of nodes (which represent devices) running in parallel and using a unique common channel to communicate with each other. All nodes are assumed to have the same transmission range (this is a quite common assumption in models for sensor networks [27]). The communication paradigm is *local broadcast*: only nodes located in the range of the transmitter may receive data. We write $n[P]^\nu$ for a node named $n$ (the device network address) executing the sequential process $P$. The tag $\nu$ contains (the names of) the neighbours of $n$ ($\nu \subseteq Nds$). In other words, $\nu$ contains all nodes in the transmission cell of $n$ (except $n$ itself), thus modelling the network topology.[2] For simplicity, when $\nu = \{m\}$ we will omit parentheses. Our wireless networks have a fixed topology as node mobility is not relevant to most sensor networks. Moreover, nodes cannot be created or destroyed.

Processes are sequential and live within the nodes. We let *Prc* be the set of all possible processes. We write nil to denote the skip process. The sender process $!\langle w \rangle.P$ allows to broadcast the message $w$, the continuation being $P$. The process $\lfloor ?(x).P \rfloor Q$ denotes a receiver with timeout. Upon successful reception, the variable $x$ of $P$ is instantiated with the received message. The process $\lfloor \sum_{i \in I} \tau.P_i \rfloor Q$ denotes internal choice with timeout. The process $\sigma.P$ models sleeping for the current time slot. The process $[w_1 = w_2]P; Q$ is the standard "if then else" construct: it behaves as $P$ if $w_1 = w_2$, and as $Q$ otherwise. The process $[w_1 \ldots w_k \vdash_r x]P; Q$ is the inference construct. It tries to infer a message $w$ from the premises $w_1 \ldots w_k$ through an application of rule $r$; if it succeeds, then it behaves as $P$ (where $w$ replaces $x$), otherwise it behaves as $Q$.

In the processes $!\langle w \rangle.P$, $\lfloor ?(x).P \rfloor Q$, $\lfloor \sum_{i \in I} \tau.P_i \rfloor Q$ and $\sigma.Q$, the occurrences of $P$, $P_i$ and $Q$ are said to be *guarded*; the occurrences of $Q$ are also said to be *time-guarded*. In the processes $\lfloor ?(x).P \rfloor Q$ and $[w_1 \ldots w_n \vdash_r x]P$ the variable $x$ is said to be *bound* in $P$. A variable which is not bound is said to be *free*. We adopt the standard notion of $\alpha$-conversion on bound variables and we identify processes up to $\alpha$-conversion. We assume there are no free variables in our networks. The absence of free variables will be maintained as networks evolve. We write $\{^w/_x\}P$ for the substitution of the variable $x$ with the message $w$ in $P$.

[2] We could have represented the topology in terms of a restriction operator à la CCS on node names; we have preferred our notation to keep at hand the neighbours of a node.

In order to deal with (guarded) recursion, we assume a set *PrcIds* of process identifiers ranged over by $H, H_1, H_2 \ldots$, and we write $H\langle w_1, \ldots, w_k\rangle$ to denote a process defined via an equation $H(x_1, \ldots, x_k) \stackrel{\text{def}}{=} P$, where (i) the tuple $x_1, \ldots, x_k$ contains all the variables that appear free in $P$, and (ii) $P$ contains only guarded occurrences of the process identifiers, such as $H$ itself. We say that recursion is *time-guarded* if $P$ contains only time-guarded occurrences of the process identifiers. We write $Prc_{\text{wt}}$ for the set of processes in which summations are finite-indexed and recursive definitions are time-guarded.

**Remark 2.1** *The recursion construct allows us to define persistent listeners, i.e., receivers which wait indefinitely for an incoming message, as* $Rcv \stackrel{\text{def}}{=} \lfloor ?(x).P \rfloor Rcv$*; similarly, internal choice (without timeout) can be defined as* $Sum \stackrel{\text{def}}{=} \lfloor \sum_{i \in I} \tau.P_i \rfloor Sum$.

We report some notational *conventions*. We write $\prod_{i \in I} M_i$ to mean the parallel composition of all $M_i$, for $i \in I$. We identify $\prod_{i \in I} M_i = \mathbf{0}$ if $I = \emptyset$. The process $[w_1 = w_2]P$ is an abbreviation for $[w_1 = w_2]P; \mathsf{nil}$. Similarly, we will write $[w_1 \ldots w_n \vdash_r x]P$ to mean $[w_1 \ldots w_n \vdash_r x]P; \mathsf{nil}$.

In the sequel, we will make use of a standard notion of structural congruence to abstract over processes that differ for minor syntactic differences.

**Definition 2.2** Structural congruence *over networks, written* $\equiv$*, is defined as the smallest equivalence relation, preserved by parallel composition, which is a commutative monoid with respect to parallel composition and internal choice, and for which* $n[H\langle \tilde{w}\rangle]^\nu \equiv n[\{^{\tilde{w}}/_{\tilde{x}}\}P]^\nu$*, if* $H(\tilde{x}) \stackrel{\text{def}}{=} P$.

Here, we provide some definitions that will be useful in the remainder of the paper. Given a network $M$, $\mathsf{nds}(M)$ returns the node names of $M$. More formally:

$$\mathsf{nds}(\mathbf{0}) \stackrel{\text{def}}{=} \emptyset; \qquad \mathsf{nds}(n[P]^\nu) \stackrel{\text{def}}{=} \{n\}; \qquad \mathsf{nds}(M_1 \mid M_2) \stackrel{\text{def}}{=} \mathsf{nds}(M_1) \cup \mathsf{nds}(M_2) \ .$$

For $m \in \mathsf{nds}(M)$, the function $\mathsf{ngh}(m, M)$ returns the set of the neighbours of $m$ in $M$. Thus, if $M \equiv m[P]^\nu \mid N$ then $\mathsf{ngh}(m, M) = \nu$. We write $\mathsf{Env}(M)$ to mean all the nodes of the environment reachable by the network $M$. Formally: $\mathsf{Env}(M) \stackrel{\text{def}}{=} \cup_{m \in \mathsf{nds}(M)} \mathsf{ngh}(m, M) \setminus \mathsf{nds}(M)$.

The syntax provided in Table 1 allows us to derive networks which are somehow ill-formed. The following definition identifies well-formed networks. Basically, it (i) rules out networks containing two nodes with the same name; (ii) imposes symmetric neighbouring relations (we recall that all nodes have the same transmission range); (iii) imposes network connectivity to allow clock synchronisation.

**Definition 2.3 (Well-formedness)** *$M$ is said to be* well-formed *if*

- *whenever* $M \equiv N \mid m_1[P_1]^{\nu_1} \mid m_2[P_2]^{\nu_2}$ *then* $m_1 \neq m_2$
- *whenever* $M \equiv N \mid m_1[P_1]^{\nu_1} \mid m_2[P_2]^{\nu_2}$*, with* $m_1 \in \nu_2$*, then* $m_2 \in \nu_1$
- *for all* $m, n \in \mathsf{nds}(M)$ *there are* $m_1, \ldots, m_k \in \mathsf{nds}(M)$*, such that* $m{=}m_1$*,* $n{=}m_k$*,* $m_i \in \mathsf{ngh}(m_{i+1}, M)$*, for* $1 {\leq} i {\leq} k{-}1$.

We let *Net* be the set of well-formed networks. Henceforth, we will always work with networks in *Net*.

**Table 2** LTS - Transmissions, internal actions and time passing.

$$(\text{Snd})\ \frac{-}{m[!\langle w\rangle.P]^{\nu} \xrightarrow{m!w\triangleright\nu} m[P]^{\nu}} \qquad (\text{Rcv})\ \frac{m \in \nu}{n[\lfloor ?(x).P\rfloor Q]^{\nu} \xrightarrow{m?w} n[\{^{w}/_{x}\}P]^{\nu}}$$

$$(\text{RcvEnb})\ \frac{m \notin \mathsf{nds}(M)}{M \xrightarrow{m?w} M} \qquad (\text{RcvPar})\ \frac{M \xrightarrow{m?w} M' \quad N \xrightarrow{m?w} N'}{M \mid N \xrightarrow{m?w} M' \mid N'}$$

$$(\text{Bcast})\ \frac{M \xrightarrow{m!w\triangleright\nu} M' \quad N \xrightarrow{m?w} N' \quad \mu := \nu \backslash \mathsf{nds}(N)}{M \mid N \xrightarrow{m!w\triangleright\mu} M' \mid N'}$$

$$(\text{Tau})\ \frac{h \in I}{m[\lfloor \sum_{i\in I} \tau.P_i\rfloor Q]^{\nu} \xrightarrow{\tau} m[P_h]^{\nu}} \qquad (\text{TauPar})\ \frac{M \xrightarrow{\tau} M'}{M \mid N \xrightarrow{\tau} M' \mid N}$$

$$(\sigma\text{-nil})\ \frac{-}{n[\mathsf{nil}]^{\nu} \xrightarrow{\sigma} n[\mathsf{nil}]^{\nu}} \qquad (\text{Sleep})\ \frac{-}{n[\sigma.P]^{\nu} \xrightarrow{\sigma} n[P]^{\nu}}$$

$$(\sigma\text{-Rcv})\ \frac{-}{n[\lfloor ?(x).P\rfloor Q]^{\nu} \xrightarrow{\sigma} n[Q]^{\nu}} \qquad (\sigma\text{-Sum})\ \frac{-}{m[\lfloor \sum_{i\in I} \tau.P_i\rfloor Q]^{\nu} \xrightarrow{\sigma} m[Q]^{\nu}}$$

$$(\sigma\text{-Par})\ \frac{M \xrightarrow{\sigma} M' \quad N \xrightarrow{\sigma} N'}{M \mid N \xrightarrow{\sigma} M' \mid N'} \qquad (\sigma\text{-}\mathbf{0})\ \frac{-}{\mathbf{0} \xrightarrow{\sigma} \mathbf{0}}$$

## 2.1 Labelled Transition Semantics

In Table 2 we provide a Labelled Transition System (LTS) for aTCWS in the SOS style of Plotkin. Intuitively, the computation proceeds in lock-step: between every global synchronisation all nodes proceeds asynchronously by performing actions with no duration, which represent either broadcast or input or internal actions. Communication proceeds even if there are no listeners: Transmission is a *non-blocking* action. Moreover, communication is *lossy* as some receivers within the range of the transmitter might not receive the message. This may be due to several reasons such as signal interferences or the presence of obstacles.

The metavariable $\lambda$ ranges over the set of labels $\{\tau, \sigma, m!w\triangleright\nu, m?w\}$ denoting internal action, time passing, broadcasting and reception. Let us comment on the transition rules of Table 2. In rule (Snd) a sender $m$ dispatches a message $w$ to its neighbours $\nu$, and then continues as $P$. In rule (Rcv) a receiver $n$ gets a message $w$ coming from a neighbour node $m$, and then evolves into process $P$, where all the occurrences of the variable $x$ are replaced with $w$. If no message is received in the current time slot, a timeout fires and the node $n$ will continue with process $Q$, according to the rule ($\sigma$-Rcv). The rule (RcvPar) models the composition of two networks receiving the same message from the same transmitter. Rule (RcvEnb) says that every node can synchronise with an external transmitter $m$. Notice that a node $n[\lfloor ?(x).P\rfloor Q]^{\nu}$ might execute rule (RcvEnb) instead of rule (Rcv). This is because a potential receiver may miss a message for several reasons (internal misbehaving, interferences, weak radio signal,

**Table 3** LTS - Matching, recursion and deduction

$$\text{(Then)}\ \frac{n[P]^{\nu} \xrightarrow{\lambda} n[P']^{\nu}}{n[[w = w]P; Q]^{\nu} \xrightarrow{\lambda} n[P']^{\nu}} \qquad \text{(Else)}\ \frac{n[Q]^{\nu} \xrightarrow{\lambda} n[Q']^{\nu} \quad w_1 \neq w_2}{n[[w_1 = w_2]P; Q]^{\nu} \xrightarrow{\lambda} n[Q']^{\nu}}$$

$$\text{(Rec)}\ \frac{n[\{\tilde{w}/\tilde{x}\}P]^{\nu} \xrightarrow{\lambda} n[P']^{\nu} \quad H(\tilde{x}) \stackrel{\text{def}}{=} P}{n[H\langle\tilde{w}\rangle]^{\nu} \xrightarrow{\lambda} n[P']^{\nu}}$$

$$\text{(DedTrue)}\ \frac{n[\{w/x\}P]^{\nu} \xrightarrow{\lambda} n[P']^{\nu} \ \ w_1 \ldots w_n \vdash_r w}{n[[w_1 \ldots w_n \vdash_r x]P; Q]^{\nu} \xrightarrow{\lambda} n[P']^{\nu}}$$

$$\text{(DedFalse)}\ \frac{n[Q]^{\nu} \xrightarrow{\lambda} n[Q']^{\nu} \ \ \nexists\, w.\ w_1 \ldots w_n \vdash_r w}{n[[w_1 \ldots w_n \vdash_r x]P; Q]^{\nu} \xrightarrow{\lambda} n[Q']^{\nu}}$$

etc); in this manner we model message loss. Rule (Bcast) models the propagation of messages on the broadcast channel. Note that this rule looses track of the neighbours of $m$ that are in $N$. Thus, in the label $m!w\triangleright\nu$ the set $\nu$ always contains the neighbours of $m$ which can receive the message $w$. Rule (Tau) models local computations within a node due to a nondeterministic internal choice. Rule (TauPar) propagates internal computations on parallel components. The remaining rules model the passage of time. Rule (Sleep) models sleeping for one time slot. Rules ($\sigma$-nil) and ($\sigma$-$\mathbf{0}$) are straightforward. Rule ($\sigma$-Rcv) models timeout on receivers, and similarly rule ($\sigma$-Sum) describes timeout on internal activities. Rule ($\sigma$-Par) models time synchronisation between parallel components. Rules (Bcast) and (TauPar) have their symmetric counterparts. Table 3 reports the straightforward rules for nodes containing matching, recursion or deduction.

Below, we report a number of basic properties of our LTS.

**Proposition 2.4** *Let $M$, $M_1$ and $M_2$ be well-formed networks.*

1. $m \notin \mathsf{nds}(M)$ *if and only if* $M \xrightarrow{m?w} N$, *for some network* $N$.
2. $M_1 \mid M_2 \xrightarrow{m?w} N$ *if and only if there are* $N_1$ *and* $N_2$ *such that* $M_1 \xrightarrow{m?w} N_1$, $M_2 \xrightarrow{m?w} N_2$ *with* $N = N_1 \mid N_2$.
3. *If* $M \xrightarrow{m!w\triangleright\mu} M'$ *then* $M \equiv m[!\langle w\rangle.P]^{\nu} \mid N$, *for some* $m$, $\nu$, $P$ *and* $N$ *such that* $m[!\langle w\rangle.P]^{\nu} \xrightarrow{m!w\triangleright\nu} m[P]^{\nu}$, $N \xrightarrow{m?w} N'$, $M' \equiv m[P]^{\nu} \mid N'$ *and* $\mu = \nu \setminus \mathsf{nds}(N)$.
4. *If* $M \xrightarrow{\tau} M'$ *then* $M \equiv m[\lfloor \sum_{i\in I} \tau.P_i \rfloor Q]^{\nu} \mid N$, *for some* $m$, $\nu$, $P_i$, $Q$ *and* $N$ *such that* $m[\lfloor \sum_{i\in I} \tau.P_i \rfloor Q]^{\nu} \xrightarrow{\tau} m[P_h]^{\nu}$, *for some* $h \in I$, *and* $M' \equiv m[P_h]^{\nu} \mid N$.
5. $M_1 \mid M_2 \xrightarrow{\sigma} N$ *if and only if there are* $N_1$ *and* $N_2$ *such that* $M_1 \xrightarrow{\sigma} N_1$, $M_2 \xrightarrow{\sigma} N_2$ *and* $N = N_1 \mid N_2$.

As the topology of our networks is static and nodes cannot be created or destroyed, it is easy to prove the following result.

**Proposition 2.5 (Well-formedness preservation)** *Let $M$ be a well-formed network. If $M \xrightarrow{\lambda} M'$ then $M'$ is a well-formed network.*

**Proof** By induction on the derivation of the transition $M \xrightarrow{\lambda} M'$. □

## 2.2 Time properties

Our calculus aTCWS enjoys some desirable time properties. Here, we outline the most significant ones. Proposition 2.6 formalises the deterministic nature of time passing: a network can reach at most one new state by executing a $\sigma$-action.

**Proposition 2.6 (Time Determinism)** *If $M$ is a well-formed network with $M \xrightarrow{\sigma} M'$ and $M \xrightarrow{\sigma} M''$, then $M'$ and $M''$ are syntactically the same.*

**Proof** By induction on the length of the proof of $M \xrightarrow{\sigma} M'$. □

Patience guarantees that a process will wait indefinitely until it can communicate [17]. In our setting, this means that if no transmissions can start then it must be possible to execute a $\sigma$-action to let time pass.

**Proposition 2.7 (Patience)** *Let $M \equiv \prod_{i \in I} m_i[P_i]^{\nu_i}$ be a well-formed network, such that for all $i \in I$ it holds that $m_i[P_i]^{\nu_i} \not\equiv m_i[!\langle w \rangle.Q_i]^{\nu_i}$, then there is a network $N$ such that $M \xrightarrow{\sigma} N$.*

**Proof** By induction on the structure of $M$. □

The maximal progress property says that processes communicate as soon as a possibility of communication arises [17]. In other words, the passage of time cannot block transmissions.

**Proposition 2.8 (Maximal Progress)** *Let $M$ be a well-formed network. If $M \equiv m[!\langle w \rangle.P]^{\nu} \mid N$ then $M \xrightarrow{\sigma} M'$ for no network $M'$.*

**Proof** By inspection on the rules that can be used to derive $M \xrightarrow{\sigma} M'$, because sender nodes cannot perform $\sigma$-actions. □

Basically, time cannot pass unless the specification itself explicitly asks for it. This approach provides a lot of power to the specification, which can precisely handle the flowing of time. Such an extra expressive power leads, as a drawback, to the possibility of abuses. For instance, infinite loops of broadcast actions or internal computations prevent time passing. The *well-timedness* (or *finite variability*) property [29] puts a limitation on the number of instantaneous actions that can fire between two contiguous $\sigma$-actions. Intuitively, well-timedness says that time passing never stops: Only a finite number of instantaneous actions can fire between two subsequent $\sigma$-actions.

**Definition 2.9 (Well-Timedness)** *A network $M$ satisfies well-timedness if there exists an upper bound $k \in \mathbb{N}$ such that whenever $M \xrightarrow{\lambda_1} \cdots \xrightarrow{\lambda_h}$ where $\lambda_j$ is not directly derived by an application of* (RcvEnb) *and $\lambda_j \neq \sigma$ (for $1 \leq j \leq h$) then $k \leq h$.*

The above definition takes into account only transitions denoting an active involvement of the network, that is why we have left out those transitions which can be derived by applying rule (RcvEnb). However, as aTCWS is basically a specification language, there is no harm in allowing specifications which do not respect well-timedness. Of course, when using our language to give a protocol implementation, then one must verify that the implementation satisfies well-timedness: No real-world service (even a attackers) can stop the passage of time.

The following proposition provides a criterion to check well-timedness. We recall that $Prc_{\rm wt}$ denotes the set of processes where summations are always finite-indexed and recursive definitions are always time-guarded.

**Proposition 2.10** *Let $M = \prod_{i \in I} m_i[P_i]^{\nu_i}$ be a network. If for all $i \in I$ we have $P_i \in Prc_{\rm wt}$ then $M$ satisfies well-timedness.*

**Proof** First notice that without an application of (RcvEnb) the network $M$ can perform only a finite number of transitions. Then proceed by induction on the structure of $M$. ◻

## 2.3 Behavioural Semantics

Based on the LTS of Section 2.1, we define a standard notion of *timed labelled bisimilarity* for aTCWS. In general, a bisimulation describes how two terms (in our case networks) can mimic each other actions. Here, we focus on weak equivalences, i.e., we abstract on internal actions of the system, thus we must distinguish between the transmissions which may be observed and those which may not be observed by the environment. We extend the set of rules of Table 2 with the following two rules:

$$\text{(Shh)}\ \frac{M \xrightarrow{m!w \triangleright \emptyset} M'}{M \xrightarrow{\tau} M'} \qquad\qquad \text{(Obs)}\ \frac{M \xrightarrow{m!w \triangleright \nu} M' \quad \nu \neq \emptyset}{M \xrightarrow{!w \triangleright \nu} M'}$$

Rule (Shh) models transmissions that cannot be observed because none of the potential receivers is in the environment. Rule (Obs) models transmissions of messages that can be received (and hence observed) by those nodes of the environment contained in $\nu$. Notice that the name of the transmitter is removed from the label. This is motivated by the fact that nodes may refuse to reveal their identities, e.g. for security reasons or limited sensory capabilities in perceiving these identities. Note also that in a derivation tree the rule (Obs) can only be applied at top-level.

In the remaining of the paper, the metavariable $\alpha$ will range over the following actions: $\tau$, $\sigma$, $!w \triangleright \nu$ and $m?w$. We adopt the standard notation for weak transitions: the relation $\Rightarrow$ denotes the reflexive and transitive closure of $\xrightarrow{\tau}$; the relation $\xRightarrow{\alpha}$ denotes $\Rightarrow \xrightarrow{\alpha} \Rightarrow$; the relation $\xRightarrow{\hat{\alpha}}$ denotes $\Rightarrow$ if $\alpha = \tau$ and $\xRightarrow{\alpha}$ otherwise.

**Definition 2.11 (Bi-similarity)** *A relation $\mathcal{R}$ over well-formed networks is a* simulation *if $M\ \mathcal{R}\ N$ implies that whenever $M \xrightarrow{\alpha} M'$ there is $N'$ such that $N \xRightarrow{\hat{\alpha}} N'$ and $M'\ \mathcal{R}\ N'$. A relation $\mathcal{R}$ is called* bisimulation *if both $\mathcal{R}$ and its converse are simulations. We say that $M$ and $N$ are* similar, *written $M \lesssim N$, if there is a simulation $\mathcal{R}$ such that $M\ \mathcal{R}\ N$. We say that $M$ and $N$ are* bisimilar, *written $M \approx N$, if there is a bisimulation $\mathcal{R}$ such that $M\ \mathcal{R}\ N$.*

The notions of similarity and bisimilarity between networks are congruences, as they are preserved by parallel composition. We give only the statement for bisimilarity. A similar statement holds for similarity.

**Theorem 2.12 ($\approx$ is a congruence)** *Let $M$ and $N$ be two well-formed networks such that $M \approx N$. Then $M \mid O \approx N \mid O$ for all networks $O$ such that $M \mid O$ and $N \mid O$ are well-formed.*

# 3 A Reformulation of *tGNDC* for Wireless Networks

In order to achieve a formal verification of key management protocols for WSNs, we adopt a general schema for the definition of timed security properties, called *timed Generalized Non-Deducibility on Compositions* (*tGNDC*) [15], a real-time generalisation of *Generalized Non-Deducibility on Compositions* (*GNDC*) [8]. The main idea is the following: a system $M$ is $tGNDC^{\rho(M)}$ if for every attacker $A$ the composed systems $M \mid A$ satisfies the specification $\rho(M)$, with respect to a given timed behavioural relation. The timed behavioural relation we will use in the following analysis is the *similarity* relation $\lesssim$ of Definition 2.11.

The *tGNDC* framework [15] was originally designed for an extension of Milner's CCS [26], where node distribution, local broadcast communication, and message loss are not primitives but codified in terms of point-to-point transmission and a (discrete) notion of time. In this section, we will reformulate *tGNDC* in our setting.

A distributed protocol involves a set of nodes $\mathcal{P} = \{m_1, \ldots m_k\}$ which may be potentially under attack, depending on the proximity to the attacker. This means that, in general, the *attacker* is a network composed by a number of, possibly colluding, nodes. In order to deal with the most general and adverse attacker we assume a set $\mathcal{A} = \{a_1, \ldots, a_k\}$ of *fresh* malicious nodes so that each node $m_i \in \mathcal{P}$ of the protocol is associated to a corresponding attacking node $a_i \in \mathcal{A}$ (for $i = 1, \ldots, k$). Every node in $\mathcal{A}$ is in touch both with the corresponding node in $\mathcal{P}$ and with the other nodes in $\mathcal{A}$.

**Definition 3.1 (Attacking Nodes)** *We say that $\mathcal{A} = \{a_1, \ldots, a_k\} \subseteq Nds$ is a set of attacking nodes for $\mathcal{P} = \{m_1, \ldots, m_k\} \subseteq Nds$ if and only if $\mathcal{A} \cap \mathcal{P} = \emptyset$. We say that $\mathcal{A}$ is a set of attacking nodes for the network $M$ if and only if $\mathcal{A}$ is a set of attacking nodes for $\mathsf{nds}(M)$ and $\mathcal{A} \cap \mathsf{Env}(M) = \emptyset$.*

In our setting, an attacker is parameterised both on the set of nodes $\mathcal{P}$ of the protocol under attack and on some initial knowledge $\phi_0$. During the execution of the protocol an attacker may increase its knowledge by grasping messages sent by the parties according to Dolev-Yao constrains.

The knowledge of a network is expressed by the set of messages that the network can manipulate. Thus, we write $\mathrm{msg}(P)$ to denote the set of the messages that appear in the process $P$. Formally, we define $\mathrm{msg}(P)$ as $\mathrm{msg}_{\emptyset}(P)$, where $\mathrm{msg}_S : Prc \to 2^{Msg}$, for $S \subseteq PrcIds$, is defined in Table 4 along the lines of [15]. Intuitively, $\mathrm{msg}_S$ is a function that visits recursively the sub-terms of $P$ and the body of the recursive definitions referred by $P$. The index $S$ is used to guarantee that the unwinding of every recursive definition is performed exactly once.

**Table 4** Function $\mathrm{msg}_S$

$$
\begin{aligned}
\mathrm{msg}_S(\mathsf{nil}) &\stackrel{\text{def}}{=} \emptyset \\
\mathrm{msg}_S(!\langle u\rangle.P) &\stackrel{\text{def}}{=} get(u) \cup \mathrm{msg}_S(P) \\
\mathrm{msg}_S(\lfloor ?(x).P \rfloor Q) &\stackrel{\text{def}}{=} \mathrm{msg}_S(P) \cup \mathrm{msg}_S(Q) \\
\mathrm{msg}_S(\lfloor \textstyle\sum_{i\in I} \tau.P_i \rfloor Q) &\stackrel{\text{def}}{=} \textstyle\bigcup_{i\in I} msg_S(P_i) \cup \mathrm{msg}_S(Q) \\
\mathrm{msg}_S(\sigma.P) &\stackrel{\text{def}}{=} \mathrm{msg}_S(P) \\
\mathrm{msg}_S([u_1 = u_2]P; Q) &\stackrel{\text{def}}{=} get(u_1) \cup get(u_2) \cup \mathrm{msg}_S(P) \cup \mathrm{msg}_S(Q) \\
\mathrm{msg}_S([u_1 \dots u_n \vdash_r x]P; Q) &\stackrel{\text{def}}{=} \textstyle\bigcup_{i=1}^{n} get(u_i) \cup \mathrm{msg}_S(P) \cup \mathrm{msg}_S(Q) \\
\mathrm{msg}_S(H\langle u_1 \dots u_r\rangle) &\stackrel{\text{def}}{=} \begin{cases} \bigcup_{i=1}^{r} get(u_i) \cup \mathrm{msg}_{S\cup\{H\}}(P) & \text{if } H(\tilde{x}) \stackrel{\text{def}}{=} P \text{ and } H \notin S \\ \bigcup_{i=1}^{r} get(u_i) & \text{otherwise} \end{cases}
\end{aligned}
$$

where $get : Val \to 2^{Msg}$ is defined as follows:

$$
\begin{aligned}
get(a) &\stackrel{\text{def}}{=} \{a\} \qquad \text{(basic message)} \\
get(x) &\stackrel{\text{def}}{=} \emptyset \qquad \text{(variable)} \\
get(\,\mathrm{F}^i(u_1, \dots, u_{k_i})\,) &\stackrel{\text{def}}{=} \begin{cases} \{\mathrm{F}^i(u_1, \dots, u_{k_i})\} \cup \{u_1 \dots u_{k_i}\} & \text{if } \mathrm{F}^i(u_1 \dots u_{k_i}) \in Msg \\ get(u_1) \cup \dots \cup get(u_{k_i}) & \text{otherwise.} \end{cases}
\end{aligned}
$$

A straightforward generalisation of $\mathrm{msg}_S$ to networks is the following:

$$
\mathrm{msg}(\mathbf{0}) \stackrel{\text{def}}{=} \emptyset\,; \quad \mathrm{msg}(n[P]^\nu) \stackrel{\text{def}}{=} \mathrm{msg}(P)\,; \quad \mathrm{msg}(M_1 \mid M_2) \stackrel{\text{def}}{=} \mathrm{msg}(M_1) \cup \mathrm{msg}(M_2)\,.
$$

Now, everything is in place to formally define our notion of attacking networks. For simplicity, in the rest of the paper, given a set of nodes $\mathcal{N}$ and a node $n$, we will write $\mathcal{N} \setminus n$ for $\mathcal{N} \setminus \{n\}$, and $\mathcal{N} \cup n$ for $\mathcal{N} \cup \{n\}$. Moreover, we will use the symbol $\uplus$ to denote *disjoint union*.

**Definition 3.2 (Attacker)** *Given a set of node names $\mathcal{P} = \{m_1, \dots, m_k\}$, a set $\mathcal{A} = \{a_1, \dots, a_k\}$ of attacking nodes for $\mathcal{P}$, and an initial knowledge $\phi_0 \subseteq Msg$, we define the set of attacking networks as follows:*

$$
\mathbb{A}^{\phi_0}_{\mathcal{A}/\mathcal{P}} \stackrel{\text{def}}{=} \left\{ \prod_{i=1}^{k} a_i[Q_i]^{\mu_i} \ : \ Q_i \in Prc_{\text{wt}},\ \mathrm{msg}(Q_i) \subseteq \mathcal{D}(\phi_0),\ \mu_i = (\mathcal{A} \setminus a_i) \cup m_i \right\} .
$$

**Remark 3.3** *By Proposition 2.10, the requirement $Q_i \in Prc_{\text{wt}}$ in the definition of $\mathbb{A}^{\phi_0}_{\mathcal{A}/\mathcal{P}}$ guarantees that our attackers respects well-timedness and hence cannot prevent the passage of time.*

Sometimes, for verification reasons, we will be interested in observing part of the protocol $M$ under examination. We will assume that the environment contains a fresh node $obs \notin \mathsf{nds}(M) \cup \mathsf{Env}(M) \cup \mathcal{A}$, that we call the 'observer', unknown to the attacker. For convenience, the observer *cannot* transmit: it can only receive messages.

**Definition 3.4** *Given a network $M = \prod_{i=1}^{k} m_i[P_i]^{\nu_i}$, picked a set $\mathcal{A} = \{a_1, \ldots, a_k\}$ of attacking nodes for $M$ and fixed a set $O \subseteq \mathsf{nds}(M)$ of nodes to be observed, we define:*

$$M_O^{\mathcal{A}} \stackrel{\text{def}}{=} \prod_{i=1}^{k} m_i[P_i]^{\nu_i'} \quad \textit{where} \quad \nu_i' \stackrel{\text{def}}{=} \begin{cases} (\nu_i \cap \mathsf{nds}(M)) \cup a_i \cup obs & \textit{if } m_i \in O \\ (\nu_i \cap \mathsf{nds}(M)) \cup a_i & \textit{otherwise.} \end{cases}$$

This definition expresses that (i) every node $m_i$ of the protocols has a dedicated attacker located at $a_i$, (ii) network and attacker are considered in *isolation*, without any external interference, (iii) only *obs* can observe the behaviour of nodes in $O$, (iv) node *obs* does not interfere with the protocol as it cannot transmit, (v) the behaviour of the nodes in $\mathsf{nds}(M) \setminus O$ is not observable. To ease the notation, whenever $O = \mathsf{nds}(M)$ we will write $M^{\mathcal{A}}$ instead of $M_{\mathsf{nds}(M)}^{\mathcal{A}}$ .

We can now formalise the *tGNDC* family properties as follows.

**Definition 3.5 (*tGNDC*)** *Given a network $M$, an initial knowledge $\phi_0$, a set $O \subseteq \mathsf{nds}(M)$ of nodes under observation and a network $\rho(M)$, representing the specification property for $M$, we write $M \in tGNDC_{\phi_0,O}^{\rho(M)}$ if and only if for all sets $\mathcal{A}$ of attacking nodes for $M$ it hods that*

$$M_O^{\mathcal{A}} \mid A \lesssim \rho(M) \quad \textit{for every } A \in \mathbb{A}_{\mathcal{A}/\mathsf{nds}(M)}^{\phi_0} .$$

It should be noticed that when showing that a system $M$ is $tGNDC_{\phi_0,O}^{\rho(M)}$, the universal quantification on attackers required by the definition makes the proof quite involved. Thus, we look for a sufficient condition for *tGNDC* which does not make use of the universal quantification. For this purpose, we rely on a timed notion of term stability [15]. Intuitively, a network $M$ is said to be *time-dependent stable* if the attacker cannot increase its knowledge in a indefinite way when $M$ runs in the space of a time slot. Thus, we can predict how the knowledge of the attacker evolves at each time slot. First, we need a formalisation of computation.

**Definition 3.6 (Execution trace)** *An* execution trace *is a sequence of labelled transitions. If $\Lambda$ is the sequence of actions $\alpha_1\alpha_2\ldots\alpha_n$, we write $M \stackrel{\Lambda}{\Longrightarrow} M'$ to mean $M \Rightarrow\xrightarrow{\alpha_1}\Rightarrow \cdots \Rightarrow\xrightarrow{\alpha_n}\Rightarrow M'$.*

In order to count how many time slots embraces an execution trace $\Lambda$, we define $\#^{\sigma}(\Lambda)$ to be the number of occurrences of $\sigma$-actions in $\Lambda$.

**Definition 3.7 (Time-dependent stability)** *A network $M$ is said to be* time-dependent stable *with respect to a sequence of knowledge $\{\phi_j\}_{j\geq 0}$ if whenever*

$$M^{\mathcal{A}} \mid A \stackrel{\Lambda}{\Longrightarrow} M' \mid A'$$

*where $\mathcal{A}$ is a set of attacking nodes for $M$, $A \in \mathbb{A}_{\mathcal{A}/\mathsf{nds}(M)}^{\phi_0}$, $\#^{\sigma}(\Lambda) = j$ and $\mathsf{nds}(M') = \mathsf{nds}(M)$, then $\mathrm{msg}(A') \subseteq \mathcal{D}(\phi_j)$.*

In other words, if $M$ is time-dependent stable with respect to $\{\phi_j\}_{j\geq 0}$ then, whenever $M^{\mathcal{A}} \mid A \stackrel{\Lambda}{\Longrightarrow} M' \mid A'$, with $\#^{\sigma}(\Lambda)=j$ and $\mathsf{nds}(M')=\mathsf{nds}(M)$, then $A' \in \mathbb{A}_{\mathcal{A}/\mathsf{nds}(M')}^{\phi_j}$. Thus, $\phi_j$ expresses the knowledge of the attacker at the end of the $j$-th time slot.

Time-dependent stability is the crucial notion that allows us to replace the universal quantification on the possible attackers with the most general attacker. Intuitively, given a sequence of knowledge $\{\phi_j\}_{j\geq 0}$ and a set $\mathcal{P} = \{m_1, \ldots, m_k\}$ of nodes we pick a set $\mathcal{A} = \{a_1, \ldots, a_k\}$ of attacking nodes for $\mathcal{P}$ and we define the top attacker $\textsc{Top}^{\phi_0}_{\mathcal{A}/\mathcal{P}}$ as the network whose initial knowledge is $\phi_0$ and which is able to manage the whole knowledge provided by $\phi_j$ after $j$ time slots.

**Definition 3.8 (Top Process)** *Given a sequence of knowledge $\{\phi_j\}_{j\geq 0}$ the set of top processes $\{\mathrm{T}_{\phi_j}\}_{j\geq 0}$ is defined as follows:*

$$\mathrm{T}_{\phi_j} \stackrel{\text{def}}{=} \lfloor \sum_{w \in \mathcal{D}(\phi_j)} \tau.!\langle w \rangle.\mathrm{T}_{\phi_j} \rfloor \mathrm{T}_{\phi_{j+1}} \ .$$

The top attacker is defined by replicating the top process in every attacking node.

**Definition 3.9 (Top Attacker)** *Given $\mathcal{P} = \{m_1, \ldots, m_k\}$ and $\mathcal{A} = \{a_1, \ldots, a_k\}$ set of attacking nodes for $\mathcal{P}$ and fixed a sequence of knowledge $\{\phi_j\}_{j\geq 0}$, the top attacker is defined as:*

$$\textsc{Top}^{\phi_j}_{\mathcal{A}/\mathcal{P}} \stackrel{\text{def}}{=} \prod_{i=1}^{k} a_i[\mathrm{T}_{\phi_j}]^{m_i} \ .$$

Basically, the network $\textsc{Top}^{\phi_j}_{\mathcal{A}/\mathcal{P}}$ can perform the following transitions:

- $\textsc{Top}^{\phi_j}_{\mathcal{A}/\mathcal{P}} \xRightarrow{a_i!w\triangleright m_i} \textsc{Top}^{\phi_j}_{\mathcal{A}/\mathcal{P}}$, for every $i \in \{1, \ldots, k\}$ and $w \in \mathcal{D}(\phi_j)$
- $\textsc{Top}^{\phi_j}_{\mathcal{A}/\mathcal{P}} \xrightarrow{\sigma} \textsc{Top}^{\phi_{j+1}}_{\mathcal{A}/\mathcal{P}}$ .

In particular, after $j$ time slots (i.e. $j$ $\sigma$-actions) $\textsc{Top}^{\phi_j}_{\mathcal{A}/\mathcal{P}}$ can *replay* any message in $\mathcal{D}(\phi_j)$ to the network under attack. Moreover, every attacking node $a_i$ can send messages to the corresponding node $m_i$, but, unlike the attackers of Definition 3.2, it does not need to communicate with the other nodes in $\mathcal{A}$ as it already owns the full knowledge of the system at time $j$.

**Remark 3.10** *Notice that the top attacker does not satisfy well-timedness (see Definition 2.9), as the process identifiers involved in the recursive definition are not time-guarded. However, this is not a problem as we are looking for a sufficient condition which ensures tGNDC with respect to well-timed attackers.*

A first compositional property that involves the top attacker is the following.

**Lemma 3.11** *Let $M_1 \mid M_2$ be time-dependent stable with respect to a sequence of knowledge $\{\phi_j\}_{j\geq 0}$. Let $\mathcal{A}_1$ and $\mathcal{A}_2$ be disjoint sets of attacking nodes for $M_1$ and $M_2$, respectively. Let $O_1 \subseteq \mathsf{nds}(M_1)$ and $O_2 \subseteq \mathsf{nds}(M_2)$. Then*

$$(M_1 \mid M_2)^{\mathcal{A}_1 \uplus \mathcal{A}_2}_{O_1 \uplus O_2} \mid \textsc{Top}^{\phi_0}_{\mathcal{A}_1 \uplus \mathcal{A}_2/\mathsf{nds}(M)} \ \lesssim \ M_1{}^{\mathcal{A}_1}_{O_1} \mid M_2{}^{\mathcal{A}_2}_{O_2} \mid \textsc{Top}^{\phi_0}_{\mathcal{A}_1/\mathsf{nds}(M_1)} \mid \textsc{Top}^{\phi_0}_{\mathcal{A}_2/\mathsf{nds}(M_2)} \ .$$

The following theorem says that $\textsc{Top}^{\phi_0}_{\mathcal{A}/\mathcal{P}}$ is the reference attacker for checking *tGNDC*.

**Theorem 3.12 (Criterion for *tGNDC*)** *If $M$ is time-dependent stable with respect to a sequence of knowledge $\{\phi_j\}_{j\geq 0}$, $\mathcal{A}$ is a set of attacking nodes for $M$ and $O \subseteq \mathsf{nds}(M)$, then*

$$M^{\mathcal{A}}_{O} \;\Big|\; \textsc{Top}^{\phi_0}_{\mathcal{A}/\mathsf{nds}(M)} \;\lesssim\; N \quad \textit{implies} \quad M \in tGNDC^{N}_{\phi_0,O} \; .$$

The notion of the most powerful attacker is eventually employed to obtain the compositional property outlined by the following proposition.

**Theorem 3.13 (Composing *tGNDC*)** *Let $M = M_1 \mid \ldots \mid M_k$ be time-dependent stable with respect to a sequence of knowledge $\{\phi_j\}_{j\geq 0}$. Let $\mathcal{A}_1, \ldots, \mathcal{A}_k$ be disjoint sets of attacking nodes for $M_1, \ldots, M_k$, respectively. Let $O_i \subseteq \mathsf{nds}(M_i)$, for $1\leq i\leq k$. Then,*

$$(M_i)^{\mathcal{A}_i}_{O_i} \;\Big|\; \textsc{Top}^{\phi_0}_{\mathcal{A}_i/\mathsf{nds}(M_i)} \;\lesssim\; N_i, \textit{ for } 1 \leq i \leq k, \textit{ implies } M \in tGNDC^{N_1|\ldots|N_k}_{\phi_0,O_1\uplus\ldots\uplus O_k} \; .$$

**Proof** By Theorem 2.12 we have

$$(M_1)^{\mathcal{A}_1}_{O_1} \;\Big|\; \ldots \;\Big|\; (M_k)^{\mathcal{A}_k}_{O_k} \;\Big|\; \textsc{Top}^{\phi_0}_{\mathcal{A}_1/\mathsf{nds}(M_1)} \;\Big|\; \ldots \;\Big|\; \textsc{Top}^{\phi_0}_{\mathcal{A}_k/\mathsf{nds}(M_k)} \;\lesssim\; N_1 \;\Big|\; \ldots \;\Big|\; N_k \; .$$

By applying Lemma 3.11 and Theorem 2.12 we obtain

$$(M_1 \mid \ldots \mid M_k)^{\mathcal{A}_1\uplus\ldots\uplus\mathcal{A}_k}_{O_1\uplus\ldots\uplus O_k} \;\Big|\; \textsc{Top}^{\phi_0}_{\mathcal{A}_1\uplus\ldots\uplus\mathcal{A}_k/\mathsf{nds}(M_1|\ldots|M_k)} \;\lesssim\; N_1 \;\Big|\; \ldots \;\Big|\; N_k \; .$$

Thus, by an application of Theorem 3.12 we can derive $M \in tGNDC^{N_1|\ldots|N_k}_{\phi_0,O_1\uplus\ldots\uplus O_k}$. □

## 3.1 Two timed security properties

We formalise two useful timed properties for security protocols as instances of $tGNDC^{\rho}_{\phi_0,O}$, by suitably defining the abstraction function $\rho$ [15]. We will focus on the two following timed properties:

- A timed notion of integrity, called *timed integrity*, which guarantees that only fresh packets are authenticated.

- A timed notion of authentication, called *timed agreement*, according to which agreement must be reached within a certain deadline, otherwise authentications does not hold.

More precisely, fixed a delay $\delta$, a protocol is said to enjoy the timed integrity property if, whenever a packet $p_i$ is authenticated during the $i$-th time interval, then this packet was sent at most $i - \delta$ time intervals before. For verification reasons, when expressing time integrity in the *tGNDC* scheme, we will introduce in the protocol under examination a special message $\mathsf{auth}_i$ which is emitted only when the packet $p_i$ is authenticated.

A protocol is said to enjoy the timed agreement property if, whenever a responder $n$ has completed a *run* of the protocol, apparently with an initiator $m$, then $m$ has initiated the protocol, apparently with $n$, at most $\delta$ time intervals before, and the two agents agreed on some set of data. When expressing time agreement in the *tGNDC* scheme, we introduce in the protocol under examination a special message $\mathsf{hello}_i$, which is emitted by the initiator at the $i$-th run of the protocol, and a special message $\mathsf{end}_i$, emitted by the responder, representing the completion of the protocol launched at run $i$.

# 4 A Security Analysis of $\mu$TESLA

The $\mu$TESLA protocol was designed by Perrig et al. [33] to provide authenticated broadcast from a base station (BS) towards all nodes of a wireless network. The protocol is based on a delayed disclosure of symmetric keys, and it requires the network to be *loosely time synchronised*. The protocol computes a MAC for every packet to be broadcast, by using different keys. The transmission time is split into *time intervals* of $\Delta_{\text{int}}$ time units each, and each key is tied to one of them. The keys belongs to a key chain $k_0, k_1, \dots, k_n$ generated by BS by means of a public *one-way* function $F$. In order to generate this chain, BS randomly chooses the last key $k_n$ and repeatedly applies $F$ to compute all the other keys, whereby $k_i := F(k_{i+i})$, for $0 \le i \le n-1$. The key-chain mechanism together with the one-way function $F$, provides two major advantages: (i) a key $k_i$ can be used to generate the beginning of the chain $k_0, \dots, k_{i-1}$, by simply applying $F$ as many time as necessary, but it cannot be used to generate any of the subsequent keys; (ii) any of the keys $k_0, \dots, k_{i-1}$ can be used to authenticate $k_i$. Moreover, each node $m_j$ is pre-loaded with a *master key* $k_{\text{BS}:m_j}$ for unicast communications with BS.

In this section, we analyse the two main phases of the protocol: *bootstrapping new receivers* and *authenticated broadcast*. The former establishes the node's initial setting in order to start receiving the authenticated packets, the latter describes the transmission of authenticated information.

## 4.1 Bootstrapping new receivers

When a new node $m$ wish to join the network, it sends a request message to the base station BS containing its name and a nonce $n_j$, where $j$ counts the number of bootstrapping requests:

$$m \rightarrow \text{BS} : n_j \mid m \;.^3$$

The base station replies with a message of initialisation of the following form:

$$\text{BS} \rightarrow m : \Delta_{\text{int}} \mid i \mid k_l \mid l \mid \text{mac}(k_{\text{BS}:m}, (n_j \mid \Delta_{\text{int}} \mid i \mid k_l \mid l))$$

where $\Delta_{\text{int}}$ is the duration of every time interval, $i$ is the current time interval of BS, $k_l$ is a key in the key chain, and $l$, with $l < i$, represents the time interval in which $k_l$ was employed for packet encryption; hence, $k_l$ can be used for authenticating the subsequent keys in the chain. The secret key $k_{\text{BS}:m}$ is used to authenticate unicast messages; the nonce $n_j$ allows the node $m$ to verify the freshness of the reply coming from BS.

**Encoding in `aTCWS`** Our encoding contains a few simplifications with respect to the original protocol. As described in Section 4.2, the authenticated-broadcast phase consists of two distinct events: packet broadcast and key disclosure. For the sake of simplicity, in our encoding these events will happen in contiguous time slots, hence the time interval $\Delta_{\text{int}}$ corresponds to two $\sigma$-actions. Moreover, we assume that $\Delta_{\text{int}}$ is already known by all nodes. Hence, BS needs to communicate just the current time interval $i$ and the time interval $l$ of the committing key $k_l$. We fix $l = i - 1$. Thus, we can simplify the reply message as follows:

$$\textsc{bs} \rightarrow m : \quad i \mid k_{i-1} \mid \mathrm{mac}(k_{\textsc{bs}:m}, (n_j \mid i \mid k_{i-1})) \ .$$

[3] Here, the "|" symbol denotes message concatenation.

When giving our specifications in `aTCWS` we will require some new deduction rules to model *Message Authentication Code* and a *pseudo-random function*:

$$(\mathrm{mac}) \ \frac{w_1 \quad w_2}{\mathrm{mac}(w_1, w_2)} \qquad (\mathrm{prf}) \ \frac{w_1 \quad w_2}{\mathrm{prf}(w_1, w_2)} \ .$$

The application $\mathrm{mac}(k, p)$ returns a unique number which represents the MAC of packet $p$, obtained from the payload of that packet and a key $k$. This will be used to authenticate the packet $p$. The application $\mathrm{prf}(m, w_i)$ returns a pseudo-random value $w_{i+1}$ associated to a node $m$ and the last generated value $w_i$.

Table 5 provides both the code running at each requesting node $m$ and the code running at the base station $\textsc{bs}$. The base station runs the process $D_i$, where $i$ represents the index of the current key as well as the current time interval. The requesting nodes run the process $A_j$, where $j$ counts the number of bootstrapping requests made by the node. At each request $j$, the receiver generates a nonce $n_j$. Upon authentication of a key $k$, the node starts the authenticated-broadcast phase, via the process $R\langle i+1, i-1, \perp, k\rangle$ defined in Table 7 of Section 4.2.

At the beginning of the bootstrapping phase the network appears as:

$$\mu\mathrm{TESLA}_{boot} \stackrel{\mathrm{def}}{=} \textsc{bs}[D_1]^{\nu_{\textsc{bs}}} \mid m_1[A_1]^{\nu_{m_1}} \mid \ldots \mid m_k[A_1]^{\nu_{m_k}}$$

where $m \in \nu_{\textsc{bs}}$ and $\textsc{bs} \in \nu_m$, for every $m \in \{m_1, \ldots, m_k\}$.

### 4.1.1 Timed Agreement

The timed agreement property for the bootstrapping phase of $\mu$TESLA requires that the base station $\textsc{bs}$ successfully replies to a request packet $p_j$, sent by the initiator node $m$, in at most $\Delta_{\mathrm{int}}$ time units (corresponding in our encoding to two $\sigma$-actions).

Here, we prove that the bootstrapping phase *does not* satisfies *timed agreement*. In particular, we show that an attacker may prevent the bootstrapping request from reaching the base station, thus the bootstrapping phase may not terminate in due time. In order to do that, we present a replay attack which can be described, without loss of generality, by focusing on a part of the protocol, called $\mu\mathrm{TESLA}'_{boot}$, consisting of a single requesting node $m$ and the base station $\textsc{bs}$. Moreover, we slightly modify the processes running at $\textsc{bs}$ to signal the end of the bootstrapping phase. Thus, we define the process $D'_i$ as a slight modification of $D_i$ (defined in Table 5) where process $E_i^4$ is replaced by

$$E_i^{4\prime} \stackrel{\mathrm{def}}{=} \sigma.[\mathsf{end}\ n \vdash_{pair} t]!\langle t\rangle.D_{i+1} \ .$$

With this modification, the encoding of the fragment under investigation of the bootstrapping phase of $\mu$TESLA becomes:

$$\mu\mathrm{TESLA}'_{boot} \stackrel{\mathrm{def}}{=} \textsc{bs}[D'_1]^{\nu_{\textsc{bs}}} \mid m[A_1]^{\nu_m} \ .$$

**Table 5** $\mu$TESLA: bootstrapping phase.

*Request at node m*

| | |
|---|---|
| $A_j \stackrel{\text{def}}{=} [m\ n_{j-1} \vdash_{prf}\ n_j]$ | Build a random nonce $n_j$ |
| $[m\ n_j \vdash_{pair}\ t]$ | Build a pair $t$ with name $m$ and nonce $n_j$ |
| $[\mathsf{req}\ t \vdash_{pair}\ p_j]$ | Build request packet using the pair $t$ |
| $!\langle p_j\rangle.\sigma.B_j$ | broadcast the request and move to $B_j$ |
| $B_j \stackrel{\text{def}}{=} \lfloor ?(w).C_j \rfloor A_{j+1}$ | Receive the bootstrapping packet |
| $C_j \stackrel{\text{def}}{=} [w \vdash_{fst}\ q]C_j^1; \sigma.A_{j+1}$ | extract the first component |
| $C_j^1 \stackrel{\text{def}}{=} [w \vdash_{\mathsf{snd}}\ h]$ | extract the MAC |
| $[n_j\ q \vdash_{pair}\ r]$ | add the nonce $n_j$ |
| $[k_{\text{BS}:m}\ r \vdash_{mac}\ h']$ | calculate MAC $h'$ on $r$ |
| $[h = h']C_j^2; \sigma.A_{j+1}$ | match the two MACs |
| $C_j^2 \stackrel{\text{def}}{=} [q \vdash_{fst}\ i]C_j^3; \sigma.A_{j+1}$ | extract the current interval $i$ |
| $C_j^3 \stackrel{\text{def}}{=} [q \vdash_{snd}\ k]\sigma.R\langle i+1, i-1, \bot, k\rangle$ | extract key $k$ and start authenticated broadcast |

*Reply at base station* BS

| | |
|---|---|
| $D_i \stackrel{\text{def}}{=} \lfloor ?(p).E_i \rfloor \sigma.D_{i+1}$ | Wait for incoming request packets |
| $E_i \stackrel{\text{def}}{=} [p \vdash_{fst}\ p_1]E_i^1; \sigma.\sigma.D_{i+1}$ | extract the first component |
| $E_i^1 \stackrel{\text{def}}{=} [p_1 = \mathsf{req}]E_i^2; \sigma.\sigma.D_{i+1}$ | check if $p$ is a request packet |
| $E_i^2 \stackrel{\text{def}}{=} [p \vdash_{\mathsf{snd}}\ t]$ | extract the second component |
| $[t \vdash_{fst}\ m]E_i^3; \sigma.\sigma.D_{i+1}$ | extract the sender name $m$ |
| $E_i^3 \stackrel{\text{def}}{=} [t \vdash_{\mathsf{snd}}\ n]$ | extract the nonce $n$ |
| $[i\ k_{i-1} \vdash_{pair}\ q_i]$ | pair up key $k_{i-1}$ with the current time interval $i$ |
| $[n\ q_i \vdash_{pair}\ r_i]$ | add the nonce $n$ |
| $[k_{\text{BS}:m}\ r_i \vdash_{mac}\ h_i]$ | calculate MAC $h_i$ on $r_i$ with $m$'s master key $k_{\text{BS}:m}$ |
| $[q_i\ h_i \vdash_{pair}\ w_i]\sigma.!\langle w_i\rangle.E_i^4$ | build packet $w_i$ with $q_i$ and MAC $h_i$ |
| $E_i^4 \stackrel{\text{def}}{=} \sigma.D_{i+1}$ | broadcast $w_i$ and go to the next requesting state |

where the request packet $p_j = \text{pair}(\mathsf{req}, \text{pair}(m, n_j))$ reports the beginning of the bootstrapping phase, while the message $\mathsf{end}_j = \text{pair}(\mathsf{end}, n_j)$ signals the end of the phase.

We define the timed agreement property as the following abstraction of the protocol:

$$\rho_{agr}(\mu\text{TESLA}'_{boot}) \stackrel{\text{def}}{=} \text{BS}[\hat{D}_1]^{obs} \mid m_1[\hat{A}_1]^{obs}$$

where

$$\hat{D}_i \stackrel{\text{def}}{=} \lfloor \tau.\sigma.!\langle w_i\rangle.\sigma.!\langle \mathsf{end}_i\rangle.\hat{D}_{i+1} \rfloor \sigma.\hat{D}_{i+1}$$
$$\hat{A}_i \stackrel{\text{def}}{=} !\langle p_i\rangle.\sigma.\lfloor \tau.\sigma.R\langle i+1, i-1, \bot, k_{i-1}\rangle \rfloor \hat{A}_{i+1}$$

with $w_i = \text{pair}(q_i, \text{mac}(k_{m:S}, \text{pair}(n_i, q_i)))$ and $q_i = \text{pair}(i, k_{i-1})$, as defined in Table 5. Basically, these processes are obtained by $D'_i$ and $A_i$ by abstracting on receptions. The node *obs* is the observing node introduced in Section 3.

The abstraction $\rho_{agr}(\mu\text{TESLA}'_{boot})$ correctly expresses the timed agreement property for the system $\mu\text{TESLA}'_{boot}$. In fact, the following proposition says that BS successfully replies to a request packet $p_i$, sent by the initiator node $m$, in exactly $\Delta_{\text{int}}$ time units (i.e. in two $\sigma$-actions).

**Proposition 4.1** *Whenever* $\rho_{agr}(\mu TESLA'_{boot}) \stackrel{\Lambda}{\Longrightarrow} \xrightarrow{!p_i \triangleright obs} \stackrel{\Omega}{\Longrightarrow} \xrightarrow{!\mathsf{end}_i \triangleright obs}$ *then* $\#^{\sigma}(\Omega) = 2$.

Now, in order to show that $\mu\text{TESLA}'_{boot}$ satisfies timed agreement, we should prove that

$$\mu\text{TESLA}'_{boot} \;\in\; tGNDC^{\rho_{agr}(\mu\text{TESLA}'_{boot})}_{\phi_0, O}$$

where $O = \{m, \text{BS}\}$ and $\phi_0 \subseteq Msg$. More precisely, given an appropriate set of attacking nodes $\mathcal{A} = \{a, b\}$, we should prove that

$$(\mu\text{TESLA}'_{boot})^{\mathcal{A}}_{O} \;\Big|\; A \;\lesssim\; \rho_{agr}(\mu\text{TESLA}'_{boot}) \quad \text{for every } A \in \mathbb{A}^{\phi_0}_{\mathcal{A}/\mathsf{nds}(\mu\text{TESLA}'_{boot})}\;.$$

This would imply that all execution traces of the system $(\mu\text{TESLA}'_{boot})^{\mathcal{A}}_{O} \mid A$ can be matched by $\rho_{agr}(\mu\text{TESLA}'_{boot})$. Unfortunately, this is not the case. The following theorem shows how an attacker $A$ can force the system $(\mu\text{TESLA}'_{boot})^{\mathcal{A}}_{O} \mid A$ to execute a trace in which the special message $\mathsf{end}_j$ is broadcast $2\Delta_{\text{int}}$ time units (that is, four $\sigma$-actions) later than $p_j$. This trace cannot be executed by $\rho_{agr}(\mu\text{TESLA}'_{boot})$, as stated in Proposition 4.1.

**Theorem 4.2 (Replay Attack to $\mu$TESLA Bootstrapping Phase)** *$\mu TESLA'_{boot}$ does not satisfy the timed agreement property.*

**Proof** We propose an attacker that delays agreement. Let us define the set of attacking nodes $\mathcal{A} = \{a, b\}$ for $\mathsf{nds}\big(\mu\text{TESLA}'_{boot}\big)$. Let us fix the initial knowledge $\phi_0 = \emptyset$, so to deal with the most general situation. We set $\nu_a = \{m, b\}$ and $\nu_b = \{\text{BS}, a\}$, and we assume that all nodes in $\mathsf{nds}\big(\mu\text{TESLA}'_{boot}\big)$ are observable, thus $\nu_m = \{\text{BS}, a, obs\}$ and $\nu_{\text{BS}} = \{m, b, obs\}$. We give an intuition of the replay attack in Table 6. Basically, the attacker delays the reception at BS of packet $p_1$ so that BS can complete the protocol only after $2\Delta_{\text{int}}$ time units. This denotes a replay attack that breaks agreement.

Formally, we define the attacker $A \in \mathbb{A}^{\phi_0}_{\mathcal{A}/\{m,\text{BS}\}}$ as follows:

$$A = a[\![X]\!]^{\nu_a} \;\Big|\; b[\![Y]\!]^{\nu_b}$$

where $X \stackrel{\text{def}}{=} \lfloor ?(x).\sigma.!\langle x\rangle.\mathsf{nil}\rfloor\mathsf{nil}$ and $Y \stackrel{\text{def}}{=} \sigma.\lfloor ?(y).\sigma.!\langle y\rangle.\mathsf{nil}\rfloor\mathsf{nil}$. We then consider the system

$$(\mu\text{TESLA}'_{boot})^{\mathcal{A}} \;\Big|\; A$$

which admits the following execution trace:

$$!p_1 \triangleright obs\,.\,\sigma\,.\,\tau\,.\,\sigma\,.\,\tau\,.\,!p_2 \triangleright obs\,.\,\sigma\,.\,!w_1 \triangleright obs\,.\,\sigma\,.\,!\mathsf{end}_1 \triangleright obs$$

**Table 6** Replay attack to $\mu$TESLA bootstrapping phase.

| | |
|---|---|
| $m \rightarrow \textsc{bs} : p_1$ | $m$ starts the protocol, but $p_1$ is grasped by $a$ and missed by BS |
| $\xrightarrow{\sigma}$ | the systems moves to the next time slot |
| $a \rightarrow b : p_1$ | $a$ sends $p_1$ to $b$ |
| $\xrightarrow{\sigma}$ | the system moves to the next time slot |
| $b \rightarrow \textsc{bs} : p_1$ | $b$ *replays* $p_1$ to BS |
| $m \rightarrow \textsc{bs} : p_2$ | $m$ sends a new request $p_2$ which gets lost |
| $\xrightarrow{\sigma}$ | the system moves to the next time slot |
| $\textsc{bs} \rightarrow m : w_1$ | BS replies to $p_1$ with $w_1$ (which is discarded by $m$) |
| $\xrightarrow{\sigma}$ | the system moves to the next time slot |
| $\textsc{bs} \rightarrow * : \mathsf{end}_1$ | BS signals the end of the protocol |

containing four $\sigma$-actions between the packets $p_1$ and $\mathsf{end}_1$ (we report the corresponding computation in the Appendix). However, by Proposition 6.1 this trace cannot be matched by $\rho_{agr}(\mu\text{TESLA}'_{boot})$. As a consequence,

$$(\mu\text{TESLA}'_{boot})^{\mathcal{A}} \;\Big|\; A \;\not\lesssim\; \rho_{agr}(\mu\text{TESLA}'_{boot})$$

and hence the timed agreement property does not hold. □

#### 4.1.2 Timed Integrity

In this section, we show that the bootstrapping phase of $\mu$TESLA *satisfies* the *timed integrity property*. In particular, we prove that nodes authenticate only keys that are associated to a nonce sent by the same node wrapped in a request packet in the previous time interval $\Delta_{\text{int}}$.

Again, without loss of generality, we focus on a part of the protocol, called $\mu\text{TESLA}''_{boot}$, consisting of the base station BS and a single node $m$. We signal authentication at the node side by broadcasting a special message. This is done by replacing the process $A_j$ of Table 5 with the process $A''_j$ which is the same as $A_j$ except for $C^3_j$ which is replaced by:

$$C^{3''}_j \stackrel{\text{def}}{=} [q \vdash_{snd} k]\sigma.[\mathsf{auth}\; n \vdash_{pair} t]!\langle t\rangle.R\langle i+1, i{-}1, \bot, k\rangle \;\; .$$

Thus, the fragment of the protocol under examination becomes:

$$\mu\text{TESLA}''_{boot} \stackrel{\text{def}}{=} \textsc{bs}[D_1]^{\nu_{\textsc{bs}}} \mid m[A''_1]^{\nu_m} \;\; .$$

The abstraction of the protocol that expresses *timed integrity* can be formalised as follows:

$$\rho_{int}(\mu\text{TESLA}''_{boot}) \stackrel{\text{def}}{=} \textsc{bs}[Tick]^{\emptyset} \mid m[\bar{A}_1]^{obs}$$

where $Tick \stackrel{\text{def}}{=} \sigma.Tick$ and $\overline{A}_i \stackrel{\text{def}}{=} !\langle p_i\rangle.\sigma.\lfloor\tau.\sigma.!\langle \mathsf{auth}_i\rangle.R\langle i+1, i{-}1, \perp, k_{i-1}\rangle\rfloor\overline{A}_{i+1}$, with $\mathsf{auth}_i = \text{pair}(\mathsf{auth}, n_i)$ and $p_i = \text{pair}(\mathsf{req}, \text{pair}(m, n_i))$. Again, the node *obs* is the observer introduced in Section 3.

In the abstraction $\rho_{int}(\mu\text{TESLA}''_{boot})$, it is straightforward to see that the action $\mathsf{auth}_i$, which authenticates a key with nonce $n_i$, occurs exactly $\Delta_{\text{int}}$ time units (that is, two $\sigma$-actions) after the request $p_i$, which carries the nonce $n_i$. This fact is stated in the following proposition.

**Proposition 4.3** *Whenever* $\rho_{int}(\mu TESLA''_{boot}) \stackrel{\Lambda}{\Longrightarrow} \xrightarrow{!p_i \triangleright obs} \stackrel{\Omega}{\Longrightarrow} \xrightarrow{!\mathsf{auth}_i \triangleright obs} M$ *then* $\#^{\sigma}(\Omega)$=2.

The previous result says that $\rho_{int}(\mu\text{TESLA}''_{boot})$ expresses correctly the timed integrity property. Thus, in order to show that the encoding of the bootstrapping phase of $\mu$TESLA satisfies the timed integrity property, we will prove that

$$\mu\text{TESLA}''_{boot} \in tGNDC^{\rho_{int}(\mu\text{TESLA}''_{boot})}_{\phi_0,\{m\}}$$

for some appropriate $\phi_0$. Notice that node $m$ signals both the begin and the end of the authentication protocol. Thus, we need to observe only the packets sent by $m$. Moreover, according to Definition 3.7, $\mu\text{TESLA}''_{boot}$ is time-dependent stable with respect to the following sequence of knowledge:

$$\begin{array}{lcll}
\phi_0 & \stackrel{\text{def}}{=} & \{p_1\} & \\
\phi_1 & \stackrel{\text{def}}{=} & \phi_0 \cup \{w_1\} & \\
\phi_2 & \stackrel{\text{def}}{=} & \phi_1 \cup \{\mathsf{auth}_1, p_2\} & \\
 & \vdots & & \\
\phi_i & \stackrel{\text{def}}{=} & \phi_{i-1} \cup \{\mathsf{auth}_j, p_{j+1}\} & \text{if } j>0 \text{ and } i=2j \\
\phi_i & \stackrel{\text{def}}{=} & \phi_{i-1} \cup \{w_{j+1}\} & \text{if } j>0 \text{ and } i=2j+1
\end{array} \qquad (1)$$

where $w_i = \text{pair}(q_i, \text{mac}(k_{m:\text{BS}}, \text{pair}(n_i, q_i))$ and $q_i = \text{pair}(i, k_{i-1})$, as defined in Table 5. Intuitively, $\phi_i$ consists of $\phi_{i-1}$ together with the set of messages an intruder can get by eavesdropping on a run of the protocol during the time slot $i$.

**Lemma 4.4** *Given two attacking nodes a and b, for m and* BS *respectively, and fixed the sequence of knowledge* $\{\phi_i\}_{i\geq 0}$ *as in* (1)*, then*

1. $\text{BS}[D_1]^b \;\big|\; \text{Top}^{\phi_0}_{b/\text{BS}} \;\lesssim\; \text{BS}[Tick]^{\emptyset}$
2. $m[A''_1]^{\{a,obs\}} \;\big|\; \text{Top}^{\phi_0}_{a/m} \;\lesssim\; m[\overline{A}_1]^{obs}.$

**Theorem 4.5 ($\mu$TESLA$_{boot}$ Timed Integrity)** *The protocol* $\mu TESLA''_{boot}$ *satisfies the timed integrity property:*

$$\mu TESLA''_{boot} \in tGNDC^{\rho_{int}(\mu TESLA''_{boot})}_{\phi_0,\{m\}} \;.$$

**Proof** By an application of Lemma 4.4 and Theorem 3.13. □

## 4.2 Authenticating broadcast packets

In the authenticated-broadcast phase, at each time interval $i$, one or more packets $p_i$ are deployed by the sender, each one containing the payload and the MAC calculated with the key $k_i$ bound to the $i$-th time interval. Thus, at time interval $i$ the BS broadcasts the authenticated message:

$$\text{BS} \rightarrow * : p_i \mid \text{mac}(p_i, k_i) \ .$$

In the same time interval $i$, the key tied to the previous time interval $i-1$ is disclosed to all receivers, so that they can authenticate all the previously received packets:

$$\text{BS} \rightarrow * : k_{i-1} \ .$$

Loose time synchronisation on the key disclosure time prevents malicious nodes to forge packets with modified payloads. Nodes discard packets containing MACs calculated with already disclosed keys, as those packets could come from an attacker. In this phase the nodes exploit the two main advantages of the key chain and the one-way function $F$: (i) the last received key $k_i$ can be authenticated by means of $F$ and the last authenticated key $k_l$; (ii) lost keys can be recovered by applying $F$ to the last received key $k_i$. For instance, suppose that BS has sent packet $p_1$ (containing a MAC with key $k_1$) in the first time interval, packet $p_2$ in the second time interval and packet $p_3$ in the third one. If the key $k_1$ is correctly received by a node $m$ while keys $k_2$ and $k_3$ get lost, then $m$ can only authenticate the packet $p_1$ but not $p_2$ or $p_3$. However, if $m$ gets the key $k_4$ then $m$ can authenticate $k_4$ by using $k_1$, and it can also recover the lost keys $k_2$ and $k_3$ to authenticate $p_2$ and $p_3$, respectively.

In Table 7 we provide an encoding of the authenticated-broadcast phase of $\mu$TESLA. Also in this case our encoding contains a few simplifications with respect to the original protocol. As said for the bootstrapping phase, we assume that the duration of the time interval $\Delta_{\text{int}}$ is fixed and it is already known by the nodes. In our encoding this time interval corresponds to two $\sigma$-actions. We assume that in each time interval $i$ the sender broadcasts alternately only one packet $p_i$ and the key $k_{i-1}$ of the previous time interval. Thus, we assume a sequence $q_1, q_2, \ldots$ of payloads to be authenticated by using the corresponding keys $k_1, k_2, \ldots$ Moreover, we do not model the recovery of lost keys, hence the payload $q_i$ can only be authenticated by receiving the key $k_i$. This simplification yields a easier to read model which can be generalised to fulfil the original requirements of the protocol.

The encoding essentially defines two kind of processes: the senders $S_i$, and the receivers $R(i, l, r, k_l)$, where $i$ is the current time interval, $r$ is the last received packet, $l$ is the time interval when the last key $k_l$ was authenticated. Since we bind one packet to one key, $i$ also refers to the index number of packets.

The authenticated-broadcast phase of $\mu$TESLA can be represented as follows:

$$\mu\text{TESLA}_{auth} \stackrel{\text{def}}{=} \text{BS}[S_1]^{\nu_{\text{BS}}} \;\big|\; m_1[R\langle 1, -1, \perp, k_{\text{BS}}\rangle]^{\nu_{m_1}} \;\big|\; \ldots \;\big|\; m_h[R\langle 1, -1, \perp, k_{\text{BS}}\rangle]^{\nu_{m_h}}$$

where $m \in \nu_{\text{BS}}$ and $\text{BS} \in \nu_m$, for every $m \in \{m_1, \ldots, m_h\}$. We use $\perp$ because at the beginning there is no packet to authenticate. We write $k_{\text{BS}}$ to denote the key transmitted by the base station

**Table 7** $\mu$TESLA: authenticated-broadcast phase.

| | | |
|---|---|---|
| *Sender:* | | |
| $S_i \stackrel{\text{def}}{=}$ | $[q_i\ k_i \vdash_{mac} u_i]$ | build MAC with payload and key |
| | $[u_i\ q_i \vdash_{pair} p_i]$ | build packet with mac and payload |
| | $!\langle p_i \rangle.\sigma.$ | broadcast packet, synchronise |
| | $!\langle k_{i-1} \rangle.\sigma.$ | broadcast previous key, synchronise |
| | $S_{i+1}$ | and go to next sending state |
| *Receiver:* | | |
| $R(i,l,r,k_l) \stackrel{\text{def}}{=}$ | $\lfloor ?(p).\sigma.P\langle i,l,p,r,k_l\rangle \rfloor$ | receive a pkt, synchronise, go to $P$ |
| | $Q\langle i,l,r,k_l\rangle$ | if timeout go to $Q$ |
| $P(i,l,p,r,k_l) \stackrel{\text{def}}{=}$ | $\lfloor ?(k).T\langle i,l,p,r,k_l,k\rangle \rfloor$ | receive a key $k$ and move to $T$ |
| | $R\langle i{+}1,l,p,k_l\rangle$ | if timeout go to next receiving state |
| $T(i,l,p,r,k_l,k) \stackrel{\text{def}}{=}$ | $[F^{i-1-l}(k) = k_l]$ | authenticate key $k$ with $F$ and $k_l$ |
| | $\quad [r \vdash_{\text{fst}} u]$ | extract MAC from previous pkt $r$ |
| | $\qquad [r \vdash_{\text{snd}} q]$ | extract payload from $r$ |
| | $\qquad [q\ k \vdash_{mac} u']$ | build MAC for $r$ with key $k$ |
| | $\qquad [u = u']$ | check MACs to authenticate $r$ |
| | $\qquad\quad \sigma.Z\langle i{+}1, i{-}1, p, r, k\rangle;$ | |
| | $\qquad \sigma.R\langle i{+}1, i{-}1, p, k\rangle;$ | |
| | $\quad \sigma.R\langle i{+}1, i{-}1, p, k\rangle;$ | |
| | $\sigma.R\langle i{+}1, l, p, k_l\rangle$ | |
| $Z(i,l,p,r,k_l) \stackrel{\text{def}}{=}$ | $R\langle i,l,p,k_l\rangle$ | authenticated-broadcast succeeded |
| $Q(i,l,r,k_l) \stackrel{\text{def}}{=}$ | $\lfloor ?(k).T\langle i,l,r,r,k_l,k\rangle \rfloor$ | receive a key, synchronise, and |
| | $R\langle i{+}1,l,r,k_l\rangle$ | go to next receiving state |

BS and authenticated at the node's site during the bootstrapping phase. Notice that, according to Table 5, $k_{\text{BS}}$ is associated to the time interval $-1$.

### 4.2.1 Timed Integrity

In this section, we show that the authenticated-broadcast phase of $\mu$TESLA *enjoys timed integrity*. In particular, we prove that receivers authenticate only packets that have been sent $2\Delta_{\text{int}}$ time units before (that is, four $\sigma$-actions before) in the correct order, even in the presence of the intruder. The crucial point is that even if the intruder acquires the shared keys then it is "too late" to break integrity, i.e., to authenticate packets which are more than $2\Delta_{\text{int}}$ time units old.

As done for $\mu\text{TESLA}_{boot}$, we signal authentication of a packet $r$ by broadcasting a special packet $\text{pair}(\mathsf{auth}, r)$. Thus, we replace the process $R(i,l,r,k_l)$ of Table 7 with $R'(i,l,r,k_l)$, where the process $Z(i, l, p, r, k_l)$ is replaced by

$$Z'(i, l, p, r, k_l) \stackrel{\text{def}}{=} [\mathsf{auth}\ r \vdash_{pair} t]!\langle t\rangle.R'\langle i, l, p, k_l\rangle \ .$$

The formalisation of the authenticated-broadcast phase for $\mu$TESLA becomes the following:

$$\mu\text{TESLA}'_{auth} \stackrel{\text{def}}{=} \text{BS}[S_1]^{\nu_{\text{BS}}} \mid m_1[R'\langle 1, -1, \bot, k_{\text{BS}}\rangle]^{\nu_{m_1}} \mid \dots \mid m_h[R'\langle 1, -1, \bot, k_{\text{BS}}\rangle]^{\nu_{m_h}} \ .$$

We define the *timed integrity* property as the following abstraction of the protocol $\mu\text{TESLA}'_{auth}$:

$$\rho_{int}(\mu\text{TESLA}'_{auth}) \stackrel{\text{def}}{=} \text{BS}[S_1]^{obs} \mid m_1[\hat{R}_1]^{obs} \mid \dots \mid m_h[\hat{R}_1]^{obs}$$

where $S_1$ is the process defined in Table 7, while $\hat{R}_i \stackrel{\text{def}}{=} \sigma.\lfloor\tau.\sigma.!\langle\mathsf{auth}_{i-1}\rangle.\hat{R}_{i+1}\rfloor\hat{R}_{i+1}$. The node *obs* is the observing node introduced in Section 3. Here, we abstract on receivers' behaviour: At time interval $i$+2 they may signal the authentication of the packet $p_i = \text{pair}(\text{mac}(k_i, q_i), q_i)$ by sending the special packet $\mathsf{auth}_i = \text{pair}(\mathsf{auth}, p_i)$.

The abstraction $\rho_{int}(\mu\text{TESLA}'_{auth})$ is a faithful representation of the timed integrity property for the authenticated-broadcast phase of $\mu$TESLA.

**Proposition 4.6** *Whenever* $\rho_{int}(\mu TESLA'_{auth}) \stackrel{\Lambda}{\Longrightarrow} \xrightarrow{!p_i \triangleright obs} \stackrel{\Omega}{\Longrightarrow} \xrightarrow{!\mathsf{auth}_i \triangleright obs}$ *M then* $\#^{\sigma}(\Omega)$=4.

In order to show that $\mu\text{TESLA}'_{auth}$ satisfies timed integrity, we will prove that

$$\mu\text{TESLA}'_{auth} \in tGNDC^{\rho_{int}(\mu\text{TESLA}'_{auth})}_{\phi_0, \{\text{BS}, m_1, \dots, m_k\}}$$

for some appropriate $\phi_0$. Notice that $\mu\text{TESLA}'_{auth}$ is time-dependent stable with respect to the following sequence of knowledge:

$$\begin{array}{lcll}
\phi_0 & \stackrel{\text{def}}{=} & \{p_1\} & \\
\phi_1 & \stackrel{\text{def}}{=} & \phi_0 \cup \{k_0\} & \\
\phi_2 & \stackrel{\text{def}}{=} & \phi_1 \cup \{p_2, \mathsf{auth}_0\} & \\
 & \vdots & & \\
\phi_i & \stackrel{\text{def}}{=} & \phi_{i-1} \cup \{p_{j+1}, \mathsf{auth}_{j-1}\} & \text{if } j > 0 \text{ and } i = 2j \\
\phi_i & \stackrel{\text{def}}{=} & \phi_{i-1} \cup \{k_j\} & \text{if } j > 0 \text{ and } i = 2j+1.
\end{array} \qquad (2)$$

Now, we choose an attacking node $a_j$ for each $m_j$, with $1 \leq j \leq h$, and an attacking node $b$ for BS. By applying the compositional criterion of Theorem 3.13, it suffices to prove a simpler integrity result for each node in isolation composed with its corresponding top attacker.

**Lemma 4.7** *Given an attacking node b for* BS *and the attacking nodes* $a_j$ *for* $m_j$, *with* $1 \leq j \leq h$, *and fixed the sequence of knowledge* $\{\phi_i\}_{i\geq 0}$ *as in* (2), *then the encoding in Table 7 satisfies the following:*

1. $\textsc{bs}[S_1]^{\{b,obs\}} \;\Big|\; \textsc{Top}^{\phi_0}_{b/\textsc{bs}} \;\lesssim\; \textsc{bs}[S_1]^{obs}$

2. $m_j[R'\langle 1,-1,\perp,\overline{k}\rangle]^{\{a_j,obs\}} \;\Big|\; \textsc{Top}^{\phi_0}_{a_j/m_j} \;\lesssim\; m_j[\hat{R}_1]^{obs}$, *for* $1 \leq j \leq h$.

**Theorem 4.8 ($\mu$TESLA$_{auth}$ Timed Integrity)** *The protocol $\mu TESLA'_{auth}$ satisfies timed integrity:*

$$\mu TESLA'_{auth} \;\in\; tGNDC^{\rho_{int}(\mu TESLA'_{auth})}_{\phi_0,\{\textsc{bs},m_1,\dots,m_k\}} \;.$$

**Proof** By applying Lemma 4.7 and Theorem 3.13. □

### 4.2.2 Timed Agreement

The timed agreement property for the authenticated-broadcast phase $\mu$TESLA$_{auth}$ requires that when the receiver $m_j$ completes the protocol, apparently with the initiator BS, then BS has initiated the protocol, apparently with $m_j$, at most two time intervals $\Delta_{\text{int}}$ before, and the two parties agree on the sent data. In other words, the packet $p_i$ is authenticated by $m_j$ exactly $2\Delta_{\text{int}}$ time units after it has been sent by BS. This says that any formulation of timed agreement for $\mu$TESLA$_{auth}$ would actually coincide with timed integrity. Thus, Proposition 4.6 demonstrates that $\rho_{int}(\mu\text{TESLA}'_{auth})$ is also a faithful abstraction of timed agreement. As a consequence, Theorem 4.8 also says that $\mu$TESLA$_{auth}$ *satisfies timed agreement*.

# 5 A Security Analysis of LEAP+

The LEAP+ protocol [43] provides a keying mechanisms to establish authenticated communications. The protocol is designed to establish four types of keys: an *individual key*, shared between a base station and a node, a *single-hop pair-wise key*, shared between two sensor nodes, a *cluster key*, shared between a node and all its neighbourhood, a *group key*, shared between a base station and all sensor nodes of the network.

In this section, we focus on the *single-hop pairwise key* mechanism as it is underlying to all other keying methods. This mechanism is aimed at establishing a pair-wise key between a sensor node and a neighbours in $\Delta_{\text{leap}}$ time units. In order to do that, LEAP+ exploits two peculiarities of sensor nodes: (i) the set of neighbours of a node is relatively static, and (ii) a sensor node that is being added to the network will discover most of its neighbours at the time of its initial deployment.

The single-hop pairwise shared key mechanism of LEAP+ consists of three phases.

*Key pre-distribution.* A network controller fixes an initial key $k_{\text{in}}$ and a computational efficient pseudo-random function prf(). Both $k_{\text{in}}$ and prf() are pre-loaded in each node, before deployment. Then, each node $r$ derives its *master key*: $k_r := \text{prf}(k_{\text{in}}, r)$.

*Neighbour discovery.* As soon as a node $m$ is scattered in the network area it tries to discover its neighbours by broadcasting a hello packet that contains its identity, $m$, and a freshly created nonce $n_i$, where $i$ counts the number of attempts:

$$m \rightarrow * : m \mid n_i \;.$$

Then each neighbour $r$ replies with an ack message which includes its identity $r$, the corresponding MAC calculated by using $r$'s master key $k_r$, to guarantee authenticity, and the nonce $n_i$, to guarantee freshness. Specifically:

$$r \rightarrow m : r \mid \text{mac}(k_r, (r \mid n_i)) \ .$$

*Pairwise Key Establishment.* When $m$ receives the packet $q$ from $r$, it tries to authenticate it by using the last created nonce $n_i$ and $r$'s master key $k_r = \text{prf}(k_{\text{in}}, r)$. Notice that $m$ can calculate $k_r$ as $k_{\text{in}}$ and prf have been pre-loaded in $m$, and $r$ is contained in $q$. If the authentication succeeds, then both nodes proceed in calculating the pairwise key $k_{m:r} := \text{prf}(k_r, m)$. Any other message between $m$ and $r$ will be authenticated by using the pairwise key $k_{m:r}$. If $m$ does not get an authenticated packet from the responder in due time, it sends a new hello packet with a fresh nonce.

In Table 8, we provide an encoding of the single-hop pairwise shared key mechanism of LEAP+. For the sake of clarity, we assume that $\Delta_{\text{leap}}$ consists of two time slots, i.e. it takes two $\sigma$-actions. To yield an easier to read model, we consider only two nodes and we define

$$\text{LEAP+} \stackrel{\text{def}}{=} m[\![S_1]\!]^{\nu_m} \ \Big| \ r[\![R]\!]^{\nu_r}$$

where $m$ is the initiator, $r$ is the responder, with $m \in \nu_r$ and $r \in \nu_m$. Moreover, we assume that $r$ has already computed its master key $k_r := \text{prf}(k_{\text{in}}, r)$. This simple model does not loose in generality with respect to the multiple nodes case.

## 5.1 Timed Agreement

The timed agreement property for LEAP+ requires that the responder $r$ successfully completes the protocol initiated by $m$, with the broadcasting of a hello packet, in at most $\Delta_{\text{leap}}$ time units (i.e. two $\sigma$-actions). We will show that LEAP+ *does not satisfy* the timed agreement property.

For our analysis, in order to make observable the completion of the protocol, we define LEAP′+ by replacing in LEAP+ the process $R$ of Table 8 with the process $R'$ defined as the same as $R$ except for process $R^6$ which is replaced by

$$R^{6\prime} \stackrel{\text{def}}{=} \sigma.[\mathsf{end}\ n \vdash_{pair} e]!\langle e\rangle.\mathsf{nil} \ .$$

We use the following abbreviations: $\mathsf{hello}_i \stackrel{\text{def}}{=} \text{pair}(\mathsf{hello}, \text{pair}(m, n_i))$ and $\mathsf{end}_i \stackrel{\text{def}}{=} \text{pair}(\mathsf{end}, n_i)$.

The timed agreement property of LEAP+ is defined by the following abstraction:

$$\rho_{agr}(\text{LEAP}'+) \stackrel{\text{def}}{=} m[\![\overline{S}_1]\!]^{obs} \mid r[\![\overline{R}_1]\!]^{obs}$$

where $\overline{S}_i \stackrel{\text{def}}{=} !\langle \mathsf{hello}_i\rangle.\sigma\lfloor\tau.\sigma.\mathsf{nil}\rfloor\overline{S}_{i+1}$ and $\overline{R}_i \stackrel{\text{def}}{=} \lfloor\tau.\sigma!\langle q_i\rangle.\sigma.!\langle\mathsf{end}_i\rangle.\mathsf{nil}\rfloor\sigma.\overline{R}_{i+1}$, with $q_i = \text{pair}(r, \text{mac}(k_r, \text{pair}(r, n_i)))$, as defined in Table 8.

The following statement says that the abstraction $\rho_{agr}(\text{LEAP}'+)$ expresses correctly the timed agreement property for LEAP+.

**Encoding in** `aTCWS`

**Table 8** LEAP+ specification

| | | |
|---|---|---|
| Sender at node $m$: | | |
| $S_i \stackrel{\text{def}}{=}$ | $[n_{i-1}\ m \vdash_{prf} n_i]$ | build a random nonce $n_i$ |
| | $[m\ n_i \vdash_{pair} t]$ | build a pair $t$ with $m$ and the nonce $n_i$ |
| | $[\mathsf{hello}\ t \vdash_{pair} p]$ | build hello packet using the pair $t$ |
| | $!\langle p \rangle.\sigma.P$ | broadcast hello, synchronise and move to P |
| $P \stackrel{\text{def}}{=}$ | $\lfloor ?(q).P^1 \rfloor S_{i+1}$ | wait for response from neighbours |
| $P^1 \stackrel{\text{def}}{=}$ | $[q \vdash_{fst} r]P^2; \sigma.S_{i+1}$ | extract node name $r$ from packet $q$, |
| $P^2 \stackrel{\text{def}}{=}$ | $[q \vdash_{\text{snd}} h]$ | extract MAC $h$ from packet $q$ |
| | $[r\ n_i \vdash_{pair} t']$ | build a pair $t'$ with $r$ and current nonce $n_i$ |
| | $[k_{\text{in}}\ r \vdash_{prf} k_r]$ | calculate $r$'s master key $k_r$ |
| | $[k_r\ t' \vdash_{mac} h']$ | calculate MAC $h'$ with $k_r$ and $t'$ |
| | $[h' = h]P^3; \sigma.S_{i+1}$ | if it matches with the received one go to $P^3$, otherwise go to next time unit and restart |
| $P^3 \stackrel{\text{def}}{=}$ | $[k_r\ m \vdash_{prf} k_{m:r}]P^4$ | calculate the pairwise key $k_{m:r}$ |
| $P^4 \stackrel{\text{def}}{=}$ | $\sigma.\mathsf{nil}$ | synchronise and conclude key establishment |
| Receiver at node $r$: | | |
| $R \stackrel{\text{def}}{=}$ | $\lfloor ?(p).R^1 \rfloor \sigma.R$ | Wait for incoming hello packets |
| $R^1 \stackrel{\text{def}}{=}$ | $[p \vdash_{fst} p_1]R^2; \sigma.\sigma.R$ | extract the first component |
| $R^2 \stackrel{\text{def}}{=}$ | $[p \vdash_{\text{snd}} p_2]$ | extract the second component |
| | $[p_1 = \mathsf{hello}]R^3; \sigma.\sigma.R$ | check if $p$ is a hello packet |
| $R^3 \stackrel{\text{def}}{=}$ | $[p_2 \vdash_{fst} m]R^4; \sigma.\sigma.R$ | extract the sender name $m$ |
| $R^4 \stackrel{\text{def}}{=}$ | $[p_2 \vdash_{\text{snd}} n]$ | extract the nonce $n$ |
| | $[r\ n \vdash_{pair} t]$ | build a pair $t$ with $n$ and $r$ |
| | $[k_r\ t \vdash_{mac} h]$ | calculate MAC $h$ on $t$ with $r$'s master key $k_r$ |
| | $[r\ h \vdash_{pair} q]$ | build packet $q$ with node name $r$ and MAC $h$ |
| | $\sigma.!\langle q \rangle.R^5$ | synchronise, broadcast $q$ and go to $R^5$ |
| $R^5 \stackrel{\text{def}}{=}$ | $[k_r\ m \vdash_{prf} k_{m:r}]R^6$ | calculate pairwise key $k_{m:r}$ |
| $R^6 \stackrel{\text{def}}{=}$ | $\sigma.\mathsf{nil}$ | synchronise and conclude key establishment |

**Proposition 5.1** *Whenever* $\rho_{agr}(\text{LEAP}'\text{+}) \stackrel{\Lambda}{\Longrightarrow} \xrightarrow{!hello_i \triangleright obs} \stackrel{\Omega}{\Longrightarrow} \xrightarrow{!\mathsf{end}_i \triangleright obs}$ *then* $\#^{\sigma}(\Omega) = 2$.

Now, in order to prove timed agreement for LEAP+ we should show that

$$\text{LEAP}'\text{+} \ \in \ tGNDC^{\rho_{agr}(\text{LEAP}'\text{+})}_{\phi_0, \{m,r\}}$$

for some appropriate $\phi_0$. This would imply that all traces of the system composed by LEAP′+ in parallel with an attacker can be mimicked by $\rho_{agr}(\text{LEAP}'\text{+})$.

**Table 9** Replay attack to LEAP+.

| | | | | | |
|---|---|---|---|---|---|
| $m$ | $\to$ | $*$ | : | $\mathsf{hello}_1$ | $m$ starts the protocol, but $\mathsf{hello}_1$ is grasped by $a$ and missed by $r$ |
| | $\xrightarrow{\sigma}$ | | | | the system moves to the next time slot |
| $a$ | $\to$ | $b$ | : | $\mathsf{hello}_1$ | $a$ sends $\mathsf{hello}_1$ to $b$ |
| | $\xrightarrow{\sigma}$ | | | | the system moves to the next time slot |
| $b$ | $\to$ | $r$ | : | $\mathsf{hello}_1$ | $b$ *replays* $\mathsf{hello}_1$ to $r$ |
| $m$ | $\to$ | $*$ | : | $\mathsf{hello}_2$ | $m$ broadcasts $\mathsf{hello}_2$ (containing a fresh nonce $n_2$), which gets lost |
| | $\xrightarrow{\sigma}$ | | | | the system moves to the next time slot |
| $r$ | $\to$ | $m$ | : | $q_1$ | $r$ replies by sending $q_1$ (which is discarded by $m$) |
| | $\xrightarrow{\sigma}$ | | | | the system moves to the next time slot |
| $r$ | $\to$ | $*$ | : | $\mathsf{end}_1$ | $r$ signals the end of the protocol |

However, this is not the case, as stated by the following theorem.

**Theorem 5.2 (Replay Attack to LEAP+)** LEAP′+ *does not satisfy the timed agreement property.*

**Proof** We define an attacker that delays agreement. Let us define the set of attacking nodes $\mathcal{A} = \{a, b\}$ for $\mathsf{nds}$(LEAP′+). Let us fix the initial knowledge $\phi_0 = \emptyset$, so to deal with the most general situation. We set $\nu_a = \{m, b\}$ and $\nu_b = \{r, a\}$, and we assume all the nodes in $\mathsf{nds}$(LEAP′+) are observable, thus $\nu_m = \{r, a, obs\}$ and $\nu_r = \{m, b, obs\}$. We give an intuition of the replay attack in Table 9. Basically, the attacker delays the reception of the packet $p_1$ at $m$ which cannot complete the protocol within two time slots, but only after four time slots, thus breaking agreement. Formally we define the attacker $A \in \mathbb{A}^{\phi_0}_{\mathcal{A}/\{m,r\}}$ as follows:

$$A = a[X]^{\nu_a} \mid b[Y]^{\nu_b}$$

where the processes $X$ and $Y$ are the same as those defined in the proof Theorem 4.2. Now, we consider the system

$$(\text{LEAP}'+)^{\mathcal{A}} \;\big|\; A \;=\; m[S_1]^{\nu_m} \;\big|\; r[R']^{\nu_r} \;\big|\; A$$

and we find that it admits the following execution trace

$$!\mathsf{hello}_1 \triangleright obs \,.\, \sigma \,.\, \tau \,.\, \sigma \,.\, \tau.!\mathsf{hello}_2 \triangleright obs \,.\, \sigma \,.\, !q_1 \triangleright obs \,.\, \sigma \,.\, !\mathsf{end}_1 \triangleright obs$$

where the packet $\mathsf{hello}_1$ and the corresponding packet $\mathsf{end}_1$ are divided by four $\sigma$-actions (we report the corresponding computation in the Appendix). Proposition 5.1 says that this trace cannot be mimicked by the specification $\rho_{agr}(\mu\text{TESLA}'_{boot})$. As a consequence, the timed agreement property for LEAP+ does not hold. □

## 5.2 Timed Integrity

The timed integrity property for LEAP+ says that hello messages and authentication messages with the same nonce must differ for at most $\Delta_{leap}$ time units. We show that LEAP+ *satisfies* the timed integrity property. For doing that, we slightly modify the specification of LEAP+ to make observable key authentication. We define

$$\text{LEAP}''+ \stackrel{\text{def}}{=} m[S_1'']^{\nu_m} \mid r[R]^{\nu_r}$$

where the process $S_i''$ is the same as process $S_i$ of Table 8, except for process $P^4$ which is replaced by

$$P^{4''} \stackrel{\text{def}}{=} \sigma.[\mathsf{auth}\ t \vdash_{pair} a]!\langle a\rangle.\mathsf{nil}\ .$$

For simplicity, we use the following abbreviation: $\mathsf{auth}_i = \text{pair}(\mathsf{auth}, \text{pair}(m, n_i))$.

In order to formally represent the timed integrity property, we define the following abstraction of the protocol:

$$\rho_{int}(\text{LEAP}''+) \stackrel{\text{def}}{=} m[\hat{S}_1]^{obs} \mid r[Tick]^{\emptyset}$$

where $\hat{S}_i \stackrel{\text{def}}{=} !\langle \mathsf{hello}_i\rangle.\sigma.\lfloor \tau.\sigma.!\langle \mathsf{auth}_i\rangle.\mathsf{nil}\rfloor \hat{S}_{i+1}$ and $Tick \stackrel{\text{def}}{=} \sigma.Tick$ .

By construction, $\rho_{int}(\text{LEAP}''+)$ is a faithful representation of timed integrity for LEAP+ (we recall that in our encoding $\Delta_{\text{leap}}$ corresponds to two $\sigma$-actions).

**Proposition 5.3** *For every $i \geq 1$, whenever $\rho_{int}(\text{LEAP}''+) \stackrel{\Lambda}{\Longrightarrow} \xrightarrow{!hello_i \triangleright obs} \stackrel{\Omega}{\Longrightarrow} \xrightarrow{!auth_i \triangleright obs}$, then $\#^{\sigma}(\Omega) = 2$.*

Now, we notice that LEAP″+ is time-dependent stable with respect to the sequence of knowledge $\{\phi_i\}_{i\geq 0}$, defined as follows:

$$\begin{array}{lcll}
\phi_0 & \stackrel{\text{def}}{=} & \{\mathsf{hello}_1\} & \\
\phi_1 & \stackrel{\text{def}}{=} & \phi_0 \cup \{mac(k_r, \text{pair}(r, n_1))\} & \\
\phi_2 & \stackrel{\text{def}}{=} & \phi_1 \cup \{\mathsf{hello}_2, \mathsf{auth}_1\} & \\
 & \vdots & & \\
\phi_i & \stackrel{\text{def}}{=} & \phi_{i-1} \cup \{\mathsf{hello}_{j+1}, \mathsf{auth}_j\} & \text{if } j > 0 \text{ and } i = 2j \\
\phi_i & \stackrel{\text{def}}{=} & \phi_{i-1} \cup \{mac(k_r, \text{pair}(r, n_{j+1}))\} & \text{if } j > 0 \text{ and } i = 2j+1\ .
\end{array} \tag{3}$$

Now, we pick two attacking nodes $a$ and $b$, for $m$ and $r$, respectively, and we focus on the observation of node $m$ as it signals both the beginning and the end of the authentication protocol. Again, by applying Theorem 3.13 it suffices to prove a simpler result for each node in isolation composed with its corresponding top attacker.

**Lemma 5.4** *Given two attacking nodes $a$ and $b$, for $m$ and $r$ respectively, and fixed the sequence of knowledge $\{\phi_i\}_{i\geq 0}$ as in (3), then*

1. $m[S_1'']^{\{a,obs\}} \mid \text{Top}^{\phi_0}_{a/m} \lesssim m[\hat{S}_1]^{obs}$

2. $r[R]^{\{b\}} \;\Big|\; \text{Top}^{\phi_0}_{b/r} \;\lesssim\; r[Tick]^{\emptyset}$ .

**Theorem 5.5 (LEAP+ Timed integrity)** LEAP″+ *satisfies the timed integrity property:*

$$\text{LEAP}''{+} \;\in\; tGNDC^{\rho_{int}(\text{LEAP}''+)}_{\phi_0,\{m\}} \;.$$

**Proof** By applying Lemma 5.4 and Theorem 3.13. □

# 6 A Security Analysis of LiSP

In order to achieve a good trade-off between resource limitations and network security, Park et al. [31] have proposed a *Lightweight Security Protocol* (LiSP) for WSNs. LiSP provides (i) an efficient *key renewal* mechanism which avoids key retransmission, (ii) authentication for each key-disclosure, and (iii) the possibility of both recovering and detecting lost keys.

A LiSP network consists of a *Key Server* (KS) and a set of *sensor nodes* $m_1, \ldots, m_k$. The protocol assumes a *one way function* $F$, pre-loaded in every node of the system, and employs two different key families: (i) a set of *temporal keys* $k_0, \ldots, k_n$, computed by KS by means of $F$, and used by all nodes to encrypt/decrypt data packets; (ii) a set of *master keys* $k_{\text{KS}:m_j}$, one for each node $m_j$, for unicast communications between $m_j$ and BS. As in $\mu$TESLA, the transmission time is split into *time intervals*, each of them is $\Delta_{\text{refresh}}$ time units long. Thus, each temporal key is tied to a time interval and renewed every $\Delta_{\text{refresh}}$ time units. At a time interval $i$, the temporal key $k_i$ is shared by all sensor nodes and it is used for data encryption. Key renewal relies on *loose node time synchronisation* among nodes. Each node stores a subset of temporal keys in a *buffer* of a fixed size, say $s$ with $s << n$. When a time interval elapses, each node removes the active key from the buffer to free a slot for the next key taken from the sequence $k_0, \ldots, k_n$.

The LiSP protocol consists of the following phases.

*Initial Setup.* At the beginning, KS randomly chooses a key $k_n$ and computes a sequence of temporal keys $k_0, \ldots, k_n$, by using the function $F$, as in $\mu$TESLA: $k_i := F(k_{i+1})$. Then, KS waits for reconfiguration requests from nodes. More precisely, when KS receives a reconfiguration request from a node $m_j$, at time interval $i$, it unicasts the packet InitKey:

$$\text{KS} \rightarrow m_j \;:\; \text{enc}(k_{\text{KS}:m_j},\, (s \mid k_{s+i} \mid \Delta_{\text{refresh}})) \;\mid\; \text{hash}(s \mid k_{s+i} \mid \Delta_{\text{refresh}})$$

where $s$ represents the buffer size, $k_{s+i}$ is the initial key and $\Delta_{\text{refresh}}$ is the duration of the refresh interval. The operator $\text{enc}(k, p)$ represents the encryption of $p$ by using the key of $k$, while $\text{hash}(p)$ generates a message digest for $p$ by means of a cryptographic hash function used to check the integrity of the packet $p$.

When $m_j$ receives the InitKey packet, it computes the sequence of keys

$$k_{s+i-1}, k_{s+i-2}, \ldots, k_i$$

by applying the function $F$ to $k_{s+i}$. Then, it activates $k_i$ for data encryption and it stores the remaining keys in its local buffer; finally it sets up a *ReKeyingTimer* to expires after $\Delta_{\text{refresh}}/2$ time units (this value applies only for the first rekeying).

*Re-Keying.* At each time interval $i$, with $i \leq n$, KS employs the active encryption key $k_i$ to encode the key $k_{s+i}$. The resulting packet is broadcast as an UpdateKey packet:

$$\text{KS} \rightarrow * \ : \ \text{enc}(k_i,\ k_{s+i}) \ .$$

When a node receives an UpdateKey packet, it tries to authenticate the key received in the packet; if the node succeeds in the authentication then it recovers all keys that have been possibly lost and updates its key buffer. When the time interval $i$ elapses, every node discards $k_i$, activates the key $k_{i+1}$ for data encryption, and sets up the *ReKeyingTimer* to expire after $\Delta_{\text{refresh}}$ time units for future key switching (after the first time, switching happens every $\Delta_{\text{refresh}}$ time units).

*Authentication and Recovery of Lost Keys.* The one-way function $F$ is used to authenticate and recover lost keys. If $s$ is the size of the key buffer and $l$, with $l \leq s$, is the number of stored keys in the buffer, then $s - l$ represents the number of keys which have been lost by the node. When a sensor node receives an UpdateKey packet carrying a new key $k$, it calculates $F^{s-l}(k)$ by applying $s - l$ times the function $F$. If the result matches with the last received temporal key, then the node stores $k$ in its buffer and recovers all lost keys.

*Reconfiguration.* When a node $m_j$ joins the network or misses more than $s$ temporal keys, then its buffer is empty. Thus, it sends a RequestKey packet in order to request the current configuration:

$$m_j \rightarrow \text{KS} \ : \ \text{RequestKey} \mid m_j \ .$$

Upon reception, node KS performs authentication of $m_j$ and, if successful, it sends the current configuration via an InitKey packet.

**Encoding in `aTCWS`** In Table 10, we provide a specification of the entire LiSP protocol in `aTCWS`. We introduce some slight simplifications with respect to the original protocol. We assume that (i) the temporal keys $k_0, \ldots, k_n$ have already been computed by KS, (ii) both the buffer size $s$ and the refresh interval $\Delta_{\text{refresh}}$ are known by each node. Thus, the broadcasting of the InitKey packet can be simplified as follows:

$$\text{KS} \rightarrow m_j \ : \ \text{enc}(k_{\text{KS}:m_j},\ k_{s+i}) \ \mid \ \text{hash}(k_{s+i}) \ .$$

Moreover, we assume that every $\sigma$-action models the passage of $\Delta_{\text{refresh}}/2$ time units. Therefore, every two $\sigma$-actions the key server broadcasts the new temporal key encrypted with the key tied to that specific interval. Finally, we do not model data encryption. Our specification can be easily generalised to fulfil the original requirements of the protocol.

When giving our encoding in `aTCWS` we will require some new deduction rules to model an hash functions and encryption/decryption of messages:

$$(\text{hash}) \ \frac{w}{\text{hash}(w)} \qquad (\text{enc}) \ \frac{w_1 \quad w_2}{\text{enc}(w_1, w_2)} \qquad (\text{dec}) \ \frac{w_1 \quad w_2}{\text{dec}(w_1, w_2)} \ .$$

**Table 10** LiSP Specification

| | | | |
|---|---|---|---|
| *Key Server:* | | | |
| $D_0$ | $\stackrel{\text{def}}{=}$ | $\sigma.D_1$ | synchronise and move to $D_1$ |
| $D_i$ | $\stackrel{\text{def}}{=}$ | $[k_i\ k_{s+i} \vdash_{enc} t_i]$ | for $i \geq 1$, encrypt $k_{s+i}$ with $k_i$ |
| | | $[\mathsf{UpdateKey}\ t_i \vdash_{pair} u_i]$ | build the UpdateKey packet $u_i$ |
| | | $!\langle u_i\rangle.\sigma.\sigma.D_{i+1}$ | broadcast $r_i$, and move to $D_{i+1}$ |
| $L_i$ | $\stackrel{\text{def}}{=}$ | $\lfloor ?(r).I_{i+1}\rfloor\sigma.L_{i+1}$ | wait for request packets |
| $I_i$ | $\stackrel{\text{def}}{=}$ | $[r \vdash_{fst} r_1]I_i^1;\sigma.\sigma.L_{i+1}$ | extract first component |
| $I_i^1$ | $\stackrel{\text{def}}{=}$ | $[r_1 = \mathsf{RequestKey}]I_i^2;\sigma.\sigma.L_{i+1}$ | check if $r_1$ is a RequestKey |
| $I_i^2$ | $\stackrel{\text{def}}{=}$ | $[r \vdash_{\mathrm{snd}} m]$ | extract node name |
| | | $[k_{\mathrm{KS}:m}\ k_{s+i} \vdash_{enc} w_i]$ | encrypt $k_{s+i}$ with $k_{\mathrm{KS}:m}$ |
| | | $[k_{s+i} \vdash_{hash} h_i]$ | calculate hash code for $k_{s+i}$ |
| | | $[w_i\ h_i \vdash_{pair} r_i]$ | build a pair $r_i$, |
| | | $[\mathsf{InitKey}\ r_i \vdash_{pair} q_i]$ | build a InitKey packet $q_i$, |
| | | $\sigma.!\langle q_i\rangle.\sigma.L_{i+1}$ | broadcast $q_i$, move to $L_{i+1}$ |
| *Receiver at node m:* | | | |
| $Z$ | $\stackrel{\text{def}}{=}$ | $[\mathsf{RequestKey}\ m \vdash_{\mathrm{pair}} r]$ | send a RequestKey packet |
| | | $!\langle r\rangle.\sigma.\lfloor ?(q).T\rfloor Z$ | wait for a reconfig. packet |
| $T$ | $\stackrel{\text{def}}{=}$ | $[q \vdash_{fst} q']T^1;\sigma.Z$ | extract fst component of $q$ |
| $T^1$ | $\stackrel{\text{def}}{=}$ | $[q' = \mathsf{InitKey}]T^2;\sigma.Z$ | check if $q$ is a InitKey packet |
| $T^2$ | $\stackrel{\text{def}}{=}$ | $[q \vdash_{snd} q'']$ | extract snd component of $q$ |
| | | $[q'' \vdash_{fst} w]$ | extract fst component of $q''$ |
| | | $[q'' \vdash_{\mathrm{snd}} h]$ | extract snd component of $q''$ |
| | | $[k_{\mathrm{KS}:m}\ w \vdash_{dec} k]T^3;\sigma.Z$ | extract the key |
| $T^3$ | $\stackrel{\text{def}}{=}$ | $[k \vdash_{hash} h'][h = h']T^4;\sigma.Z$ | verify hash codes |
| $T^4$ | $\stackrel{\text{def}}{=}$ | $\sigma.\sigma.R\langle F^{s-1}(k), k, s-1\rangle$ | synchronise and move to $R$ |
| $R(k_{\mathrm{C}}, k_{\mathrm{L}}, l)$ | $\stackrel{\text{def}}{=}$ | $\lfloor ?(u).E\rfloor F$ | wait for incoming packets |
| $E$ | $\stackrel{\text{def}}{=}$ | $[u \vdash_{fst} u']E^1;\sigma.F$ | extract fst component of $u$ |
| $E^1$ | $\stackrel{\text{def}}{=}$ | $[u' = \mathsf{UpdateKey}]E^2;\sigma.F$ | check UpdateKey packet |
| $E^2$ | $\stackrel{\text{def}}{=}$ | $[u \vdash_{\mathrm{snd}} u'']$ | extract snd component of $u$ |
| | | $[k_{\mathrm{C}}\ u'' \vdash_{dec} k]E^3;\sigma.F$ | decrypt $u''$ by using $k_{\mathrm{C}}$ |
| $E^3$ | $\stackrel{\text{def}}{=}$ | $[F^{s-l}(k) = k_{\mathrm{L}}]E^4;\sigma.F$ | authenticate $k$ |
| $E^4$ | $\stackrel{\text{def}}{=}$ | $\sigma.\sigma.R\langle F^{s-1}(k), k, s-1\rangle$ | synchronise and move to $R$ |
| $F$ | $\stackrel{\text{def}}{=}$ | $[l = 0]Z;\sigma.R\langle F^{l-1}(k_{\mathrm{L}}), k_{\mathrm{L}}, l-1\rangle$ | check if buffer key is empty |

The protocol executed by the key server is expressed by the following two threads: a key distributor $D_i$ and a listener $L_i$ waiting for reconfiguration requests from the sensor nodes, with $i$ being the current time interval. Every $\Delta_{\text{refresh}}$ time units (that is, every two $\sigma$-actions) the process $D_i$ broadcasts the new temporal key $k_{s+i}$ encrypted with the key $k_i$ of the current time interval $i$. The listener process $L_i$ replies to reconfiguration requests coming from sensor nodes by sending an initialisation packet.

At the beginning of the protocol, a sensor node runs the process $Z$, which broadcasts a request packet to KS, waits for a reconfiguration packet $q$, and then checks authenticity by verifying the hash code. If the verification is successful then the node starts the broadcasting new keys phase. This phase is formalised by the process $R(k_{\text{c}}, k_{\text{L}}, l)$, where $k_{\text{c}}$ represents the current temporal key, $k_{\text{L}}$ is the last authenticated temporal key, and the integer $l$ counts the number of keys that are actually stored in the buffer. This process waits for a new UpdateKey packet $u$, which is sent by the key server and carries the new temporal key in the key chain. If $u$ is correctly received, the process $E$ decrypts the packet, by using the current key $k_{\text{c}}$, and authenticates the received key by applying the function $F$. If the key authentication is successful, then the sensor node synchronises and moves to the next receiving process by updating its state: $k_{\text{c}}$ is discarded and replaced by the first key in the buffer, $k_{\text{L}}$ is replaced by the key just authenticated, and $l := s-1$, as the function $F$ allows the recovery of lost keys. In case of either packet loss or authentication/decryption failure, the process checks if the buffer still contains keys. If so, the process switches the keys and moves into the next receiving state with a new current key and $l := l-1$. Otherwise, if the buffer is empty, the node needs a reconfiguration as authentication and recovery are not longer possible. Therefore, the process moves into $Z$, and restarts the initial setup phase.

To simplify the exposition of our security analysis, we formalise the key server as a pair of nodes: a key disposer KD, which executes the process $D_i$, and a listener KL, which executes the process $L_i$. Thus, the LiSP protocol, in its initial configuration, can be represented as:

$$\text{LiSP} \stackrel{\text{def}}{=} \prod_{j\in J} m_i[Z]^{\nu_{m_i}} \;|\; \text{KS}[D_0]^{\nu_{\text{KS}}} \;|\; \text{KL}[L_0]^{\nu_{\text{KL}}}$$

where $\cup_{j\in J}\{m_j\}$ is the set of sensor nodes, and for every $j \in J$ node $m_j \in \nu_{\text{KD}} \cap \nu_{\text{KL}}$ and $\{\text{KD}, \text{KL}\} \subseteq \nu_{m_j}$.

## 6.1 Timed Integrity

The timed integrity property for LiSP says that a node $m$ must authenticate only keys sent by the key server in the previous $\Delta_{\text{refresh}}$ time units (that is, every two $\sigma$-actions). Otherwise, a node needing a reconfiguration would authenticate an obsolete temporal key and it would not be synchronised with the rest of the network. In this section, we show that LiSP *does not satisfy* the timed integrity property because a time span of more than $\Delta_{\text{refresh}}$ time units may elapses between the transmission of a message by the key server and the authentication of that message by the node.

For our analysis, without loss of generality, it suffices to focus on a part of the protocol composed by the KL node of the key server and a single sensor node $m$. Moreover, in order to make observable a successful reconfiguration, we replace the process $Z$ of Table 10 with a process $Z'$ which is defined as the same as $Z$ except for process $T^4$ which is replaced by

$$T^{4'} \stackrel{\text{def}}{=} \sigma.[\mathsf{auth}\ k \vdash_{pair} a]!\langle a\rangle.\sigma.R\langle F^{s-1}(k), k, s-1\rangle \ .$$

Thus, the part of the protocol under examination is defined as follows:

$$\text{LiSP}' \stackrel{\text{def}}{=} m[\![Z']\!]^{\nu_m} \mid \text{KL}[\![L_0]\!]^{\nu_{\text{KL}}} \ .$$

The timed integrity property can be expressed by the following abstraction of the protocol:

$$\rho_{int}(\text{LiSP}') \stackrel{\text{def}}{=} m[\![\hat{Z}_0]\!]^{obs} \mid \text{KL}[\![\hat{L}_0]\!]^{obs}$$

where

- $\hat{Z}_i \stackrel{\text{def}}{=} !\langle r\rangle.\sigma.\lfloor\tau.\sigma.!\langle \mathsf{auth}_i\rangle.\sigma.R(k_{i+1}, k_{s+i}, s-1)\rfloor\hat{Z}_{i+1}$, with $r = \text{pair}(\mathsf{RequestKey},\ m)$ and $\mathsf{auth}_i = \text{pair}(\mathsf{auth},\ k_{s+i})$ as defined in Table 10;
- $\hat{L}_i \stackrel{\text{def}}{=} \lfloor\tau.\sigma.!\langle q_i\rangle.\sigma.\hat{L}_{i+1}\rfloor\sigma.\hat{L}_{i+1}$, and $q_i$ is defined as in Table 10: $q_i = \text{pair}(\mathsf{InitKey}\ r_i)$ with $r_i = \text{pair}(\text{enc}(k_{\text{KS}:m_j},\ k_{s+i}), \text{hash}(k_{s+i}))$.

The next result says that $\rho_{int}(\text{LiSP}')$ is a faithful representation of the timed integrity property of LiSP$'$.

**Proposition 6.1** *Whenever* $\rho_{int}(\text{LiSP}') \stackrel{\Lambda}{\Longrightarrow} \xrightarrow{!q_i \triangleright obs} \stackrel{\Omega}{\Longrightarrow} \xrightarrow{!\mathsf{auth}_i \triangleright obs}$ *then* $\#^{\sigma}(\Omega) = 2$.

In order to show that LiSP$'$ satisfies timed integrity, we should prove that

$$\text{LiSP}' \in tGNDC^{\rho_{int}(\text{LiSP}')}_{\phi_0, obs}$$

for some appropriate $\phi_0$.

Unfortunately, this is not the case. The following theorem describes an attacker which obliges LiSP$'$ to perform a trace in which $\mathsf{auth}_i$ occurs $2\Delta_{\text{refresh}}$ time units (that is, four $\sigma$-actions) after $q_i$. Proposition 6.1 says that such a trace cannot be mimicked by $\rho_{int}(\text{LiSP}')$.

**Theorem 6.2 (Replay Attack to LiSP)** LiSP$'$ *does not satisfy the timed integrity property.*

**Proof** We propose an attacker that delays authentication. Let us define the set of attacking nodes $\mathcal{A} = \{a, b\}$ for $\mathsf{nds}\,(\text{LiSP}')$. Let us fix the initial knowledge $\phi_0 = \emptyset$ so to deal with the most general situation. We set $\nu_a = \{m, b\}$ and $\nu_b = \{\text{KL}, a\}$, and we assume that all nodes in $\mathsf{nds}\,(\text{LiSP}')$ are observable, thus $\nu_m = \{\text{KL}, a, obs\}$ and $\nu_{\text{KL}} = \{m, b, obs\}$. We give an intuition of the replay attack in Table 11. Basically, the attacker prevents the node $m$ to receive the $\mathsf{InitKey}$ packet within $\Delta_{\text{refresh}}$ time units. Thus $m$ completes the protocol only after $2\Delta_{\text{refresh}}$ time units, and it authenticates an old key. This denotes a replay attack that breaks integrity. Formally, we define the attacker $A \in \mathbb{A}^{\phi_0}_{\mathcal{A}/\{m,\text{KL}\}}$ as follows:

$$A = a[\![X]\!]^{\nu_a} \;\Big|\; b[\![Y]\!]^{\nu_b}$$

**Table 11** Replay attack to LiSP.

| | | | | |
|---|---|---|---|---|
| $m$ | $\longrightarrow$ | KL | : $r$ | $m$ sends a RequestKey and KL correctly receives the packet |
| | $\xrightarrow{\sigma}$ | | | the system moves to the next time slot |
| KL | $\longrightarrow$ | $m$ | : $q_1$ | KL replies with an InitKey which is lost by $m$ and grasped by $b$ |
| | $\xrightarrow{\sigma}$ | | | the system moves to the next time slot |
| $b$ | $\rightarrow$ | $a$ | : $q_1$ | $b$ sends $q_1$ to $a$ |
| $m$ | $\rightarrow$ | KL | : $r$ | $m$ sends a new RequestKey which gets lost |
| | $\xrightarrow{\sigma}$ | | | the system moves to the next time slot |
| $a$ | $\rightarrow$ | $m$ | : $q_1$ | *a replays* $q_1$ to $m$ |
| | $\xrightarrow{\sigma}$ | | | the system moves to the next time slot |
| $m$ | $\rightarrow$ | $*$ | : $\mathsf{auth}_1$ | $m$ authenticates $q_1$ and signals the end of the protocol |

where $X \stackrel{\text{def}}{=} \sigma.\sigma.\lfloor ?(x).\sigma.!\langle x\rangle.\mathsf{nil}\rfloor\mathsf{nil}$ and $Y \stackrel{\text{def}}{=} \sigma.\lfloor ?(y).\sigma.!\langle y\rangle.\mathsf{nil}\rfloor\mathsf{nil}$. We then consider the system $(\text{LiSP}')^{\mathcal{A}} \mid A$ which admits the following execution trace:

$$!r\triangleright obs\,.\,\sigma\,.\,!q_1\triangleright obs\,.\,\sigma\,.\,\tau\,.\,!r\triangleright obs\,.\,\sigma\,.\,\tau\,.\,\sigma\,.\,!\mathsf{auth}_1\triangleright obs$$

where the packet $q_1$ and the corresponding $\mathsf{auth}_1$ packet are divided by three $\sigma$-actions (we report the corresponding computation in the Appendix). By Proposition 6.1, this trace cannot be matched by $\rho_{int}(\text{LiSP}')$. As a consequence, $(\text{LiSP}')^{\mathcal{A}} \mid A \not\lesssim \rho_{int}(\text{LiSP}')$. Hence the timed integrity property does not hold. □

### 6.2 Timed Agreement

The timed agreement property for LiSP requires that when a sensor node $m$ completes the protocol, apparently with the initiator KL, then KL has initiated the protocol $\Delta_{\text{refresh}}$ time units before and the two nodes agree on the transmitted data. In other words: the packet $q_i$ must be received and authenticated by $m$ exactly $\Delta_{\text{refresh}}$ time units after it has been sent by BS. This suggests that, as seen for $\mu$TESLA in Section 4.2, any formulation of timed agreement for LiSP would actually coincide with timed integrity. As a consequence, Theorem 6.2 also says that LiSP *does not satisfies timed agreement.*

## 7 Conclusions, Related and Future Work

We have proposed a times broadcasting calculus, called `aTCWS`, to formalise and verify real-world key management protocols for WSNs. Our calculus comes with a well-defined operational semantics and a (bi)simulation-based behavioural semantics. We have provided formal specifications in `aTCWS` of three well-known key management protocols for WSNs: LiSP [31], $\mu$TESLA [32] and LEAP+ [43]. Our specifications meet the requirements of Proposition 2.10, thus they all satisfy well-timedness. We have revised Gorrieri and Martinelli's *tGNDC* [15] framework in such a way that it can be applied to WSNs. In particular, we have expressed two timed security properties as instances of tGNDC: timed integrity and timed agreement.

We have formally proved that the bootstrapping phase of $\mu$TESLA and the single-hop pairwise shared key mechanism of LEAP+ enjoy timed integrity, and that the authenticated-broadcast phase of $\mu$TESLA enjoys both timed integrity and timed agreement. On the other hand, we have provided three different *replay attacks* showing that the bootstrapping phase of $\mu$TESLA and the single-hop pairwise shared key mechanism of LEAP+ do not enjoy timed agreement, and LiSP does not satisfy neither timed integrity nor timed agreement. The two attacks for $\mu$TESLA and LEAP+ are somehow similar as they both delay the reception of the initial packets of the protocols. The attack on LiSP delays the reception of an intermediate packet which is required for the completion of the protocol.

The present work is the continuation and generalisation of [4], where a slight variant of the calculus was introduced, and an early security analysis for the authenticated-broadcast phase of $\mu$TESLA and the single-hop pairwise shared key mechanism of LEAP+ was performed. In [38] the calculus `aTCWS` has been used by the last author to analyse the LiSP protocol. The design of our calculus is strongly inspired by `tCryptoSPA` [15], a timed "cryptographic" variant of Milner's CCS [26].

The *tGNDC* schema for `tCryptoSPA`, has already been used by Gorrieri et al. [16] to study several security protocols, for both wired and wireless networks. In particular, they studied the authenticated-broadcast phase of $\mu$TESLA, proving timed integrity. The formalisation for $\mu$TESLA we have proposed here is much less involved than the one of [16] thanks to the specific features of our calculus for broadcast communications.

Several process calculi for wireless systems have been recently proposed. Mezzetti and Sangiorgi [21] have introduced a calculus to describe interferences in wireless systems. Nanz and Hankin [28] have proposed a calculus for mobile ad hoc networks for specification and security analysis of communication protocols. They provide a decision procedure to check security against fixed intruders known in advance. Merro [24] has proposed a behavioural theory for mobile ad hoc networks. Godskesen [14] has proposed a calculus for mobile ad hoc networks with a formalisation of an attack on the cryptographic routing protocol ARAN. Singh et al. [35] have proposed the $\omega$-calculus for modelling the AODV routing protocol. Ghassemi et al. [11, 12] have proposed a process algebra, provided with model checking and equational reasoning, which models topology changes implicitly in the semantics. Merro and Sibilio [25] have proposed a timed calculus for wireless systems focusing on the notion of communication collision. Godskesen and Nanz [13] have proposed a simple timed calculus for wireless systems to express a wide range of mobility models. Gallina and Rossi [9] have proposed a calculus for the analysis of energy-aware communications in mobile ad hoc networks. Song and Godskesen [36] have proposed the first probabilistic un-timed calculus for mobile wireless systems in which connection probabilities may change due to node mobility. Kouzapas and Philippou [20] have proposed a process calculus for dynamic networks which contains features for broadcasting at multiple transmission ranges and for viewing networks at different levels of abstraction.

Recently, Arnaud et al. [3] have proposed a calculus for modelling and reasoning about security protocols, including secure routing protocols, for a bounded number of sessions. They provide two NPTIME decision procedures for analysing routing protocols for any network topology, and apply their framework to analyse the protocol SRP [30] applied to DSR [18].

The AVISPA model checker [2] has been used in [40] for an analysis of TinySec [19], LEAP [42], and TinyPK [41], three wireless sensor network security protocols, and in [39] for an analysis of the Sensor Network Encryption Protocol SNEP [32]. In particular, in [40] the authors considered the communication between immediate neighbour nodes which use the pairwise shared key already established by LEAP. In this case AVISPA found a man-in-the-middle attack where the intruder may play at the same time the role of two nodes in order to obtain real information from one of them, thus loosing confidentiality.

It is our intention to apply our framework to study the correctness of a wide range of wireless network security protocols, as for instance, MiniSec [23], and evolutions of LEAP+, such as R-LEAP+ [6] and LEAP++ [22].

# A Proofs

**Proof of Proposition 2.4** We single out each item of the proposition.
*Item 1.* The forward direction is an instance of rule (RcvEnb), the converse is proved by a straightforward rule induction.

*Item 2.* The forward direction follows by noticing that only rules (RcvEnb) and (RcvPar) are suitable for deriving the action $m?w$ from $M_1 \mid M_2$; in the case of rule (RcvEnb) we just apply rule (RcvEnb) both on $M_1$ and on $M_2$, in the case of rule (RcvPar) the premises require both $M_1$ and $M_2$ to perform an action $m?w$ and to move to $N_1$ and $N_2$ with $N = N_1 \mid N_2$. The converse is an instance of rule ($\sigma$-Par).

*Item 3.* The result is a consequence of the combination of rules (Snd) and (Bcast) and it is proved by a straightforward rule induction.

*Item 4.* Again, the proof is done by a straightforward rule induction.

*Item 5.* The forward direction follows by noticing that the only rule for deriving the action $\sigma$ from $M_1 \mid M_2$ is ($\sigma$-Par) which, in the premises, requires both $M_1$ and $M_2$ to perform an action $\sigma$. The converse is an instance of rule ($\sigma$-Par). □

**Proposition A.1** *If $M \approx N$ then* $\mathsf{nds}(M) = \mathsf{nds}(N)$.

**Proof** By contradiction. Assume there exists a node $m$ such that $m \in \mathsf{nds}(M)$ and $m \notin \mathsf{nds}(N)$. Then, by rule (RcvEnb), $N \xrightarrow{m?w} N$. Since $M \approx N$ there must be $M'$ such that $M \overset{m?w}{\Longrightarrow} M'$ with $M' \approx N'$. However, since $m \in \mathsf{nds}(M)$, by inspection on the transition rules, there is no way to deduce a weak transition of the form $M \overset{m?w}{\Longrightarrow} M'$. □

**Proof of Theorem 2.12** We prove that the relation

$$\mathcal{R} = \{\, (M \mid O,\ N \mid O) \text{ s.t. } M \approx N \text{ and } M \mid O,\ N \mid O \text{ are well-formed} \,\}$$

is a bisimulation. We proceed by case analysis on why $M \mid O \xrightarrow{\alpha} Z$. The interesting cases are when the transition is due to an interaction between $M$ and $O$. The remaining cases are more elementary.

Let $M \mid O \xrightarrow{!w\triangleright\nu} M' \mid O'$ ($\nu \neq \emptyset$) by an application of rule (Obs), because $M \mid O \xrightarrow{m!w\triangleright\nu} M' \mid O'$, by an application of rule (Bcast). There are two possible ways to derive this transition, depending on where the sender node is located in the network.

1. $M \xrightarrow{m!w\triangleright\mu} M'$ and $O \xrightarrow{m?w} O'$, with $m \in \mathsf{nds}(M)$ and $\nu = \mu \setminus \mathsf{nds}(O)$. By an application of rule (Obs) we obtain that $M \xrightarrow{!w\triangleright\mu} M'$. Since $M \approx N$, it follows that there is $N'$ such that $N \overset{!w\triangleright\mu}{\Longrightarrow} N'$ with $M' \approx N'$. This implies that there exists $h \in \mathsf{nds}(N)$ such that $N \overset{h!w\triangleright\mu}{\Longrightarrow} N'$. Moreover:

   (a) $h \notin \mathsf{nds}(O)$, as $N \mid O$ is well-formed and it cannot contain two nodes with the same name;

   (b) $\mu \subseteq \mathsf{ngh}(h, N)$, by Proposition 2.4(3);

   (c) If $k \in \mu \cap \mathsf{nds}(O)$ then $h \in \mathsf{ngh}(k, O)$, as the neighbouring relation is symmetric.

   Now, in case $O \xrightarrow{m?w} O'$ exclusively by rule (RcvEnb) then also $O \xrightarrow{h?w} O'$ by rule (RcvEnb) and item (a). In case the derivation of $O \xrightarrow{m?w} O'$ involves some applications of the rule (Rcv) then the concerned nodes have the form $k[\![\lfloor ?(x).P \rfloor Q]\!]^{\eta}$ with $k \in \mu$, hence $h \in \mathsf{ngh}(k, O)$ by item (c), and so we can derive $O \xrightarrow{h?w} O'$ by applying the rules (RcvEnb) and (RcvPar).

   Thus we have $O \xrightarrow{h?w} O'$ in any case. Then by an application of rule (Bcast) and several applications of rule (TauPar) we have $N \mid O \overset{h!w\triangleright\nu}{\Longrightarrow} N' \mid O'$. As $\nu \neq \emptyset$, by an application of rule (Obs) and several applications of rule (TauPar) it follows that $N \mid O \overset{!w\triangleright\nu}{\Longrightarrow} N' \mid O'$. Since $M' \approx N'$, we obtain $(M' \mid O',\ N' \mid O') \in \mathcal{R}$.

2. $M \xrightarrow{m?w} M'$ and $O \xrightarrow{m!w\triangleright\mu} O'$, with $m \in \mathsf{nds}(O)$ and $\nu = \mu \setminus \mathsf{nds}(M)$. Since $M \approx N$, it follows that there is $N'$ such that $N \overset{m?w}{\Longrightarrow} N'$ with $M' \approx N'$. By an application of rule (Bcast) and several applications of rule (TauPar) we have $N \mid O \overset{m!w\triangleright\nu'}{\Longrightarrow} N' \mid O'$, with $\nu' = \mu \setminus \mathsf{nds}(N)$. Since $M \approx N$, by Proposition A.1 it follows that $\nu' = \nu \neq \emptyset$. Thus, by an application of rule (Obs) it follows that $N \mid O \xRightarrow{!w \triangleright \nu} N' \mid O'$. Since $M' \approx N'$, we obtain $(M' \mid O',\ N' \mid O') \in \mathcal{R}$.

Let $M \mid O \xrightarrow{\tau} M' \mid O'$ by an application of rule (Shh) because $M \mid O \xrightarrow{m!w \triangleright \emptyset} M' \mid O'$. This case is similar to the previous one.

Let $M \mid O \xrightarrow{m?w} M' \mid O'$ by an application of rule (RcvPar) because $M \xrightarrow{m?w} M'$ and $O \xrightarrow{m?w} O'$. Since $M \approx N$, it follows that there is $N'$ such that $N \xRightarrow{m?w} N'$ with $M' \approx N'$. By an application of rule (RcvPar) and several applications of rule (TauPar) we have $N \mid O \xRightarrow{m?w} N' \mid O'$. Since $M' \approx N'$, we obtain $(M' \mid O',\ N' \mid O') \in \mathcal{R}$.

Let $M \mid O \xrightarrow{\sigma} M' \mid O'$ by an application of rule ($\sigma$-Par) because $M \xrightarrow{\sigma} M'$ and $O \xrightarrow{\sigma} O'$. This case is similar to the previous one. □

**Proof of Lemma 3.11** We first note that a straightforward consequence of Definition 3.9 is:

$$\textsc{Top}^{\phi_0}_{(\mathcal{A}_1 \uplus \mathcal{A}_2)/\mathsf{nds}(M)} = \textsc{Top}^{\phi_0}_{\mathcal{A}_1/\mathsf{nds}(M_1)} \mid \textsc{Top}^{\phi_0}_{\mathcal{A}_2/\mathsf{nds}(M_2)} \ .$$

Then, in order to prove the result, we just need to show that

$$(M_1 \mid M_2)^{\mathcal{A}_1 \uplus \mathcal{A}_2}_{O_1 \uplus O_2} \mid \textsc{Top}^{\phi_0}_{\mathcal{A}_1 \uplus \mathcal{A}_2/\mathsf{nds}(M)} \ \lesssim\ (M_1)^{\mathcal{A}_1}_{O_1} \mid (M_2)^{\mathcal{A}_2}_{O_2} \mid \textsc{Top}^{\phi_0}_{\mathcal{A}_1 \uplus \mathcal{A}_2/\mathsf{nds}(M)} \ .$$

To improve readability, we consider the most general case, that is $O_1 = \mathsf{nds}(M_1)$ and $O_2 = \mathsf{nds}(M_2)$. Moreover, we assume $M_1 = m_1[P_1]^{\nu_1}$, $M_2 = m_2[P_2]^{\nu_2}$ and therefore $\mathcal{A}_1 = \{a_1\}$, $\mathcal{A}_2 = \{a_2\}$. The generalisation is straightforward. Then we have:

- $(M_1 \mid M_2)^{\mathcal{A}_1 \uplus \mathcal{A}_2} = m_1[P_1]^{\nu'_1} \mid m_2[P_2]^{\nu'_2}$
  with $\{a_1, obs\} \subseteq \nu'_1 \subseteq \{a_1, m_2, obs\}$ and $\{a_2, obs\} \subseteq \nu'_2 \subseteq \{a_2, m_1, obs\}$;
- $M_1^{\mathcal{A}_1} = m_1[P_1]^{\nu''_1}$ with $\nu''_1 = \{a_1, obs\}$;
- $M_2^{\mathcal{A}_2} = m_2[P_2]^{\nu''_2}$ with $\nu''_2 = \{a_2, obs\}$.

We define $\mathcal{P} = \{m_1, m_2\}$ and $\mathcal{A} = \{a_1, a_2\}$. We need to prove

$$m_1[P_1]^{\nu'_1} \mid m_2[P_2]^{\nu'_2} \mid \textsc{Top}^{\phi_0}_{\mathcal{A}/\mathcal{P}} \lesssim m_1[P_1]^{\nu''_1} \mid m_2[P_2]^{\nu''_2} \mid \textsc{Top}^{\phi_0}_{\mathcal{A}/\mathcal{P}} \ .$$

We prove that the following binary relation is a simulation:

$$\begin{array}{rcl} \mathcal{R} & \stackrel{\text{def}}{=} & \bigcup_{j \geq 0} \{ (\, m_1[Q_1]^{\nu'_1} \mid m_2[Q_2]^{\nu'_2} \mid N \,,\ m_1[Q_1]^{\nu''_1} \mid m_2[Q_2]^{\nu''_2} \mid \textsc{Top}^{\phi_j}_{\mathcal{A}/\mathcal{P}}) \\ & & \quad \text{such that } m_1[P_1]^{\nu'_1} \mid m_2[P_2]^{\nu'_2} \mid \textsc{Top}^{\phi_0}_{\mathcal{A}/\mathcal{P}} \xRightarrow{\Lambda} m_1[Q_1]^{\nu'_1} \mid m_2[Q_2]^{\nu'_2} \mid N \\ & & \quad \text{for some } \Lambda \text{ with } \#^{\sigma}(\Lambda) = j \,\} \ . \end{array}$$

We consider $(\, m_1[Q_1]^{\nu'_1} \mid m_2[Q_2]^{\nu'_2} \mid N \,,\ m_1[Q_1]^{\nu''_1} \mid m_2[Q_2]^{\nu''_2} \mid \textsc{Top}^{\phi_j}_{\mathcal{A}/\mathcal{P}}) \in \mathcal{R}$ and we proceed by case analysis on why $m_1[Q_1]^{\nu'_1} \mid m_2[Q_2]^{\nu'_2} \mid N \xrightarrow{\alpha} m_1[\hat{Q}_1]^{\nu'_1} \mid m_2[\hat{Q}_2]^{\nu'_2} \mid \hat{N}$ .

$\alpha = m?w$. This case is straightforward. In fact, the environment of the system contains exclusively the node *obs* which cannot transmit; thus the rule (Rcv) cannot be applied. We can consider just the rules (RcvEnb) and (RcvPar), which do not modify the network.

$\alpha = \sigma$. Then $m_i[Q_i]^{\nu'_i} \xrightarrow{\sigma} m_i[\hat{Q}_i]^{\nu'_i}$ (for $i = 1, 2$) and $N \xrightarrow{\sigma} \hat{N}$. Now also $\textsc{Top}^{\phi_j}_{\mathcal{A}/\mathcal{P}} \xrightarrow{\sigma} \textsc{Top}^{\phi_{j+1}}_{\mathcal{A}/\mathcal{P}}$, hence $m_1[Q_1]^{\nu''_1} \mid m_2[Q_2]^{\nu''_2} \mid \textsc{Top}^{\phi_j}_{\mathcal{A}/\mathcal{P}} \xrightarrow{\sigma} m_1[\hat{Q}_1]^{\nu''_1} \mid m_2[\hat{Q}_2]^{\nu''_2} \mid \textsc{Top}^{\phi_{j+1}}_{\mathcal{A}/\mathcal{P}}$.

$\alpha = !w \rhd \nu$. We observe: (i) the environment of the system contains just the node *obs* and (ii) $\mathsf{Env}(N) = \{m_1, m_2\}$. Thus there exists $i \in \{1, 2\}$ such that the transition has been derived just by rule (Obs) from the following premise

$$m_1[Q_1]^{\nu'_1} \mid m_2[Q_2]^{\nu'_2} \mid N \xrightarrow{m_i!w \rhd obs} m_1[\hat{Q}_1]^{\nu'_1} \mid m_2[\hat{Q}_2]^{\nu'_2} \mid \hat{N} \ .$$

Without loss of generality we assume $i = 1$, then we have $m_1[Q_1]^{\nu'_1} \xrightarrow{m_1!w \rhd \nu'_1} m_1[\hat{Q}_1]^{\nu'_1}$, $m_2[Q_2]^{\nu'_2} \xrightarrow{m_1?w} m_2[\hat{Q}_2]^{\nu'_2}$ and $N \xrightarrow{m_1?w} \hat{N}$. Now, to prove the similarity, we need to simulate the $m_1?w$-action at the node $m_2[Q_2]^{\nu''_2}$ which cannot actually receive packets from $m_1 \notin \nu''_2$. We first observe that the message $w$ can be eavesdropped by an attacker at the time interval $j$, thus $w \in \mathcal{D}(\phi_j)$ thanks to time-dependent stability. Then $\textsc{Top}^{\phi_j}_{\mathcal{A}/\mathcal{P}} \xRightarrow{a_2!w \rhd m_2} \textsc{Top}^{\phi_j}_{\mathcal{A}/\mathcal{P}}$. Since $a_2 \in \nu''_2$ we have $m_2[Q_2]^{\nu''_2} \xrightarrow{a_2?w} m_2[\hat{Q}_2]^{\nu''_2}$. Finally $m_1[Q_1]^{\nu''_1} \xrightarrow{a_2?w} m_1[Q_1]^{\nu''_1}$ by rule (RcvEnb). Thus by applying rule (Bcast) we obtain

$$m_1[Q_1]^{\nu''_1} \mid m_2[Q_2]^{\nu''_2} \mid \textsc{Top}^{\phi_j}_{\mathcal{A}/\mathcal{P}} \xRightarrow{a_2!w \rhd \emptyset} m_1[Q_1]^{\nu''_1} \mid m_2[\hat{Q}_2]^{\nu''_2} \mid \textsc{Top}^{\phi_j}_{\mathcal{A}/\mathcal{P}}$$

and by rule (Shh) $m_1[Q_1]^{\nu''_1} \mid m_2[Q_2]^{\nu''_2} \mid \textsc{Top}^{\phi_j}_{\mathcal{A}/\mathcal{P}} \xRightarrow{\tau} m_1[Q_1]^{\nu''_1} \mid m_2[\hat{Q}_2]^{\nu''_2} \mid \textsc{Top}^{\phi_j}_{\mathcal{A}/\mathcal{P}}$. Now $m_1[Q_1]^{\nu''_1} \xrightarrow{m_1!w \rhd \nu''_1} m_1[\hat{Q}_1]^{\nu''_1}$ and by rule (RcvEnb) we have both $m_2[\hat{Q}_2]^{\nu''_2} \xrightarrow{m_1?w} m_2[\hat{Q}_2]^{\nu''_2}$ and $\textsc{Top}^{\phi_j}_{\mathcal{A}/\mathcal{P}} \xrightarrow{m_1?w} \textsc{Top}^{\phi_j}_{\mathcal{A}/\mathcal{P}}$. Thus

$$m_1[Q_1]^{\nu''_1} \mid m_2[\hat{Q}_2]^{\nu''_2} \mid \textsc{Top}^{\phi_j}_{\mathcal{A}/\mathcal{P}} \xrightarrow{m_1!w \rhd obs} m_1[\hat{Q}_1]^{\nu''_1} \mid m_2[\hat{Q}_2]^{\nu''_2} \mid \textsc{Top}^{\phi_j}_{\mathcal{A}/\mathcal{P}}.$$

$\alpha = \tau$. The most significant case is an application of rule (Shh), from the premise $m_1[Q_1]^{\nu'_1} \mid m_2[Q_2]^{\nu'_2} \mid N \xrightarrow{m_1!w \rhd \emptyset} m_1[\hat{Q}_1]^{\nu'_1} \mid m_2[\hat{Q}_2]^{\nu'_2} \mid \hat{N}$. Since $obs \in \nu'_1 \cap \nu'_2$, the broadcast action must be performed by $N$; thus there exists $i \in \{1, 2\}$ such that $N \xrightarrow{a_i!w \rhd m_i} \hat{N}$ and $m_l[Q_l]^{\nu'_l} \xrightarrow{a_i?w} m_l[\hat{Q}_l]^{\nu'_l}$, for $l = 1, 2$. Now also $\textsc{Top}^{\phi_j}_{\mathcal{A}/\mathcal{P}} \xRightarrow{a_i!w \rhd m_i} \textsc{Top}^{\phi_j}_{\mathcal{A}/\mathcal{P}}$ and $m_l[Q_l]^{\nu''_l} \xrightarrow{a_i?w} m_l[\hat{Q}_l]^{\nu''_l}$, for $l = 1, 2$. Thus $m_1[Q_1]^{\nu''_1} \mid m_2[Q_2]^{\nu''_2} \mid \textsc{Top}^{\phi_j}_{\mathcal{A}/\mathcal{P}} \xrightarrow{\tau} m_1[\hat{Q}_1]^{\nu''_1} \mid m_2[\hat{Q}_2]^{\nu''_2} \mid \textsc{Top}^{\phi_j}_{\mathcal{A}/\mathcal{P}}$ . □

**Lemma A.2** *If $M$ is time-dependent stable with respect to a sequence of knowledge $\{\phi_j\}_{j\geq 0}$, $\mathcal{A}$ is a set of attacking nodes for $M$ and $O \subseteq \mathsf{nds}(M)$ then*

$$M^{\mathcal{A}}_{O} \,\big|\, A \;\lesssim\; M^{\mathcal{A}}_{O} \,\big|\, \mathrm{Top}^{\phi_0}_{\mathcal{A}/\mathsf{nds}(M)} \quad \textit{for every } A \in \mathbb{A}^{\phi_0}_{\mathcal{A}/\mathsf{nds}(M)} \ .$$

**Proof** We prove the lemma in the most general case, that is $O = \mathsf{nds}(M)$. Then we fix an arbitrary $A \in \mathbb{A}^{\phi_0}_{\mathcal{A}/\mathsf{nds}(M)}$ and we define the proper simulation as follows:

$$\mathcal{R} \stackrel{\text{def}}{=} \bigcup_{j\geq 0} \{ (M' \mid A',\ M' \mid \mathrm{Top}^{\phi_j}_{\mathcal{A}/\mathsf{nds}(M)}) \text{ s.t. } M^{\mathcal{A}} \mid A \stackrel{\Lambda}{\Longrightarrow} M' \mid A' \text{ with } \mathsf{nds}(M') = \mathsf{nds}\big(M^{\mathcal{A}}\big) \text{ and } \#^{\sigma}(\Lambda) = j \}$$

We let $(M' \mid A',\ M' \mid \mathrm{Top}^{\phi_j}_{\mathcal{A}/\mathsf{nds}(M)}) \in \mathcal{R}$. We make a case analysis on why $M' \mid A' \xrightarrow{\alpha} N$.

$\alpha = m?w$. As for Lemma 3.11, this case is straightforward.

$\alpha = \sigma$. Then $N = M'' \mid A''$ with $M' \xrightarrow{\sigma} M''$ and $A' \xrightarrow{\sigma} A''$. Now also $\mathrm{Top}^{\phi_j}_{\mathcal{A}/\mathsf{nds}(M)} \xrightarrow{\sigma} \mathrm{Top}^{\phi_{j+1}}_{\mathcal{A}/\mathsf{nds}(M)}$ by rule ($\sigma$-Sum), hence by rule ($\sigma$-Par) we have $M' \mid \mathrm{Top}^{\phi_j}_{\mathcal{A}/\mathsf{nds}(M)} \xrightarrow{\sigma} M'' \mid \mathrm{Top}^{\phi_{j+1}}_{\mathcal{A}/\mathsf{nds}(M)}$.

$\alpha = !w \triangleright v$. Since the environment of the system contains just the node *obs*, the transition has to be derived by the rule (Obs) whose premise is $M' \mid A' \xrightarrow{m!w\triangleright obs} N$. Since $obs \notin \mathsf{Env}(A')$ then $m \in \mathsf{nds}(M')$ and $N = M'' \mid A''$ with $M' \xrightarrow{m!w\triangleright v'} M''$, $\{obs\} = v' \setminus \mathsf{nds}(A')$ and $A' \xrightarrow{m?w} A''$. Now we have $\mathrm{Top}^{\phi_j}_{\mathcal{A}/\mathsf{nds}(M)} \xrightarrow{m?w} \mathrm{Top}^{\phi_j}_{\mathcal{A}/\mathsf{nds}(M)}$ by rule (RcvEnb). Hence $M' \mid \mathrm{Top}^{\phi_j}_{\mathcal{A}/\mathsf{nds}(M)} \xrightarrow{m!w\triangleright obs} M'' \mid \mathrm{Top}^{\phi_j}_{\mathcal{A}/\mathsf{nds}(M)}$ by rule (Bcast) and the fact that $\mathsf{nds}(A') = \mathcal{A} = \mathsf{nds}\big(\mathrm{Top}^{\phi_j}_{\mathcal{A}/\mathsf{nds}(M)}\big)$. Finally, by rule (Obs): $M' \mid \mathrm{Top}^{\phi_j}_{\mathcal{A}/\mathsf{nds}(M)} \xrightarrow{!w\triangleright obs} M'' \mid \mathrm{Top}^{\phi_j}_{\mathcal{A}/\mathsf{nds}(M)}$.

$\alpha = \tau$. The most significant case is when $\tau$ is derived by an application of rule (Shh), then we have $M' \mid A' \xrightarrow{a!w\triangleright\emptyset} N$ and $a \in \mathsf{nds}(A') = \mathcal{A}$ since the broadcast from any of the nodes in $\mathsf{nds}(M') = \mathsf{nds}\big(M^{\mathcal{A}}\big)$ can be observed by the node *obs*. In this case we have $M' \xrightarrow{a?w} M''$ and $A' \xrightarrow{a!w\triangleright m} A''$ where $m$ is the single node of $M$ attacked by $a$. Now also $\mathrm{Top}^{\phi_j}_{\mathcal{A}/\mathsf{nds}(M)} \xrightarrow{\tau} \xrightarrow{a!w\triangleright m} \mathrm{Top}^{\phi_j}_{\mathcal{A}/\mathsf{nds}(M)}$ by rules (Tau) and (Snd) since the attacking node associated to $m$ does not change and $\mathrm{msg}(A') \subseteq \mathcal{D}(\phi_j)$. Hence, by rule (Bcast): $M' \mid \mathrm{Top}^{\phi_j}_{\mathcal{A}/\mathsf{nds}(M)} \stackrel{a!w\triangleright\emptyset}{\Longrightarrow} M'' \mid \mathrm{Top}^{\phi_j}_{\mathcal{A}/\mathsf{nds}(M)}$. Thus $M' \mid \mathrm{Top}^{\phi_j}_{\mathcal{A}/\mathsf{nds}(M)} \stackrel{\tau}{\Longrightarrow} M'' \mid \mathrm{Top}^{\phi_j}_{\mathcal{A}/\mathsf{nds}(M)}$ by rule (Shh). □

**Proof of Theorem 3.12** By Lemma A.2 we have $M^{\mathcal{A}}_{O} \mid A \lesssim M^{\mathcal{A}}O \mid \mathrm{Top}^{\phi_0}_{\mathcal{A}/\mathsf{nds}(M)}$ for every $A \in \mathbb{A}^{\phi_0}_{\mathcal{A}/\mathsf{nds}(M)}$. Then by transitivity of $\lesssim$ we have $M^{\mathcal{A}}O \mid A \lesssim N$ for every $A \in \mathbb{A}^{\phi_0}_{\mathcal{A}/\mathsf{nds}(M)}$ and we conclude that $M$ is $tGNDC^{N}_{\phi_0,O}$. □

**Proof of Proposition 4.1** By induction on $i$ we show that whenever $\textsc{bs}[\hat{D}_1]^{\nu_{\textsc{bs}}} \stackrel{\Lambda}{\Longrightarrow} \textsc{bs}[\hat{D}_i]^{obs}$ or $m[\hat{A}_1]^{obs} \stackrel{\Lambda}{\Longrightarrow} m[\hat{A}_i]^{obs}$ then $\#^{\sigma}(\Lambda) = 2(i-1)$. Moreover, we observe that $!p_i \triangleright obs$ can be performed exclusively because $m[\hat{A}_i]^{obs} \xrightarrow{!p_i \triangleright obs}$. While $!\mathsf{end}_i \triangleright obs$ can be performed exclusively because $\textsc{bs}[\hat{D}_i]^{obs} \stackrel{\Omega}{\Longrightarrow} \xrightarrow{!\mathsf{end}_i \triangleright obs}$ with $\#^{\sigma}\Omega = 2$. Hence we deduce that:

1. if $m[\hat{A}_1]^{obs} \stackrel{\Lambda}{\Longrightarrow} \xrightarrow{!p_i \triangleright obs}$ then $\#^{\sigma}(\Lambda) = 2(i-1)$.
2. if $\textsc{bs}[\hat{D}_1]^{obs} \stackrel{\Lambda}{\Longrightarrow} \xrightarrow{!\mathsf{end}_i \triangleright obs}$ then $\#^{\sigma}(\Lambda) = 2i$.

Now, the result is a straightforward consequence of these two properties. □

**Proof of Theorem 4.2** The system $(\mu\text{TESLA}'_{boot})^{\mathcal{A}} \mid A$ performs the following computation:

$$\begin{array}{lr}
\textsc{bs}[D'_1]^{\nu_{\textsc{bs}}} \mid m[A_1]^{\nu_m} \mid A & \xrightarrow{!p_1 \triangleright obs} \\
\textsc{bs}[D'_1]^{\nu_{\textsc{bs}}} \mid m[\sigma.B_1]^{\nu_m} \mid a[\sigma.!\langle p_1\rangle.\mathsf{nil}]^{\nu_a} \mid b[Y]^{\nu_b} & \xrightarrow{\sigma} \\
\textsc{bs}[\sigma.D'_2]^{\nu_{\textsc{bs}}} \mid m[B_1]^{\nu_m} \mid a[!\langle p_1\rangle.\mathsf{nil}]^{\nu_a} \mid b[\lfloor ?(y).\sigma.!\langle y\rangle.\mathsf{nil}\rfloor \mathsf{nil}]^{\nu_b} & \xrightarrow{\tau} \\
\textsc{bs}[\sigma.D'_2]^{\nu_{\textsc{bs}}} \mid m[B_1]^{\nu_m} \mid a[\mathsf{nil}]^{\nu_a} \mid b[\sigma.!\langle p_1\rangle.\mathsf{nil}]^{\nu_b} & \xrightarrow{\sigma} \\
\textsc{bs}[D'_2]^{\nu_{\textsc{bs}}} \mid m[A_2]^{\nu_m} \mid a[\mathsf{nil}]^{\nu_a} \mid b[!\langle p_1\rangle.\mathsf{nil}]^{\nu_b} & \xrightarrow{\tau} \\
\textsc{bs}[E'_2\{^{p_1}/_{p}\}]^{\nu_{\textsc{bs}}} \mid m[A_2]^{\nu_m} \mid a[\mathsf{nil}]^{\nu_a} \mid b[\mathsf{nil}]^{\nu_b} & \xrightarrow{!p_2 \triangleright obs} \\
\textsc{bs}[E'_2\{^{p_1}/_{p}\}]^{\nu_{\textsc{bs}}} \mid m[\sigma.B_2]^{\nu_m} \mid a[\mathsf{nil}]^{\nu_a} \mid b[\mathsf{nil}]^{\nu_b} & \xrightarrow{\sigma} \\
\textsc{bs}[!\langle w_i\rangle.G'_2\{^{p_1}/_{p}\}]^{\nu_{\textsc{bs}}} \mid m[B_2]^{\nu_m} \mid a[\mathsf{nil}]^{\nu_a} \mid b[\mathsf{nil}]^{\nu_b} & \xrightarrow{!w_1 \triangleright obs} \\
\textsc{bs}[E^{4\prime}_2\{^{n_1}/_{n}\}]^{\nu_{\textsc{bs}}} \mid m[C_2\{^{w_1}/_{w}\}]^{\nu_m} \mid a[\mathsf{nil}]^{\nu_a} \mid b[\mathsf{nil}]^{\nu_b} & \xrightarrow{\sigma} \\
\textsc{bs}[!\langle \mathsf{end}_1\rangle.D'_3]^{\nu_{\textsc{bs}}} \mid m[A_3]^{\nu_m} \mid a[\mathsf{nil}]^{\nu_a} \mid b[\mathsf{nil}]^{\nu_b} & \xrightarrow{!\mathsf{end}_1 \triangleright obs} \quad .
\end{array}$$

Hence agreement is not reached. □

**Proof of Lemma 4.4** We provide the proper simulation in both cases.
*Case 1: Base Station.* To show that $\textsc{bs}[D_1]^{b} \mid \textsc{Top}^{\phi_0}_{b/\textsc{bs}} \lesssim \textsc{bs}[Tick]^{\emptyset}$ we define the relation

$$\mathcal{R} \stackrel{\text{def}}{=} \left\{ \left( M, \textsc{bs}[Tick]^{\emptyset} \right) \quad \text{such that } \textsc{bs}[D_1]^{\{b\}} \mid \textsc{Top}^{\phi_0}_{b/\textsc{bs}} \stackrel{\Lambda}{\Longrightarrow} M \right\} .$$

We first notice that for every $(M, \textsc{bs}[Tick]^{\emptyset}) \in \mathcal{R}$ we have $\mathsf{Env}(M) = \emptyset$ . Thus the most significant actions can only be $M \xrightarrow{\tau}$ or $M \xrightarrow{\sigma}$ or input actions that can be derived without applying rule (Rcv). Then it is straightforward to prove that $\mathcal{R}$ is a simulation.
*Case 2: Node.* To show that $m[A''_1]^{\{a,obs\}} \mid \textsc{Top}^{\phi_0}_{a/m} \lesssim m[\overline{A}_1]^{obs}$ we pick an index $i \geq 1$, the messages $\overline{w}, w', w''$ and we build a relation $\mathcal{R}_i(w', \overline{w}, w'')$ containing the pair $(m[A''_i]^{\{a,obs\}} \mid \textsc{Top}^{\phi_{2(i-1)}}_{a/m},\ m[\overline{A}_i]^{obs})$ along with its derivatives which may be generated when $m$ receives $\overline{w}$ from the attacker. To improve readability: (i) we abbreviate the process $R\langle i+1, i-1, \bot, k_{i-1}\rangle$ simply as $R_i$, (ii) we employ the structural congruence $\equiv$ to rewrite $\overline{A}_i$ as:

$$\overline{A}_i \stackrel{\text{def}}{=} !\langle p_i\rangle.\sigma.\overline{B}_i \qquad \overline{B}_i \stackrel{\text{def}}{=} \lfloor \tau.\overline{C}_i \rfloor \overline{A}_{i+1} \qquad \overline{C}_i \stackrel{\text{def}}{=} \sigma.!\langle \mathsf{auth}_i\rangle.R_i \quad .$$

Then we define $\mathcal{R}_i(w', \bar{w}, w'')$ to be the following binary relation:

$$\Big\{ \Big( m[A''_i]^{\{a,obs\}} \mid \text{Top}^{\phi_{2(i-1)}}_{a/m},\ m[\bar{A}_i]^{obs} \Big),\ \Big( m[A''_i]^{\{a,obs\}} \mid a[!\langle w' \rangle.\mathrm{T}_{\phi_{2(i-1)}}]^m,\ m[\bar{A}_i]^{obs} \Big)\ ,$$
$$\Big( m[\sigma.B''_i]^{\{a,obs\}} \mid \text{Top}^{\phi_{2(i-1)}}_{a/m},\ m[\sigma.\bar{B}_i]^{obs} \Big),\ \Big( m[\sigma.B''_i]^{\{a,obs\}} \mid a[!\langle w' \rangle.\mathrm{T}_{\phi_{2(i-1)}}]^m,\ m[\sigma.\bar{B}_i]^{obs} \Big)\ ,$$
$$\Big( m[B''_i]^{\{a,obs\}} \mid \text{Top}^{\phi_{2i-1}}_{a/m},\ m[\bar{B}_i]^{obs} \Big),\ \Big( m[B''_i]^{\{a,obs\}} \mid a[!\langle \bar{w} \rangle.\mathrm{T}_{\phi_{2i-1}}]^m,\ m[\bar{B}_i]^{obs} \Big)\ ,$$
$$\Big( m[\{^{\bar{w}}/_{w}\}C''_i]^{\{a,obs\}} \mid \text{Top}^{\phi_{2i-1}}_{a/m},\ m[\bar{B}_i]^{obs} \Big),\ \Big( m[\{^{\bar{w}}/_{w}\}C''_i]^{\{a,obs\}} \mid a[!\langle \bar{w} \rangle.\mathrm{T}_{\phi_{2i-1}}]^m,\ m[\bar{B}_i]^{obs} \Big)\ ,$$
$$\Big( m[!\langle \mathsf{auth}_i \rangle.R_i]^{\{a,obs\}} \mid \text{Top}^{\phi_{2i}}_{a/m},\ m[!\langle \mathsf{auth}_i \rangle.R_i]^{obs} \Big)\ ,$$
$$\Big( m[!\langle \mathsf{auth}_i \rangle.R_i]^{\{a,obs\}} \mid a[!\langle w'' \rangle.\mathrm{T}_{\phi_{2i}}]^m,\ m[!\langle \mathsf{auth}_i \rangle.R_i]^{obs} \Big)\ ,$$
$$\Big( m[R_i]^{\{a,obs\}} \mid \text{Top}^{\phi_{2i}}_{a/m},\ m[R_i]^{obs} \Big),\ \Big( m[R_i]^{\{a,obs\}} \mid a[!\langle w'' \rangle.\mathrm{T}_{\phi_{2i}}]^m !\langle w'' \rangle.,\ m[R_i]^{obs} \Big) \Big\}\ .$$

Now we notice that both the process $R_i$ and its derivatives cannot perform any broadcast action. Moreover, the network $m[R_i]^{\{a,obs\}} \mid \text{Top}^{\phi_{2i}}_{a/m}$, along with its derivatives, can perform just $\tau$-actions, $\sigma$-actions or input actions derived without applying rule (Rcv). Then it is straightforward to prove that there exists a simulation $\overline{\mathcal{R}}_i$ containing the pair $(m[R_i]^{\{a,obs\}} \mid \text{Top}^{\phi_{2i}}_{a/m},\ m[R_i]^{obs})$.

We show that the required simulation is given by the following relation

$$\mathcal{R} \stackrel{\text{def}}{=} \bigcup_{i \geq 1} \Big( \overline{\mathcal{R}}_i \ \cup \bigcup_{\substack{w' \in \mathcal{D}(\phi_{2(i-1)}) \\ \bar{w} \in \mathcal{D}(\phi_{2i-1}) \\ w'' \in \mathcal{D}(\phi_{2i})}} \mathcal{R}_i(w', \bar{w}, w'') \Big)\ .$$

We outline the most significant cases. We omit input actions since the environment contains exclusively the node *obs* which cannot transmit, thus all input actions can be derived just by combining rules (RcvEnb) and (RcvPar). We also omit internal choices of the attacker.
In the pair $(\ m[A''_i]^{\{a,obs\}} \mid \text{Top}^{\phi_{2(i-1)}}_{a/m}\ ,\ m[\bar{A}_i]^{obs}\ )$ we have a single significant action:

- $m[A''_i]^{\{a,obs\}} \mid \text{Top}^{\phi_{2(i-1)}}_{a/m} \xrightarrow{!p_i \triangleright obs} m[\sigma.B''_i]^{\{a,obs\}} \mid \text{Top}^{\phi_{2(i-1)}}_{a/m}$. Then the second network replies with $m[\bar{A}_i]^{obs} \xrightarrow{!p_i \triangleright obs} m[\sigma.\bar{B}_i]^{obs}$.

In the pair $(\ m[\sigma.B''_i]^{\{a,obs\}} \mid \text{Top}^{\phi_{2(i-1)}}_{a/m}\ ,\ m[\sigma.\bar{B}_i]^{obs}\ )$ we have a single significant action:

- $m[\sigma.B''_i]^{\{a,obs\}} \mid \text{Top}^{\phi_{2(i-1)}}_{a/m} \xrightarrow{\sigma} m[B''_i]^{\{a,obs\}} \mid \text{Top}^{\phi_{2i-1}}_{a/m}$. Then $m[\sigma.\bar{B}_i]^{obs} \xrightarrow{\sigma} m[\bar{B}_i]^{obs}$.

In the pair $(m[B''_i]^{\{a,obs\}} \mid \text{Top}^{\phi_{2i-1}}_{a/m}\ ,\ m[\sigma.\bar{B}_i]^{obs}\ )$ we have a significant action:

- $m[B''_i]^{\{a,obs\}} \mid \text{Top}^{\phi_{2i-1}}_{a/m} \xrightarrow{\sigma} m[A''_{i+1}]^{\{a,obs\}} \mid \text{Top}^{\phi_{2i}}_{a/m}$ where $m$ does not receive anything thus performs a timeout. Then $m[\bar{B}_i]^{obs} \xrightarrow{\sigma} m[\bar{A}_{i+1}]^{obs}$.

In the pair $(m[B''_i]^{\{a,obs\}} \mid a[!\langle \bar{w} \rangle.\mathrm{T}_{\phi_{2i-1}}]^m\ ,\ m[\sigma.\bar{B}_i]^{obs}\ )$ we have two significant actions:

- $m[B''_i]^{\{a,obs\}} \mid a[!\langle \bar{w} \rangle.\mathrm{T}_{\phi_{2i-1}}]^m \xrightarrow{\tau} m[\{^{\bar{w}}/_{w}\}C''_i]^{\{a,obs\}} \mid \text{Top}^{\phi_{2i-1}}_{a/m}$ where $m$ receives $\bar{w}$. Then $m[\bar{B}_i]^{obs} \Longrightarrow m[\sigma.\bar{B}_i]^{obs}$ .

- $m[B''_i]^{\{a,obs\}} \mid a[!\langle\bar{w}\rangle.\mathrm{T}_{\phi_{2i-1}}]^m \xrightarrow{\tau} m[B''_i]^{\{a,obs\}} \mid \textsc{Top}^{\phi_{2i-1}}_{a/m}$ where $\bar{w}$ gets lost. Then the second network $m[\sigma.\bar{B}_i]^{obs} \Rightarrow m[\sigma.\bar{B}_i]^{obs}$.

In the pair $(m[\{^{\bar{w}}/_{w}\}C''_i]^{\{a,obs\}} \mid \textsc{Top}^{\phi_{2i-1}}_{a/m}\ ,\ m[\bar{B}_i]^{obs})$ we have two significant transitions

- $m[\{^{\bar{w}}/_{w}\}C''_i]^{\{a,obs\}} \mid \textsc{Top}^{\phi_{2i-1}}_{a/m} \xrightarrow{\sigma} m[!\langle \mathsf{auth}_i\rangle.R_i]^{\{a,obs\}} \mid \textsc{Top}^{\phi_{2i}}_{a/m}$ where $m$ verifies the MAC of the message $\bar{w}$, checks that the nonce included in $\bar{w}$ is actually $n_i$ and then it authenticates the key $k_{i-1}$. Then $m[\bar{B}_i]^{obs} \stackrel{\sigma}{\Longrightarrow} m[!\langle \mathsf{auth}_i\rangle.R_i]^{obs}$.
- $m[\{^{\bar{w}}/_{w}\}C''_i]^{\{a,obs\}} \mid \textsc{Top}^{\phi_{2i-1}}_{a/m} \xrightarrow{\sigma} m[A''_{i+1}]^{\{a,obs\}} \mid \textsc{Top}^{\phi_{2i}}_{a/m}$ where $m$ does not verify the MAC of the message $\bar{w}$, thus it cannot check that the nonce included in $\bar{w}$ is actually $n_i$, or in general it finds out that the message is corrupted. Then $m[\bar{B}_i]^{obs} \stackrel{\sigma}{\Longrightarrow} m[\bar{A}_{i+1}]^{obs}$. □

**Proof of Proposition 4.6** Similar to that of Proposition 4.1. □

**Proof of Lemma 4.7** We provide the proper simulation in both the cases.
*Case 1: Base Station.* We notice that the process $S_i$, along with its derivatives, cannot receive any message. Thus an attacker in $b$ cannot affect the behaviour of $\textsc{bs}[S_1]^{\{b,obs\}}$. Hence it is straightforward to prove that $\textsc{bs}[S_1]^{\{b,obs\}} \mid \textsc{Top}^{\phi_0}_{b/\textsc{bs}} \lessapprox \textsc{bs}[S_1]^{obs}$.
*Case 2: Nodes.* We fix a node $m \in \{m_1, \ldots, m_h\}$, we let $a \in \{a_1, \ldots, a_h\}$ denote its corresponding attacking place and we show that

$$m[R'\langle 1, -1, \perp, \bar{k}\rangle]^{\{a,obs\}} \mid \textsc{Top}^{\phi_0}_{a/m} \quad \lessapprox \quad m[\hat{R}_1]^{obs} \ .$$

To uniform the notation, we define $k_{-1} \stackrel{\text{def}}{=} \bar{k}$. We pick the indexes $i \geq 1$ and $-1 \geq l \geq i-2$, and the messages $\hat{r}$, $\hat{p}$, $\hat{k}$ and $\hat{q}$. Then we build the relation $\mathcal{R}^{l,\hat{r}}_i(\hat{p}, \hat{k}, \hat{q})$ which contains the pair

$$\Big(m[R'\langle i, l, \hat{r}, k_l\rangle]^{\{a,obs\}} \mid \textsc{Top}^{\phi_{2(i-1)}}_{a/m}\ ,\ m[\hat{R}_i]^{obs}\Big)$$

along with its derivatives which may be generated when $m$ first receives $\hat{p}$ and then $\hat{k}$ from the attacker. To improve the readability: (i) we define $\nu'_m \stackrel{\text{def}}{=} \{a, obs\}$, (ii) we employ the structural congruence $\equiv$ to rewrite the process $\hat{R}_i$ as:

$$\hat{R}_i \stackrel{\text{def}}{=} \sigma.\hat{P}_i \qquad \hat{P}_i \stackrel{\text{def}}{=} \lfloor \tau.\sigma.\hat{Z}_{i+1}\rfloor \hat{R}_{i+1} \qquad \hat{Z}_i \stackrel{\text{def}}{=} !\langle \mathsf{auth}_{i-2}\rangle.\hat{R}_i \ .$$

Then we define

$$\begin{aligned}
\mathcal{R}_i^{l,\hat{r}}(\hat{p},\hat{k},\hat{q}) \stackrel{\text{def}}{=} \Big\{ & \Big( m[R'\langle i,l,\hat{r},k_l\rangle]^{\nu'_m} \mid \textsc{Top}^{\phi_{2(i-1)}}_{a/m},\ m[\hat{R}_i]^{obs} \Big) \ , \\
& \Big( m[R'\langle i,l,\hat{r},k_l\rangle]^{\nu'_m} \mid a[!\langle \hat{p}\rangle.\mathrm{T}_{\phi_{2(i-1)}}]^m,\ m[\hat{R}_i]^{obs} \Big) \ , \\
& \Big( m[\sigma.P'\langle i,l,\hat{p},\hat{r},k_l\rangle]^{\nu'_m} \mid \textsc{Top}^{\phi_{2(i-1)}}_{a/m},\ m[\hat{R}_i]^{obs} \Big) \ , \\
& \Big( m[\sigma.P'\langle i,l,\hat{p},\hat{r},k_l\rangle]^{\nu'_m} \mid a[!\langle \hat{p}\rangle.\mathrm{T}_{\phi_{2(i-1)}}]^m,\ m[\hat{R}_i]^{obs} \Big) \ , \\
& \Big( m[P'\langle i,l,\hat{p},\hat{r},k_l\rangle]^{\nu'_m} \mid \textsc{Top}^{\phi_{2i-1}}_{a/m},\ m[\hat{P}_i]^{obs} \Big) \ , \\
& \Big( m[P'\langle i,l,\hat{p},\hat{r},k_l\rangle]^{\nu'_m} \mid a[!\langle \hat{k}\rangle.\mathrm{T}_{\phi_{2i-1}}]^m,\ m[\hat{P}_i]^{obs} \Big) \ , \\
& \Big( m[T'\langle i,l,\hat{p},\hat{r},k_l,\hat{k}\rangle]^{\nu'_m} \mid \textsc{Top}^{\phi_{2i-1}}_{a/m},\ m[\hat{P}_i]^{obs} \Big) \ , \\
& \Big( m[T'\langle i,l,\hat{p},\hat{r},k_l,\hat{k}\rangle]^{\nu'_m} \mid a[!\langle \hat{k}\rangle.\mathrm{T}_{\phi_{2i-1}}]^m,\ m[\hat{P}_i]^{obs} \Big) \ , \\
& \Big( m[Q'\langle i,l,\hat{r},k_l\rangle]^{\nu'_m} \mid \textsc{Top}^{\phi_{2i-1}}_{a/m},\ m[\hat{P}_i]^{obs} \Big)\Big\} \ , \\
& \Big( m[Q'\langle i,l,\hat{r},k_l\rangle]^{\nu'_m} \mid a[!\langle \hat{k}\rangle.\mathrm{T}_{\phi_{2i-1}}]^m,\ m[\hat{P}_i]^{obs} \Big)\Big\} \ , \\
& \Big( m[Z'\langle i{+}1,i{-}1,\hat{p},\hat{r},k_{i-1}\rangle]^{\nu'_m} \mid \textsc{Top}^{\phi_{2i}}_{a/m},\ m[\hat{Z}_{i+1}]^{obs} \Big) \ , \\
& \Big( m[Z'\langle i{+}1,i{-}1,\hat{p},\hat{r},k_{i-1}\rangle]^{\nu'_m} \mid a[!\langle \hat{q}\rangle.\mathrm{T}_{\phi_{2i}}]^m,\ m[\hat{Z}_{i+1}]^{obs} \Big)\Big\} \ .
\end{aligned}$$

and we show that the required simulation is given by the following relation

$$\mathcal{R} \stackrel{\text{def}}{=} \bigcup_{i\geq 1} \ \bigcup_{\substack{-1\leq l\leq i-2 \\ \hat{r}\in\mathcal{D}(\phi_{2(i-2)})}} \ \bigcup_{\substack{\hat{p}\in\mathcal{D}(\phi_{2(i-1)}) \\ \hat{k}\in\mathcal{D}(\phi_{2i-1}) \\ \hat{q}\in\mathcal{D}(\phi_{2i})}} \mathcal{R}_i^{l,\hat{r}}(\hat{p},\hat{k},\hat{q})$$

As done for Lemma 4.4, we outline the most significant cases. Again, we omit input actions and internal choices of the attacker.
In the pair ( $m[R'\langle i,l,\hat{r},k_l\rangle]^{\nu'_m} \mid \textsc{Top}^{\phi_{2(i-1)}}_{a/m},\ m[\hat{R}_i]^{obs}$ ) we have a significant action:

- $m[R'\langle i,l,\hat{r},k_l\rangle]^{\nu'_m} \mid \textsc{Top}^{\phi_{2(i-1)}}_{a/m} \xrightarrow{\sigma} m[Q'\langle i,l,\hat{r},k_l\rangle]^{\nu'_m} \mid \textsc{Top}^{\phi_{2i-1}}_{a/m}$, where $m$ does not receive anything. Then $m[\hat{R}_i]^{obs} \stackrel{\sigma}{\Longrightarrow} m[\hat{P}_i]^{obs}$.

In the pair ( $m[R'\langle i,l,\hat{r},k_l\rangle]^{\nu'_m} \mid a[!\langle \hat{p}\rangle.\mathrm{T}_{\phi_{2(i-1)}}]^m,\ m[\hat{R}_i]^{obs}$ ) we have two significant actions:

- $m[R'\langle i,l,\hat{r},k_l\rangle]^{\nu'_m} \mid a[!\langle \hat{p}\rangle.\mathrm{T}_{\phi_{2(i-1)}}]^m \xrightarrow{\tau} m[\sigma.P'\langle i,l,\hat{p},\hat{r},k_l\rangle]^{\nu'_m} \mid \textsc{Top}^{\phi_{2(i-1)}}_{a/m}$, where $m$ receives $\hat{p}$ from the attacker. Then $m[\hat{R}_i]^{obs} \Longrightarrow m[\hat{R}_i]^{obs}$.
- $m[R'\langle i,l,\hat{r},k_l\rangle]^{\nu'_m} \mid a[!\langle \hat{p}\rangle.\mathrm{T}_{\phi_{2(i-1)}}]^m \xrightarrow{\tau} m[R'\langle i,l,\hat{r},k_l\rangle]^{\nu'_m} \mid \textsc{Top}^{\phi_{2(i-1)}}_{a/m}$, where $\hat{p}$ gets lost. Then $m[\hat{R}_i]^{obs} \Longrightarrow m[\hat{R}_i]^{obs}$.

In the pair ( $m[\sigma.P'\langle i,l,\hat{p},\hat{r},k_l\rangle]^{\nu'_m} \mid \textsc{Top}^{\phi_{2(i-1)}}_{a/m},\ m[\hat{R}_i]^{obs}$ ) we have just a significant action:

- $m[\sigma.P'\langle i,l,\hat{p},\hat{r},k_l\rangle]^{\nu'_m} \mid \textsc{Top}^{\phi_{2(i-1)}}_{a/m} \xrightarrow{\sigma} m[P'\langle i,l,\hat{p},\hat{r},k_l\rangle]^{\nu'_m} \mid \textsc{Top}^{\phi_{2i-1}}_{a/m}$. Then the second network replies with $m[\hat{R}_i]^{obs} \stackrel{\sigma}{\Longrightarrow} m[\hat{P}_i]^{obs}$.

In the pair $( m[P'\langle i,l,\hat{p},\hat{r},k_l\rangle]^{\nu'_m} \mid \mathrm{TOP}^{\phi_{2i-1}}_{a/m},\ m[\hat{P}_i]^{obs} )$ we have just a significant action:

- $m[P'\langle i,l,\hat{p},\hat{r},k_l\rangle]^{\nu'_m} \mid \mathrm{TOP}^{\phi_{2i-1}}_{a/m} \xrightarrow{\sigma} m[R'\langle i{+}1,l,\hat{p},k_l\rangle]^{\nu'_m} \mid \mathrm{TOP}^{\phi_{2i}}_{a/m}$, where $m$ does not receive anything. Then $m[\hat{P}_i]^{obs} \xRightarrow{\sigma} m[\hat{R}_{i+1}]^{obs}$.

In the pair $( m[P'\langle i,l,\hat{p},\hat{r},k_l\rangle]^{\nu'_m} \mid a[!\langle\hat{k}\rangle.\mathrm{T}_{\phi_{2i-1}}]^{m},\ m[\hat{P}_i]^{obs} )$ we have two significant actions:

- $m[P'\langle i,l,\hat{p},\hat{r},k_l\rangle]^{\nu'_m} \mid a[!\langle\hat{k}\rangle.\mathrm{T}_{\phi_{2i-1}}]^{m} \xrightarrow{\tau} m[T'\langle i,l,\hat{p},\hat{r},k_l,\hat{k}\rangle]^{\nu'_m} \mid \mathrm{TOP}^{\phi_{2i-1}}_{a/m}$, where $m$ receives $\hat{k}$. Then the second network replies with $m[\hat{P}_i]^{obs} \Rightarrow m[\hat{P}_i]^{obs}$.

- $m[P'\langle i,l,\hat{p},\hat{r},k_l\rangle]^{\nu'_m} \mid a[!\langle\hat{k}\rangle.\mathrm{T}_{\phi_{2i-1}}]^{m} \xrightarrow{\tau} m[P'\langle i,l,\hat{p},\hat{r},k_l\rangle]^{\nu'_m} \mid \mathrm{TOP}^{\phi_{2i-1}}_{a/m}$, where $\hat{k}$ gets lost. The second network replies with $m[\hat{P}_i]^{obs} \Rightarrow m[\hat{P}_i]^{obs}$.

In the pair $( m[T'\langle i,l,\hat{p},\hat{r},k_l,\hat{k}\rangle]^{\nu'_m} \mid \mathrm{TOP}^{\phi_{2i-1}}_{a/m},\ m[\hat{P}_i]^{obs} )$ we have three significant actions:

- $m[T'\langle i,l,\hat{p},\hat{r},k_l,\hat{k}\rangle]^{\nu'_m} \mid \mathrm{TOP}^{\phi_{2i-1}}_{a/m} \xrightarrow{\sigma} m[Z'\langle i{+}1,i{-}1,\hat{p},\hat{r},k_{i-1}\rangle]^{\nu'_m} \mid \mathrm{TOP}^{\phi_{2i}}_{a/m}$ where $m$ checks that $k_l = F^{i-1-l}(\hat{h})$ and authenticates $\hat{r} = p_{i-1}$. Then $m[\hat{P}_i]^{obs} \xRightarrow{\sigma} m[\hat{Z}_{i+1}]^{obs}$.

- $m[T'\langle i,l,\hat{p},\hat{r},k_l,\hat{k}\rangle]^{\nu'_m} \mid \mathrm{TOP}^{\phi_{2i-1}}_{a/m} \xrightarrow{\sigma} m[R'\langle i{+}1,i{-}1,\hat{p},k_{i-1}\rangle]^{\nu'_m} \mid \mathrm{TOP}^{\phi_{2i}}_{a/m}$ where $m$ checks that $k_l = F^{i-l}(\hat{h})$ without but it does not authenticate $\hat{r}$. Then $m[\hat{P}_i]^{obs} \xRightarrow{\sigma} m[\hat{R}_{i+1}]^{obs}$.

- $m[T'\langle i,l,\hat{p},\hat{r},k_l,\hat{k}\rangle]^{\nu'_m} \mid \mathrm{TOP}^{\phi_{2i-1}}_{a/m} \xrightarrow{\sigma} m[R'\langle i{+}1,l,\hat{p},k_l\rangle]^{\nu'_m} \mid \mathrm{TOP}^{\phi_{2i}}_{a/m}$ where $m$ verifies $k_l \neq F^{i-l}(\hat{h})$. Then again $m[\hat{P}_i]^{obs} \xRightarrow{\sigma} m[\hat{R}_{i+1}]^{obs}$ by timeout.

In the pair $( m[Q'\langle i,l,\hat{r},k_l\rangle]^{\nu'_m} \mid \mathrm{TOP}^{\phi_{2i-1}}_{a/m},\ m[\hat{P}_i]^{obs} )$ we have a significant action

- $m[Q'\langle i,l,\hat{r},k_l\rangle]^{\nu'_m} \mid \mathrm{TOP}^{\phi_{2i-1}}_{a/m} \xrightarrow{\sigma} m[R\langle i{+}1,l,\hat{r},k_l\rangle]^{\nu'_m} \mid \mathrm{TOP}^{\phi_{2i}}_{a/m}$, where $m$ does not receive anything and thus performs a timeout. Then $m[\hat{P}_i]^{obs} \xRightarrow{\sigma} m[\hat{R}_{i+1}]^{obs}$.

In the pair $( m[Q'\langle i,l,\hat{r},k_l\rangle]^{\nu'_m} \mid a[!\langle\hat{k}\rangle.\mathrm{T}_{\phi_{2i-1}}]^{m},\ m[\hat{P}_i]^{obs} )$ the first network can perform two significant actions

- $m[Q'\langle i,l,\hat{r},k_l\rangle]^{\nu'_m} \mid a[!\langle\hat{k}\rangle.\mathrm{T}_{\phi_{2i-1}}]^{m} \xrightarrow{\tau} m[T\langle i,l,\hat{r},\hat{r},k_l,\hat{k}\rangle]^{\nu'_m} \mid \mathrm{TOP}^{\phi_{2i-1}}_{a/m}$, where $m$ receives $\hat{k}$. Then the second network replies $m[\hat{P}_i]^{obs} \Rightarrow m[\hat{P}_i]^{obs}$.

- $m[Q'\langle i,l,\hat{r},k_l\rangle]^{\nu'_m} \mid a[!\langle\hat{k}\rangle.\mathrm{T}_{\phi_{2i-1}}]^{m} \xrightarrow{\tau} m[R\langle i{+}1,l,\hat{r},k_l\rangle]^{\nu'_m} \mid \mathrm{TOP}^{\phi_{2i-1}}_{a/m}$, where $\hat{k}$ gets lost. Then the second network replies $m[\hat{P}_i]^{obs} \Rightarrow m[\hat{P}_i]^{obs}$. □

**Proof of Proposition 5.1** Similar to that of Proposition 4.1. □

**Proof of Theorem 5.2** The system $(\text{LEAP}'+)^{\mathcal{A}} \mid A$ admits the following computation:

$$\begin{array}{lr}
m[S_1]^{\nu_m} \mid r[R']^{\nu_r} \mid A & \xrightarrow{!\mathsf{hello}_1 \triangleright obs} \\
m[\sigma.P]^{\nu_m} \mid r[R']^{\nu_r} \mid a[\sigma.!\langle \mathsf{hello}_1 \rangle.\mathsf{nil}]^{\nu_a} \mid b[Y]^{\nu_b} & \xrightarrow{\sigma} \\
m[P]^{\nu_m} \mid r[\sigma.R']^{\nu_r} \mid a[!\langle \mathsf{hello}_1 \rangle.\mathsf{nil}]^{\nu_a} \mid b[\lfloor ?(y).\sigma.!\langle y \rangle.\mathsf{nil}]^{\nu_b} & \xrightarrow{\tau} \\
m[\{^{\mathsf{hello}_1}/_q\}P^1]^{\nu_m} \mid r[\sigma.R']^{\nu_r} \mid a[\mathsf{nil}]^{\nu_a} \mid b[\sigma.!\langle \mathsf{hello}_1 \rangle.\mathsf{nil}]^{\nu_b} & \xrightarrow{\sigma} \\
m[S_2]^{\nu_m} \mid r[R']^{\nu_r} \mid a[\mathsf{nil}]^{\nu_a} \mid b[!\langle \mathsf{hello}_1 \rangle.\mathsf{nil}]^{\nu_b} & \xrightarrow{\tau} \\
m[S_2]^{\nu_m} \mid r[\sigma.!\langle q_1 \rangle.R^{8'}]^{\nu_r} \mid a[\mathsf{nil}]^{\nu_a} \mid b[\mathsf{nil}]^{\nu_b} & \xrightarrow{!\mathsf{hello}_2 \triangleright obs} \\
m[\sigma.P]^{\nu_m} \mid r[\sigma.!\langle q_1 \rangle.R^{8'}]^{\nu_r} \mid a[\mathsf{nil}]^{\nu_a} \mid b[\mathsf{nil}]^{\nu_b} & \xrightarrow{\sigma} \\
m[P]^{\nu_m} \mid r[!\langle q_1 \rangle.R^{8'}]^{\nu_r} \mid a[\mathsf{nil}]^{\nu_a} \mid b[\mathsf{nil}]^{\nu_b} & \xrightarrow{!q_1 \triangleright obs} \\
m[\{^{q_1}/_q\}P_1]^{\nu_m} \mid r[R^{8'}]^{\nu_r} \mid a[\mathsf{nil}]^{\nu_a} \mid b[\mathsf{nil}]^{\nu_b} & \xrightarrow{\sigma} \\
m[S^3]^{\nu_m} \mid r[!\langle \mathsf{end}_1 \rangle.\mathsf{nil}]^{\nu_r} \mid a[\mathsf{nil}]^{\nu_a} \mid b[\mathsf{nil}]^{\nu_b} & \xrightarrow{!\mathsf{end}_1 \triangleright obs}
\end{array}$$

Then agreement is not reached. □

**Proof of Lemma 5.4** We prove this lemma by showing the appropriate simulations.
*Case 1: Sender.* We define $\nu'_m = \{a, obs\}$. We need to prove $m[S_1]^{\nu'_m} \mid \text{Top}^{\phi_0}_{a/m} \lesssim m[\hat{S}_1]^{obs}$. Thus we fix an index $i = 1, 2, \ldots$, we pick the messages $q' \in \mathcal{D}(\phi_{2(i-1)})$ and $\hat{q} \in \mathcal{D}(\phi_{2i-1})$, and we build the relation $\mathcal{R}_i(q', \hat{q})$ containing $(\, m[S''_i]^{\nu'_m} \mid \text{Top}^{\phi_{2(i-1)}}_{a/m},\ m[\hat{S}_i]^{obs}\,)$ along with its derivatives which may be generated when $m$ receives $\hat{q}$ from the attacker.

$$\begin{array}{rcl}
\mathcal{R}_i(q', \hat{q}) & \stackrel{\text{def}}{=} & \Big\{ \Big( m[S''_i]^{\nu'_m} \mid \text{Top}^{\phi_{2(i-1)}}_{a/m},\ m[\hat{S}_i]^{obs} \Big)\ , \\
& & \Big( m[S''_i]^{\nu'_m} \mid a[!\langle q' \rangle.\text{T}_{\phi_{2(i-1)}}]^{m},\ m[\hat{S}_i]^{obs} \Big)\ , \\
& & \Big( m[\sigma.P'']^{\nu'_m} \mid \text{Top}^{\phi_{2(i-1)}}_{a/m},\ m[\sigma.\lfloor \tau.\sigma.!\langle \mathsf{auth}_i \rangle.\mathsf{nil} \rfloor \hat{S}_{i+1}]^{obs} \Big)\ , \\
& & \Big( m[\sigma.P'']^{\nu'_m} \mid a[!\langle q' \rangle.\text{T}_{\phi_{2(i-1)}}]^{m},\ m[\sigma.\lfloor \tau.\sigma.!\langle \mathsf{auth}_i \rangle.\mathsf{nil} \rfloor \hat{S}_{i+1}]^{obs} \Big)\ , \\
& & \Big( m[P'']^{\nu'_m} \mid \text{Top}^{\phi_{2i-1}}_{a/m},\ m[\lfloor \tau.\sigma.!\langle \mathsf{auth}_i \rangle.\mathsf{nil} \rfloor \hat{S}_{i+1}]^{obs} \Big)\ , \\
& & \Big( m[P'']^{\nu'_m} \mid a[!\langle \hat{q} \rangle.\text{T}_{\phi_{2i-1}}]^{m},\ m[\lfloor \tau.\sigma.!\langle \mathsf{auth}_i \rangle.\mathsf{nil} \rfloor \hat{S}_{i+1}]^{obs} \Big)\ , \\
& & \Big( m[\{^{\hat{q}}/_q\}P^{1''}]^{\nu'_m} \mid \text{Top}^{\phi_{2i-1}}_{a/m},\ m[\lfloor \tau.\sigma.!\langle \mathsf{auth}_i \rangle.\mathsf{nil} \rfloor \hat{S}_{i+1}]^{obs} \Big)\ , \\
& & \Big( m[\{^{\hat{q}}/_q\}P^{1''}]^{\nu'_m} \mid a[!\langle \hat{q} \rangle.\text{T}_{\phi_{2i-1}}]^{m},\ m[\lfloor \tau.\sigma.!\langle \mathsf{auth}_i \rangle.\mathsf{nil} \rfloor \hat{S}_{i+1}]^{obs} \Big)\ , \\
& & \Big( m[!\langle \mathsf{auth}_i \rangle.\mathsf{nil}]^{\nu'_m} \mid \text{Top}^{\phi_{2i}}_{a/m},\ m[!\langle \mathsf{auth}_i \rangle.\mathsf{nil}]^{obs} \Big) \Big\}\ .
\end{array}$$

Moreover, it is straightforward to prove that there exists a simulation $\overline{\mathcal{R}}_i$ containing the pair $(\, m[!\langle \mathsf{auth}_i \rangle.\mathsf{nil}]^{\nu'_m} \mid \text{Top}^{\phi_{2i}}_{a/m},\ m[!\langle \mathsf{auth}_i \rangle.\mathsf{nil}]^{obs}\,)$.

Then we show that the required simulation is given by the following relation

$$\mathcal{R}_i \stackrel{\text{def}}{=} \bigcup_{i \geq 1} \Big( \overline{\mathcal{R}}_i \ \cup \bigcup_{\substack{q' \in \mathcal{D}(\phi_{2(i-1)}) \\ \hat{q} \in \mathcal{D}(\phi_{2i-1})}} \mathcal{R}_i(q', \hat{q}) \Big)\ .$$

As done for Lemma 4.4, we outline the most significant cases. Again, we omit input actions and internal choices of the attacker.

In the pair ( $m[S_i'']^{\nu'_m} \mid \textsc{Top}^{\phi_{2(i-1)}}_{a/m}$, $m[\hat{S}_i]^{obs}$ ) we have a significant action:

- $m[S_i'']^{\nu'_m} \mid \textsc{Top}^{\phi_{2(i-1)}}_{a/m} \xrightarrow{!\mathsf{hello}_i \triangleright obs} m[\sigma.P'']^{\nu'_m} \mid \textsc{Top}^{\phi_{2(i-1)}}_{a/m}$, where $m$ broadcasts the packet $\mathsf{hello}_i$. Then $m[\hat{S}_i]^{obs} \xRightarrow{!\mathsf{hello}_i \triangleright obs} m[\sigma.\lfloor\tau.\sigma.!\langle\mathsf{auth}_i\rangle.\mathsf{nil}\rfloor\hat{S}_{i+1}]^{obs}$.

In the pair ( $m[\sigma.P'']^{\nu'_m} \mid \textsc{Top}^{\phi_{2(i-1)}}_{a/m}, m[\sigma.\lfloor\tau.\sigma.!\langle\mathsf{auth}_i\rangle.\mathsf{nil}\rfloor\hat{S}_{i+1}]^{obs}$ ) we have a significant action:

- $m[\sigma.P'']^{\nu'_m} \mid \textsc{Top}^{\phi_{2(i-1)}}_{a/m} \xrightarrow{\sigma} m[P'']^{\nu'_m} \mid \textsc{Top}^{\phi_{2i-1}}_{a/m}$. Then $m[\sigma.\lfloor\tau.\sigma.!\langle\mathsf{auth}_i\rangle.\mathsf{nil}\rfloor\hat{S}_{i+1}]^{obs} \xRightarrow{\sigma} m[\lfloor\tau.\sigma.!\langle\mathsf{auth}_i\rangle.\mathsf{nil}\rfloor\hat{S}_{i+1}]^{obs}$.

In the pair ( $m[P'']^{\nu'_m} \mid \textsc{Top}^{\phi_{2i-1}}_{a/m}$, $m[\lfloor\tau.\sigma.!\langle\mathsf{auth}_i\rangle.\mathsf{nil}\rfloor\hat{S}_{i+1}]^{obs}$ ) we have a significant action:

- $m[P'']^{\nu'_m} \mid \textsc{Top}^{\phi_{2i-1}}_{a/m} \xrightarrow{\sigma} m[S''_{i+1}]^{\nu'_m} \mid \textsc{Top}^{\phi_{2i}}_{a/m}$, where $m$ does not receive anything and performs a timeout. Then $m[\lfloor\tau.\sigma.!\langle\mathsf{auth}_i\rangle.\mathsf{nil}\rfloor\hat{S}_{i+1}]^{obs} \xRightarrow{\sigma} m[\hat{S}_{i+1}]^{obs}$.

In the pair ( $m[P'']^{\nu'_m} \mid a[!\langle\hat{q}\rangle.\mathrm{T}_{\phi_{2i-1}}]^m$, $m[\lfloor\tau.\sigma.!\langle\mathsf{auth}_i\rangle.\mathsf{nil}\rfloor\hat{S}_{i+1}]^{obs}$ ) we consider two actions:

- $m[P'']^{\nu'_m} \mid a[!\langle\hat{q}\rangle.\mathrm{T}_{\phi_{2i-1}}]^m \xrightarrow{\tau} m[\{^{\hat{q}}/_{q}\}P^{1''}]^{\nu'_m} \mid \textsc{Top}^{\phi_{2i-1}}_{a/m}$, where $m$ receives $\hat{q}$. Then the second network replies $m[\lfloor\tau.\sigma.!\langle\mathsf{auth}_i\rangle.\mathsf{nil}\rfloor\hat{S}_{i+1}]^{obs} \Rightarrow m[\lfloor\tau.\sigma.!\langle\mathsf{auth}_i\rangle.\mathsf{nil}\rfloor\hat{S}_{i+1}]^{obs}$.

- $m[P'']^{\nu'_m} \mid a[!\langle\hat{q}\rangle.\mathrm{T}_{\phi_{2i-1}}]^m \xrightarrow{\tau} m[P'']^{\nu'_m} \mid \textsc{Top}^{\phi_{2i-1}}_{a/m}$, where $\hat{q}$ gets lost. Then the second network replies $m[\lfloor\tau.\sigma.!\langle\mathsf{auth}_i\rangle.\mathsf{nil}\rfloor\hat{S}_{i+1}]^{obs} \Rightarrow m[\lfloor\tau.\sigma.!\langle\mathsf{auth}_i\rangle.\mathsf{nil}\rfloor\hat{S}_{i+1}]^{obs}$.

In ( $m[\{^{\hat{q}}/_{q}\}P^{1''}]^{\nu'_m} \mid \textsc{Top}^{\phi_{2i-1}}_{a/m}$, $m[\lfloor\tau.\sigma.!\langle\mathsf{auth}_i\rangle.\mathsf{nil}\rfloor\hat{S}_{i+1}]^{obs}$ ) we have two significant actions:

- $m[\{^{\hat{q}}/_{q}\}P^{1''}]^{\nu'_m} \mid \textsc{Top}^{\phi_{2i-1}}_{a/m} \xrightarrow{\sigma} m[!\langle\mathsf{auth}_i\rangle.\mathsf{nil}]^{\nu'_m} \mid \textsc{Top}^{\phi_{2i}}_{a/m}$, where $m$ verifies that $\hat{q}$ refers to the nonce $n_i$. Then $m[\lfloor\tau.\sigma.!\langle\mathsf{auth}_i\rangle.\mathsf{nil}\rfloor\hat{S}_{i+1}]^{obs} \xRightarrow{\sigma} m[!\langle\mathsf{auth}_i\rangle.\mathsf{nil}]^{obs}$.

- $m[\{^{\hat{q}}/_{q}\}P^{1''}]^{\nu'_m} \mid \textsc{Top}^{\phi_{2i-1}}_{a/m} \xrightarrow{\sigma} m[S''_{i+1}]^{\nu'_m} \mid \textsc{Top}^{\phi_{2i}}_{a/m}$, where $m$ verifies that $\hat{q}$ does not refer to $n_i$, or it finds out that $\hat{q}$ is corrupted. Then $m[\lfloor\tau.\sigma.!\langle\mathsf{auth}_i\rangle.\mathsf{nil}\rfloor\hat{S}_{i+1}]^{obs} \xRightarrow{\sigma} m[\hat{S}_{i+1}]^{obs}$.

*Case 2: Receiver.* Similar to *Case 1* of Lemma 4.4 □

**Proof of Proposition 6.1** Similar to that of Proposition 4.1. □

**Proof of Theorem 6.2** The system $(\text{LiSP}')^{\mathcal{A}} \mid A$ admits the following computation:

$$
\begin{array}{lr}
m[\overline{Z}]^{\nu_m} \mid \textsc{kl}[L_0]^{\nu_{\textsc{kl}}} \mid A & \xrightarrow{!r \triangleright obs} \\
m[\sigma.\overline{W}]^{\nu_m} \mid \textsc{kl}[\sigma.\{^r/_r\}I_1]^{\nu_{\textsc{kl}}} \mid A & \xrightarrow{\sigma} \\
m[\overline{W}]^{\nu_m} \mid \textsc{kl}[\{^r/_r\}I_1]^{\nu_{\textsc{kl}}} \mid a[\sigma.\lfloor ?(x).\sigma.!\langle x\rangle.\mathsf{nil}\rfloor \mathsf{nil}]^{\nu_a} \mid b[\lfloor ?(y).\sigma.!\langle y\rangle.\mathsf{nil}\rfloor \mathsf{nil}]^{\nu_b} & \xrightarrow{!q_1 \triangleright obs} \\
m[\overline{W}]^{\nu_m} \mid \textsc{kl}[\sigma.L_1]^{\nu_{\textsc{kl}}} \mid a[\sigma.\lfloor ?(x).\sigma.!\langle x\rangle.\mathsf{nil}\rfloor \mathsf{nil}]^{\nu_a} \mid b[\sigma.!\langle q_1\rangle.\mathsf{nil}]^{\nu_b} & \xrightarrow{\sigma} \\
m[\overline{Z}]^{\nu_m} \mid \textsc{kl}[L_1]^{\nu_{\textsc{kl}}} \mid a[\lfloor ?(x).\sigma.!\langle x\rangle.\mathsf{nil}\rfloor \mathsf{nil}]^{\nu_a} \mid b[!\langle q_1\rangle.\mathsf{nil}]^{\nu_b} & \xrightarrow{\tau} \\
m[\overline{Z}]^{\nu_m} \mid \textsc{kl}[\sigma.\{^{q_1}/_r\}I_2]^{\nu_{\textsc{kl}}} \mid a[\sigma.!\langle q_1\rangle.\mathsf{nil}]^{\nu_a} \mid b[\mathsf{nil}]^{\nu_b} & \xrightarrow{!r \triangleright obs} \\
m[\sigma.\overline{W}]^{\nu_m} \mid \textsc{kl}[\sigma.\{^{q_1}/_r\}I_2]^{\nu_{\textsc{kl}}} \mid a[\sigma.!\langle q_1\rangle.\mathsf{nil}]^{\nu_a} \mid b[\mathsf{nil}]^{\nu_b} & \xrightarrow{\sigma} \\
m[\overline{W}]^{\nu_m} \mid \textsc{kl}[\{^{q_1}/_r\}I_2]^{\nu_{\textsc{kl}}} \mid a[!\langle q_1\rangle.\mathsf{nil}]^{\nu_a} \mid b[\mathsf{nil}]^{\nu_b} & \xrightarrow{\tau} \\
m[\sigma.\{^{q_1}/_q\}\overline{T^5}]^{\nu_m} \mid \textsc{kl}[\{^{q_1}/_r\}I_2]^{\nu_{\textsc{kl}}} \mid a[\mathsf{nil}]^{\nu_a} \mid b[\mathsf{nil}]^{\nu_b} & \xrightarrow{\sigma} \\
m[!\langle \mathsf{auth}_1\rangle.\sigma.R(k_2, k_{s+1}, s-1)]^{\nu_m} \mid \textsc{kl}[L_2]^{\nu_{\textsc{kl}}} \mid a[\mathsf{nil}]^{\nu_a} \mid b[\mathsf{nil}]^{\nu_b} & \xrightarrow{!\mathsf{auth}_1 \triangleright obs}
\end{array}
$$

Then $m$ signals the correct reconfiguration based on an old packet. □